\documentclass{rmf-d}
\usepackage{nopageno,rmfbib,multicol,times,epsf,amsmath,amssymb,cite}
\usepackage[T1]{fontenc} %Especial para español (for spanish)
\usepackage[]{caption2}
\usepackage{graphicx}
\usepackage{blindtext}
\usepackage{xcolor}
\usepackage{hyperref}
\usepackage{listings}
\usepackage{float}
\newcommand{\BiBi}         {\mbox{Bi+Bi}\xspace}

\newcommand{\AuAu}         {\mbox{Au+Au}\xspace}

\newcommand{\snn}          {\ensuremath{\sqrt{s_{\mathrm{NN}}}}\xspace}
\newcommand{\pT}           {\ensuremath{p_{\rm T}}\xspace}
\newcommand{\mT}           {\ensuremath{m_{\rm T}}\xspace}
\newcommand{\kT}           {\ensuremath{k_{\rm T}}\xspace}

\newcommand{\dndy}         {\ensuremath{\mathrm{d}N_\mathrm{ch}/\mathrm{d}y}\xspace}

\newcommand{\Npart}        {\ensuremath{N_\mathrm{part}}\xspace}
\newcommand{\Ncoll}        {\ensuremath{N_\mathrm{coll}}\xspace}
\newcommand{\dEdx}         {\ensuremath{\textrm{d}E/\textrm{d}x}\xspace}

\newcommand{\Qn}           {\ensuremath{Q_{\rm n}}\xspace}
\newcommand{\vn}           {\ensuremath{v_{\rm n}}\xspace}
\newcommand{\Nev}          {\ensuremath{N_\mathrm{ev}}\xspace}

\newcommand{\VZ}           {\ensuremath{V^{\rm 0}}\xspace}

\newcommand{\nineH}        {$\sqrt{s}~=~0.9$~Te\kern-.1emV\xspace}
\newcommand{\seven}        {$\sqrt{s}~=~7$~Te\kern-.1emV\xspace}
\newcommand{\twoH}         {$\sqrt{s}~=~0.2$~Te\kern-.1emV\xspace}
\newcommand{\twosevensix}  {$\sqrt{s}~=~2.76$~Te\kern-.1emV\xspace}
\newcommand{\five}         {$\sqrt{s}~=~5.02$~Te\kern-.1emV\xspace}
\newcommand{\twosevensixnn}{$\sqrt{s_{\mathrm{NN}}}~=~2.76$~Te\kern-.1emV\xspace}
\newcommand{\fivenn}       {$\sqrt{s_{\mathrm{NN}}}~=~5.02$~Te\kern-.1emV\xspace}

\newcommand{\GeVc}         {Ge\kern-.1emV/$c$\xspace}
\newcommand{\GeVcsq}       {Ge\kern-.1emV/$c^2$\xspace}
\newcommand{\MeVc}         {Me\kern-.1emV/$c$\xspace}
\newcommand{\TeV}          {Te\kern-.1emV\xspace}
\newcommand{\GeV}          {Ge\kern-.1emV\xspace}
\newcommand{\MeV}          {Me\kern-.1emV\xspace}
\newcommand{\GeVmass}      {Ge\kern-.2emV/$c^2$\xspace}
\newcommand{\MeVmass}      {Me\kern-.2emV/$c^2$\xspace}

\newcommand{\MPD}          {\rm{MPD}\xspace}
\newcommand{\FFD}          {\rm{FFD}\xspace}
\newcommand{\FHCAL}        {\rm{FHCAL}\xspace}
\newcommand{\TOF}          {\rm{TOF}\xspace}
\newcommand{\TPC}          {\rm{TPC}\xspace}
\newcommand{\ECAL}          {\rm{ECAL}\xspace}

\newcommand{\pip}          {\ensuremath{\pi^{+}}\xspace}
\newcommand{\pim}          {\ensuremath{\pi^{-}}\xspace}
\newcommand{\pipm}          {\ensuremath{\pi^{\pm}}\xspace}
\newcommand{\kap}          {\ensuremath{\rm{K}^{+}}\xspace}
\newcommand{\kam}          {\ensuremath{\rm{K}^{-}}\xspace}
\newcommand{\kapm}          {\ensuremath{\rm{K}^{\pm}}\xspace}

\newcommand{\lmb}          {\ensuremath{\Lambda}\xspace}
\newcommand{\almb}         {\ensuremath{\overline{\Lambda}}\xspace}

\newcommand{\sigmaZ}       {\ensuremath{\sigma_\mathrm{z, vertex}}} 
\newcommand{\zvtx}         {\ensuremath{z_\mathrm{vertex}}} 
\newcommand{\tffde}        {\ensuremath{t_\mathrm{FFD}^\mathrm{E}}} 
\newcommand{\tffdw}        {\ensuremath{t_\mathrm{FFD}^\mathrm{W}}} 
\newcommand{\RCPV}         {\ensuremath{R_\mathrm{CPV}}} 

\def\rmfcornisa{ }
\def\rmfcintilla{ }

\clearpage \rmfcaptionstyle 
\begin{document}
%\markboth{ RMF Editorial Team    }{}% A \LaTeX template for the RMF, RMF-E, SRMF }

%%%%%
%
% Please provide the following information
%
%%%%%
\title{MPD physics performance studies in \BiBi collisions at $\snn=9.2$ GeV
\vspace{-6pt}}

\newcommand*{\aanlarm}{A.I.~Alikhanyan National Lab of Armenia, Yerevan, ARMENIA}
\newcommand*{\aanlarmindex}{1}

\newcommand*{\bnrrus}{Belgorod National Research University, Belgorod, RUSSIA}
\newcommand*{\bnrrusindex}{2}

\newcommand*{\uammex}{Departamento de F\'isica, Universidad Aut\'onoma Metropolitana-Iztapalapa, Mexico City, MEXICO}
\newcommand*{\uammexindex}{3}

\newcommand*{\umsnhdpfiemex}{Divisi\'on de Posgrado, Facultad de Ingenier\'\i a El\'ectrica, Universidad Michoacana de San Nicol\'as de Hidalgo, Morelia, MEXICO}
\newcommand*{\umsnhdpfiemexindex}{4}

\newcommand*{\udcmex}{Facultad de Ciencias - CUICBAS, Universidad de Colima, Colima, MEXICO}
\newcommand*{\udcmexindex}{5}

\newcommand*{\uadsmex}{Facultad de Ciencias F\'i{}sico-Matem\'a{}ticas, Universidad Aut\'o{}noma de Sinaloa, Culiac\'a{}n, MEXICO}
\newcommand*{\uadsmexindex}{6}

\newcommand*{\fuchn}{Fudan University, Shanghai, CHINA}
\newcommand*{\fuchnindex}{7}

\newcommand*{\hseurus}{HSE University, Moscow, RUSSIA}
\newcommand{\hseurusindex}{8}

\newcommand*{\huchn}{Huzhou University, Huzhou, CHINA}
\newcommand*{\huchnindex}{9}

\newcommand*{\inrrus}{Institute for Nuclear Research of the Russian Academy of Sciences, Moscow, RUSSIA}
\newcommand*{\inrrusindex}{10}

\newcommand*{\impchn}{Institute of Modern Physics of the Chinese Academy of Science, Langzhou, CHINA}
\newcommand*{\impchnindex}{11}

\newcommand*{\iptsukaz}{Institute of Physics and Technology, Satbayev University, Almaty, KAZAKHSTAN}
\newcommand{\iptsukazindex}{12}

\newcommand*{\ipenasbelarus}{Institute of Power Engineering of the National Academy of Sciences of Belarus, Minsk, BELARUS}
\newcommand{\ipenasbelarusindex}{13}

\newcommand*{\icnmex}{Instituto de Ciencias Nucleares, Universidad Nacional Autónoma de México, Mexico City, MEXICO}
\newcommand*{\icnmexindex}{14}

\newcommand*{\umsnhifmmex}{Instituto de F\'\i sica y Matemáticas, Universidad Michoacana de San Nicol\'as de Hidalgo, Morelia, MEXICO}
\newcommand*{\umsnhifmmexindex}{15}

\newcommand*{\jinrru}{Joint Institute for Nuclear Research, Dubna, RUSSIA}
\newcommand*{\jinrruindex}{16}

\newcommand*{\ccnuchn}{Key Laboratory of Quark and Lepton Physics of the Ministry of Education, Central China Normal University, Wuhan, CHINA}
\newcommand*{\ccnuchnindex}{17}

\newcommand*{\miptrus}{Moscow Institute of Physics and Technology, Moscow, RUSSIA}
\newcommand*{\miptrusindex}{18}

\newcommand*{\nrcmoscowrus}{National Research Center \lq\lq Kurchatov Institute", Moscow, RUSSIA}
\newcommand*{\nrcmoscowrusindex}{19}

\newcommand*{\nrcrus}{Petersburg Nuclear Physics Institute named by B.P. Konstantinov of NRC "Kurchatov Institute", Gatchina, RUSSIA}
\newcommand*{\nrcrusindex}{20}

\newcommand*{\mephirus}{National Research Nuclear University MEPhI (Moscow Engineering Physics Institute), Moscow, RUSSIA}
\newcommand*{\mephirusindex}{21}

\newcommand*{\nosurus}{North Ossetian State University, Vladikavkaz, RUSSIA}
\newcommand*{\nosurusindex}{22}

\newcommand*{\pjsslov}{Pavol Jozef \v{S}af\'arik University, Ko\v{s}ice, SLOVAKIA}
\newcommand*{\pjsslovindex}{23}

\newcommand*{\ptgppurus}{Peter the Great St. Petersburg Polytecnic University, Saint Petersburg, RUSSIA}
\newcommand*{\ptgppurusindex}{24}

\newcommand*{\pleruerus}{Plekhanov Russian University of Economics, Moscow, RUSSIA}
\newcommand*{\pleruerusindex}{25}

\newcommand*{\uopbul}{Plovdiv University Paisii Hilendarski, Plovdiv, BULGARIA}
\newcommand*{\uopbulindex}{26}

\newcommand*{\inpbsubelarus}{Research Institute for Nuclear Problems of Belarusian State University (INP BSU), Minsk, BELARUS}
\newcommand*{\inpbsubelarusindex}{27}

\newcommand*{\spsurus}{Saint Petersburg State University, Saint Petersburg, RUSSIA}
\newcommand*{\spsurusindex}{28}

\newcommand*{\ihepchn}{School of Nuclear Science and Technology, University of the Chinese Academy of Sciences, Beijing, CHINA}
\newcommand*{\ihepchnindex}{29}

\newcommand*{\shuchn}{Shandong University, Qingdao, CHINA}
\newcommand*{\shuchnindex}{30}

\newcommand*{\msusinprus}{Skobeltsyn Institute of Nuclear Physics (SINP) of the Moscow State University, Moscow, RUSSIA}
\newcommand*{\msusinprusindex}{31}

\newcommand*{\sosnyrus}{State Scientific Institution \lq\lq Joint Institute for Energy and Nuclear Research – Sosny” of the National Academy of Sciences of Belarus, Minsk, BELARUS}
\newcommand{\sosnyrusindex}{32}

\newcommand*{\tguchn}{Three Gorges University, Yichang, CHINA}
\newcommand*{\tguchnindex}{33}

\newcommand*{\tuchn}{Tsinghua University, Beijing, CHINA}
\newcommand*{\tuchnindex}{34}

\newcommand*{\ustuchn}{University of Science and Technology of China, Hefei, CHINA}
\newcommand*{\ustuchnindex}{35}

\newcommand*{\uscchn}{University of South China, Hengyang, CHINA}
\newcommand*{\uscchnindex}{36}

\newcommand*{\vincaserbia}{\lq\lq VIN\v{C}A" Institute of Nuclear Science - National Institute of the Republic of Serbia, University of Belgrade, SERBIA}
\newcommand*{\vincaserbiaindex}{37}

%\author{The MPD Collaboration}
\author{R.~Abdulin$^{\nrcrusindex}$,
V.~Abgaryan$^{\aanlarmindex}$,
R.~Adhikary$^{\jinrruindex}$,
K.G.~Afanaciev$^{\jinrruindex,\ipenasbelarusindex}$,
S.V.~Afanaciev$^{\jinrruindex}$,
G.~Agakishiev$^{\jinrruindex}$,
E.I.~Alexandrov$^{\jinrruindex}$,
I.N.~Alexandrov$^{\jinrruindex}$,
M. Alvarado-Hern\'andez$^{\icnmexindex}$, 
D.I.~Andreev$^{\jinrruindex}$
% ...
}
\address{}

\address{$^{\aanlarmindex}$\aanlarm}

\address{$^{\bnrrusindex}$\bnrrus}

\address{$^{\uammexindex}$\uammex}

\address{$^{\umsnhdpfiemexindex}$\umsnhdpfiemex}

\address{$^{\udcmexindex}$\udcmex}

\address{$^{\uadsmexindex}$\uadsmex}

\address{$^{\fuchnindex}$\fuchn}

\address{$^{\hseurusindex}$\hseurus}

\address{$^{\huchnindex}$\huchn}

\address{$^{\inrrusindex}$\inrrus}

\address{$^{\impchnindex}$\impchn}

\address{$^{\iptsukazindex}$\iptsukaz}

\address{$^{\ipenasbelarusindex}$\ipenasbelarus}

\address{$^{\icnmexindex}$\icnmex}

\address{$^{\umsnhifmmexindex}$\umsnhifmmex}

\address{$^{\jinrruindex}$\jinrru}

\address{$^{\ccnuchnindex}$\ccnuchn}

\address{$^{\miptrusindex}$\miptrus}

\address{$^{\nrcmoscowrusindex}$\nrcmoscowrus}

\address{$^{\nrcrusindex}$\nrcrus}

\address{$^{\mephirusindex}$\mephirus}

\address{$^{\nosurusindex}$\nosurus}

\address{$^{\pjsslovindex}$\pjsslov}

\address{$^{\ptgppurusindex}$\ptgppurus}

\address{$^{\pleruerusindex}$\pleruerus}

\address{$^{\uopbulindex}$\uopbul}

\address{$^{\inpbsubelarusindex}$\inpbsubelarus}

\address{$^{\spsurusindex}$\spsurus}

\address{$^{\ihepchnindex}$\ihepchn}

\address{$^{\shuchnindex}$\shuchn}

\address{$^{\msusinprusindex}$\msusinprus}

\address{$^{\sosnyrusindex}$\sosnyrus}

\address{$^{\tguchnindex}$\tguchn}

\address{$^{\tuchnindex}$\tuchn}

\address{$^{\ustuchnindex}$\ustuchn}

\address{$^{\uscchnindex}$\uscchn}

\address{$^{\vincaserbiaindex}$\vincaserbia}

\maketitle
%%%%%
%
% To be filled by the Editorial Team of RMF, RMF-E 
% and SRMF
%
%%%%%

%\tableofcontents

%\recibido{}{
%\vspace{-12pt}}
\section*{\centering{Abstract}}
\begin{abstract}
\vspace{-0.2cm} 
The Multi-Purpose Detector (MPD) is one of the three experiments of the Nuclotron Ion Collider-fAcility (NICA) complex, which is currently under construction at the Joint Institute for Nuclear Research in Dubna. With collisions of heavy ions in the collider mode, the MPD will cover the energy range $\snn=4-11$~GeV to scan the high baryon-density region of the QCD phase diagram. 
With expected statistics of 50--100 million events collected during the first run, MPD will be able to study a number of observables, including measurements of light hadrons and (hyper)nuclei production, particle flow, correlations and fluctuations, have a first look at dielectron production, and modification of vector-meson properties in dense matter. 
In this paper, we present selected results of the physics feasibility studies for the \MPD experiment in \BiBi collisions at $\snn=9.2$~GeV, the system considered as one of the first available at the NICA collider. 
%\vspace{1em}
\end{abstract}
\keys{ \bf{\textit{Heavy-ion collision experiments, Quark-Gluon matter}} 
\vspace{-4pt}}
\begin{multicols}{2}

\section{Introduction}
Heavy-ion collisions have been used to study QCD matter under extreme conditions of high temperatures and baryon densities for over 30 years. The main goal of this research has been to better understand the rich structure of the QCD phase diagram and to search for the phase transition into a new state of matter, the quark-gluon plasma (QGP), and the existence of a Critical End-Point (CEP)~\cite{CBM:2016kpk,STAR:2005gfr,Luo:2020pef}.  The research program started in the late 80s at the AGS ($\snn \sim 5$~\GeV) and the SPS ($\snn \sim 17$~\GeV). It was followed later by detailed studies of the hot matter at much higher energies at RHIC (up to $\snn = 200$~\GeV) and LHC (up to $\snn = 5$~\TeV). All these studies revealed the existence of a transition from hadronic matter to a QGP at a temperature $T_\mathrm{c} \sim 160$~\MeV and near-zero net baryon densities, which is consistent with the lattice QCD predictions of a cross-over transition~\cite{Borsanyi:2013bia}.
 
%The first evidence for the formation of a new state of matter, a quark-gluon plasma (QGP), was found in heavy-ion collisions at SPS ($\snn \sim 17$~\GeV)~\cite{PRESSCUT-2000-210,Jacak:2000sc}. Later, a detailed study of the properties of hot matter was continued in heavy-ion experiments at RHIC~\cite{Akiba:2015jwa} and LHC~\cite{ALICE:2022wpn} energies  ($\snn \sim 100-1000$~\GeV). 

%These studies revealed the existence of a crossover transition from hadronic matter to a QGP at a temperature $T_\mathrm{c} \sim 160$~\MeV and near-zero net baryon densities, which is consistent with the predictions of lattice QCD~\cite{Borsanyi:2013bia}.

%The state of matter produced in such collisions is believed to exist during the first microseconds after the Big Bang.

Heavy-ion collisions at lower energies ($\snn = 2-10$~\GeV) provide the means to study a different region of the QCD phase diagram, which is characterized by lower temperatures but higher net baryon densities. Models predict that a first-order phase transition and a CEP may exist under such conditions, which remain to be proven experimentally~\cite{Stephanov:1998dy}. 
%Similar conditions may exist in the modern Universe in the cores of the compact neutron stars and in neutron star mergers. 
The Beam Energy Scan (BES) programs carried out by the STAR experiment at RHIC, the NA61 experiment at SPS, the BM@N experiment at the Nuclotron and the HADES experiment at SIS18, studied the corresponding region of the QCD phase diagram \cite{Chen:2024aom,Gazdzicki:2008kk,Luo:2022mtp}. So far, there is no evidence of the CEP nor signs of the first-order phase transition were observed in these experiments.

%The Nuclotron Ion Collider-fAcility (NICA), which is in the final stage of construction at the Joint Institute for Nuclear Research (JINR) in Dubna, Russia, will provide an excellent opportunity to extend these studies to the range of energies $\sqrt{s_{NN}} = 2.4 - 11$~\GeV by providing high-luminosity scans both in collision energy and in system size~\cite{Golovatyuk:2019rkb}. The Multi-Purpose Detector (\MPD) is a collider experiment at NICA to study heavy-ion collisions in the energy range $\snn = 4-11$~\GeV. 

The Multi-Purpose Detector (\MPD) at the Nuclotron Ion Collider-fAcility (NICA), which is in the final stage of construction at the Joint Institute for Nuclear Research (JINR) in Dubna, Russia, will provide an excellent opportunity to extend these studies to the range of energies $\sqrt{s_{NN}} = 4 - 11$~\GeV by providing high-luminosity scans both in collision energy and in system size~\cite{Golovatyuk:2019rkb}. The search for the phase transition and CEP will be done by measuring a wide variety of observables, including production of light-flavor hadrons and     
\begin{figure}[H]
\centering
\includegraphics[width=0.9\linewidth]{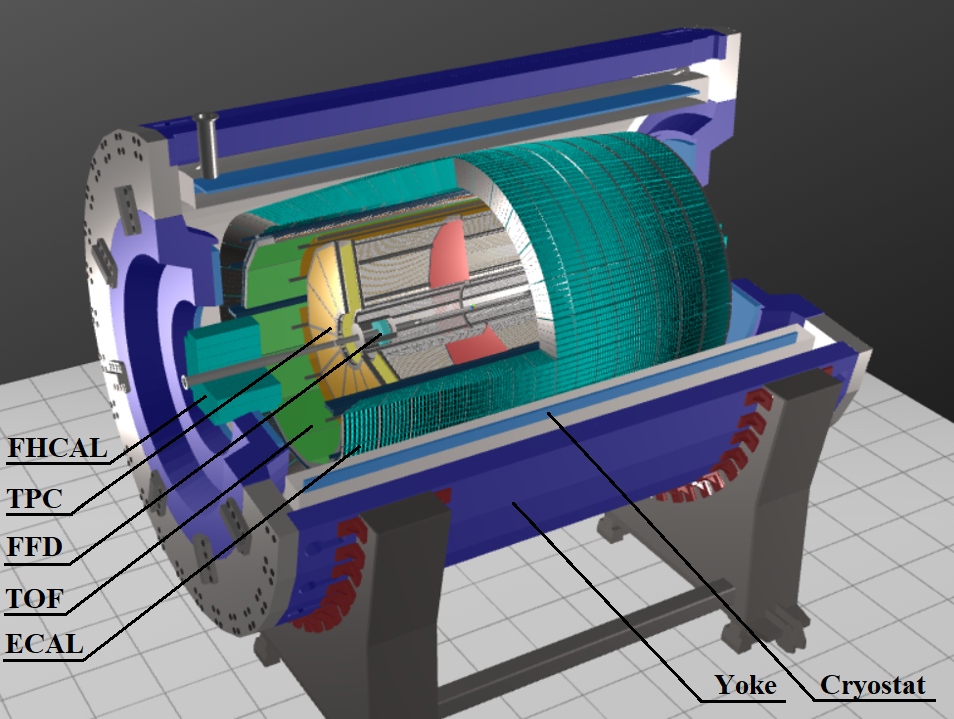}
\caption{Schematic view of the \MPD apparatus in the first stage of operation. The central barrel subsystems from inside to outside: \TPC, \TOF and \ECAL and the forward subsystems: \FFD  and \FHCAL.\label{fig:fig01}}
\end{figure} 
\noindent (hyper)nuclei, electromagnetic probes such as (direct) photons and dielectrons, by studying the particle flow, correlations and fluctuations. 
%The \MPD is well equipped to measure a wide variety of observables, including production of light-flavor hadrons and (hyper)nuclei, electromagnetic probes such as (direct) photons and dielectrons, and to study particle flow, correlations and fluctuations. 
First tests with a beam at the NICA collider are expected to start in the summer of 2025.
%The commissioning of the NICA collider is expected to start at the end of 2025. 
\BiBi collisions at $\snn = 9.2$~\GeV are among the first systems to be studied in the NICA collider. The choice of nuclei is determined by the ion source capabilities in the initial configuration. The energy was picked to be close to one of the energies studied in \AuAu collisions by the STAR experiment during the BES program to provide some basic comparison. 

In this paper, we present selected results of physics feasibility studies for the \MPD experiment in \BiBi collisions at $\snn = 9.2$~\GeV with a focus on observables that will become available with 50--100 M collected events. The paper is organized as follows: In Sec.~\ref{sec:detector}, we briefly describe the setup of the \MPD experiment. In Sec.~\ref{sec:framework}, we describe the data analysis framework, which was used to produce the presented results. In Sec.~\ref{sec:global}, we discuss the global characterization of heavy-ion collisions and in Sec.~\ref{sec:physics}, we present physics feasibility and performance studies for selected physics observables that can be carried out with the MPD in the first run. A summary is provided in Sec.~\ref{sec:conclusions}.

\section{MPD setup}
\label{sec:detector}

The design of the MPD experimental setup and preliminary results of the \MPD performance with heavy-ion beams have been published in~\cite{MPD:2022qhn}. The MPD is designed as a magnetic spectrometer capable of measuring and identifying charged hadrons, electrons, and photons over a wide range of momentum and rapidity.  In this section, we give a short description of the first stage set-up of the MPD~\cite{MPD:TDR}. A schematic view of the MPD is shown in Figure~\ref{fig:fig01}. The superconducting magnet generates a magnetic field up to  $B = 0.57$~T with a nominal field for regular operation of $B = 0.5$~T. Reduced and reversed-field runs are also expected to provide a better coverage for lower-momentum particles and systematic studies, respectively. 

The central barrel detectors are mounted inside the magnet, cover full
%$|\Delta\varphi| < 2\pi$ in 
azimuthal angle and a pseudorapidity range $|\eta| < 1.5$. A detailed description of the MPD is presented in Ref.~\cite{MPD:2022qhn}.

\begin{figure*}[t]
\centering
\includegraphics[width=0.49\textwidth]{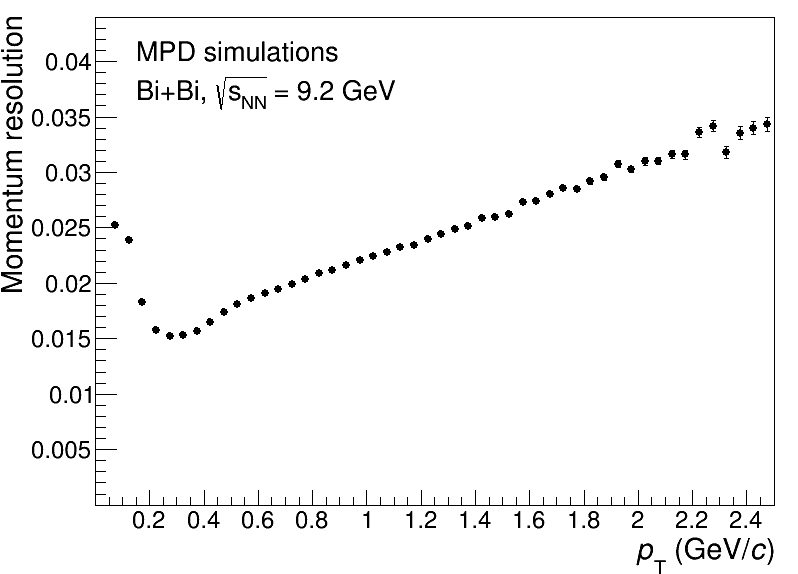}
\hfill
\includegraphics[width=0.49\textwidth]{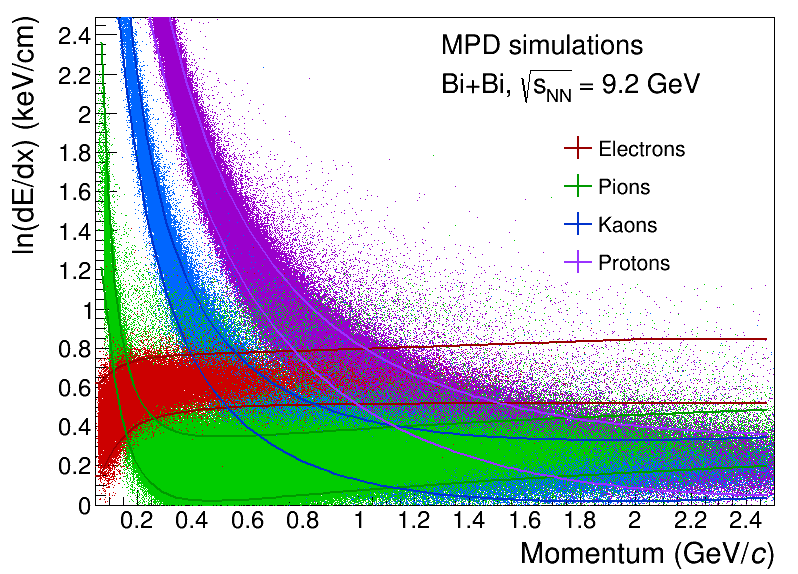}
\caption{Left: momentum resolution for primary charged particles reconstructed in the \TPC with number of points $n_\mathrm{hits}^{\TPC} > 20$. Right: \dEdx signals for primary charged particles reconstructed in the \TPC with number of points $n_\mathrm{hits}^{\TPC} > 20$. The bands of different colors correspond to 2$\sigma_{\TPC}$ selections for electrons, pions, kaons and (anti)protons. Simulation results are shown for \BiBi collisions at $\snn = 9.2$~GeV.\label{fig:TPC}}
\end{figure*}

The trajectories and momenta of charged particles are measured in a large volume Time Projection Chamber (\TPC). 
%The \TPC covers the full azimuthal angle $|\Delta\varphi| < 2\pi$
%and pseudorapidity $|\Delta\eta| < 1.6$. 
The \TPC also provides particle identification by measuring their energy loss (\dEdx) in the operational gas (90\% Ar and 10\% CH$_4$), with a typical resolution of $\sim$ 6.5\% achieved in heavy-ion collisions. Up to 53 points are measured along the track trajectory to provide reliable momentum reconstruction and particle identification. The left panel of Figure~\ref{fig:TPC} shows the momentum resolution for primary particles with more than 20 measured points in the \TPC. In a wide momentum range, the resolution is $\sim 2 - 3 \%$, deteriorating at lower momentum due to multiple scattering and at higher momentum due to limited spatial resolution. The right panel of the same figure shows the distribution of $\dEdx$ signals reconstructed for charged particles as a function of momentum, where one can identify bands corresponding to electrons, pions, kaons and protons. The solid curves show the $\pm2\sigma_{\TPC}$ selections for different particle species. The \TPC provides $\pi$/K and K/p separations within $2\sigma$ in the momentum range up to 0.7 \GeVc and 1.2 \GeVc, respectively.

\begin{figure*}[b]
\centering
\includegraphics[width=0.49\textwidth]{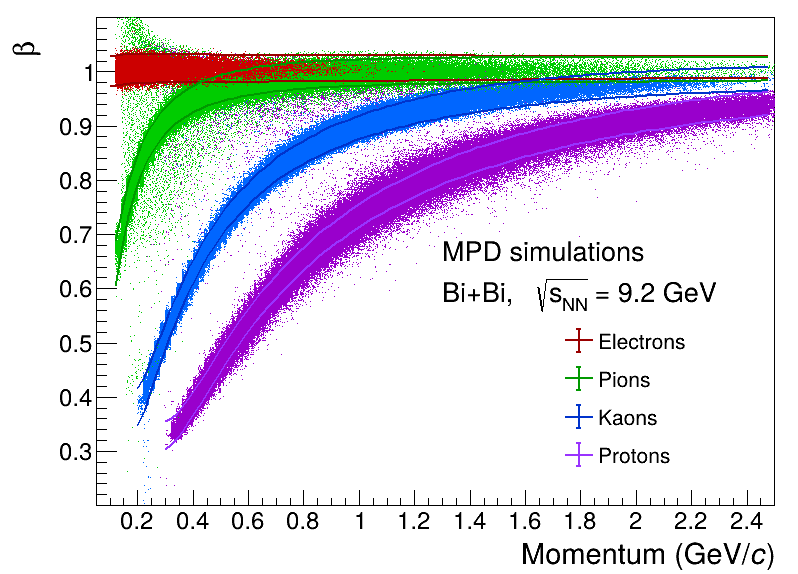}
\hfill
\includegraphics[width=0.49\textwidth]{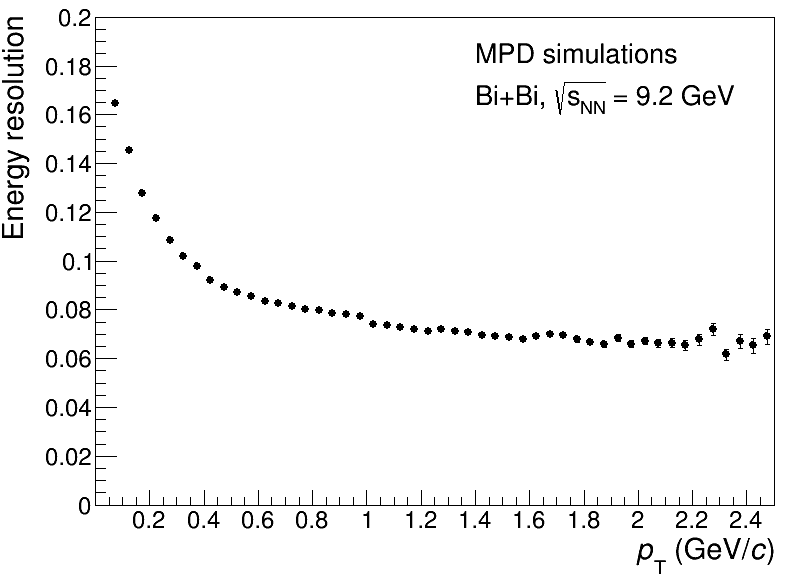}
\caption{Left: particle velocities evaluated using combined measurements of momentum and track length in the \TPC and time-of-flight in the \TOF. The bands of different colors correspond to 2$\sigma_{\TOF}$ selections for electrons, pions, kaons and (anti)protons. Right: energy resolution of the \ECAL for primary photons. Results are shown for \BiBi collisions at $\snn = 9.2$~GeV.\label{fig:TOFECAL}}
\end{figure*}

A wall of \TOF detectors follows the \TPC in radius and consists of 28 modules (14 modules in $\varphi$ and two modules in $z$-direction), each made of 10 Multi-gap Resistive Plate Chambers (MRPC). 
%It covers full azimuthal angle and pseudorapidity range $|\Delta\eta| < 1.4$.
The \TOF detector provides time-of-flight measurements for charged particles  with a typical resolution of $\sim$ 80 ps. Along with the momentum and track-length measurements in the \TPC, it provides particle separation by mass$^2$ or velocity $\beta$, as shown in the left panel of Figure~\ref{fig:TOFECAL}. The \TOF detector extends the particle identification capabilities of the \TPC to higher momenta, providing $2\sigma$ separation of $\pi$/K and K/p up to 1.5 \GeVc and 2.5 \GeVc, respectively. Only pion (proton) tracks with transverse momentum $\pT > 150\, (350)$ \MeVc can reach the \TOF at the nominal magnetic field. At lower momenta, charged particle identification is only possible with the \TPC.

The electromagnetic calorimeter (\ECAL) is the outermost detector, consisting of 38,400 shashlyk-type towers packed into 50 half-sectors (25 half-sectors in $\varphi$ and 2 half-sectors in the $z$-direction). It spans full azimuthal angle and $|\eta| < 1.4$ in pseudorapidity. It is built with projective geometry, i.e., the tower orientation varies in the $z$-direction to ensure that the towers point approximately to the nominal interaction point (IP). 
The projective geometry of the \ECAL is important for efficient registration of low-energy showers, which are the majority at NICA energies. 
The \ECAL energy resolution estimated for photons in heavy-ion collisions is shown in the right panel of Figure~\ref{fig:TOFECAL}. It is defined by the intrinsic resolution of the detector and is degraded by the cluster reconstruction procedure, which takes care of the shower reconstruction and of splitting of merged showers in high-multiplicity events. The electromagnetic calorimeter is the primary detector for measuring photons. It also helps to identify electrons at higher momenta, where \TPC and \TOF become less effective, by requiring the $E/p$ ratio to be close to unity, and where $E$ and $p$ are the measured electron energy and momentum, respectively.

The \MPD is also equipped with two forward detectors for event triggering, measurement of event starting time ($t_{0}$) and estimation of collision centrality and geometry. The Fast Forward Detector (\FFD) consists of two identical detectors located at $\pm 140$ cm from the nominal interaction point (IP). The detector covers full azimuthal angle and $2.9 < |\eta| < 3.3$ in pseudorapidity. Each FFD consists of 80 \v{C}erenkov quartz counters surrounding the beam pipe. Each counter has a 1 cm thick lead radiator to induce showers from photons produced in $\pi^{0}$-decays. Besides the photons, the FFD detects fast charged particles. The time resolution of each counter is $\sim 50$ ps. By measuring the arrival times of the fastest particles (photons for most of the time) in the two arms ($\tffde$ and $\tffdw$), one can determine the event starting time and the event vertex 
\begin{eqnarray}
t_{0}^{\rm FFD} &=& (t^{\rm E}_{\rm FFD}+t^{\rm W}_{\rm FFD})/2- L/c\nonumber\\
z_{\rm vertex}^{\rm FFD} &=& c(t_{\rm FFD}^{\rm E}-t_{\rm FFD}^{\rm W})/2,
\end{eqnarray}
respectively,
%(\tffde + \tffdw ) /2  –  L/c
%and the event vertex $\zvtx^{\FFD}$ = $c$ ( $\tffde$ - $\tffdw$ ) /2,
where $L$ is the distance from the nominal IP to the \FFD along the beam axis and $c$ is the speed of light. The $t_{0}^{\FFD}$ resolution of the \FFD depends on the number of channels $N$ fired on each side by fast particles and is better than 50/$\sqrt{N}$ ps. However, the measured time resolution  degrades to $\sim$ 70 ps in peripheral events due to the spread in the arrival times of incoming particles and becomes comparable to the \TOF time resolution. The vertex resolution varies from 0.5 to 2~cm from central to peripheral collision, respectively. 

The Forward Hadron Calorimeters (\FHCAL) are designed to measure fragments produced in the forward direction. They are located at a distance of $\pm$ 3.5 m from the nominal IP and cover 2$\pi$ in 
azimuthal angle and $2 < |\eta| < 5$ in pseudorapidity. Each \FHCAL calorimeter consists of 44 towers with a transverse size of $15\times 15$ cm$^2$, covering in total about 1 m$^2$. Similar to the \FFD, the \FHCAL can provide the start time and the event vertex position of each event. The typical time resolution of \FHCAL modules is $\sim 1$ ns, making the $t_{0}^{\FHCAL}$ resolution inferior to that of the \TOF and \ECAL. Due to the hole occupied by the beam pipe, a significant part of the fragments escape detection, resulting in an ambiguity between the measured energy deposition and the event centrality. Various methods are being developed to resolve this ambiguity and to relate the measured energy deposition to the event centrality. The \FHCAL is mainly used for event plane measurements at forward rapidity.

%By the end of 2025, the \MPD is expected to be fully assembled and ready for commissioning with cosmic rays and beams at the NICA collider in accordance with the accelerator construction schedule. 
Collisions of \BiBi at $\snn = 9.2$~\GeV are proposed as one of the first systems to be studied at NICA. The collider luminosity at start-up is expected to be two orders of magnitude lower than the nominal one, corresponding to an event rate of $\sim$ 50~Hz. With a realistic estimate of the first run duration, we may expect around 50-100 million collected events. Due to the incomplete optics of the collider rings, the vertex distribution in the \MPD interaction region will be quite broad along the beam direction with $\sigmaZ \sim$ 50~cm. This poses challenges for the trigger system and effective track reconstruction, but at the same time provides access to a wider rapidity coverage of the detector.

\section{Data analysis framework}
\label{sec:framework}
Physics feasibility studies were carried out using centralized Monte Carlo (MC) productions (listed in Table~\ref{EvGens}) to ensure consistency of the results obtained by different groups and to provide a test of the existing computing and software infra\-structure in preparation for real data analysis. Despite limited statistics, these productions are used to address a large number of observables using realistic data analysis techniques. A centralized data analysis framework, the so-called Data Analysis Train, was developed and implemented to process the simulated data samples with minimal load on disks, network, and CPU resources.

\subsection{Event generators and centralized productions} 
A list of MC productions for physics feasibility studies is presented in Table~\ref{EvGens}.  Various event generators, such as the cascade version of UrQMD~\cite{Bleicher:1999xi,Bass:1998ca}, the fragmentation model DCM-QGSM-SMM~\cite{Baznat:2019iom}, the microscopic transport model PHQMD~\cite{Aichelin:2019tnk}, hybrid models with QGP formation and hadronic phase PHSD~\cite{Cassing:2008sv,Cassing:2009vt} and  vHLLE+UrQMD~\cite{Karpenko:2013wva,Karpenko:2015xea} were used to generate \BiBi collisions at $\snn = 9.2$~GeV. These models provide physically well-motivated scenarios for heavy-ion collisions at NICA energies. 
The choice of the event generator for a particular study was driven by the physics observable of interest and the range of measurements. For example, the UrQMD, PHSD and vHLLE+UrQMD event generators were used to study the production of light hadrons and (hyper)nuclei at midrapidity, while PHQMD and DCM-QGSM-SMM were also used to study the response of forward detectors, where realistic simulation of fragment production is important. 

The generated events were used as input for the complete chain of realistic simulations of particle propagation through the detector materials, based on GEANT-4 ~\cite{GEANT4}. The simulations of the detector subsystems and global tracking were performed using the MpdRoot~\cite{mpdroot} code, which is the official software of the MPD Collaboration. For all generators, the event vertex along the beam axis was smeared by a Gaussian function with $\sigmaZ$ = 50 cm. The impact parameter 
\begin{table}[H]
\centering
\caption{The list of centralized MC productions for physics feasibility studies}
\begin{tabular}{c|c|c|c}
No. & Generator & Events & Purpose \\
\hline
1&UrQMD&50 M&General purpose\\
2&DCM-QGSM-SMM&1 M&Trigger\\
%3&PHQMD&1M&Trigger\\
3&PHQMD&20 M&(Hyper)nuclei\\
4&PHSD&15 M&Global polarization\\
5&vHLLE+UrQMD&15 M&Flow, correlations\\
\end{tabular}\label{EvGens}
\end{table}
\noindent ranged within 0-16 fm, except for productions numbers 3 and 4,  where it was set to 0-12 fm to enhance the statistics for (semi)central events. 

The simulations were carried out using computational resources of the MLIT Multifunctional Information and Computing Complex (MICS), including the "Govorun"
supercomputer and VBLHEP computing farm "NICA" at JINR, united by the DIRAC platform~\cite{Baginyan:2021aii, Kutovskiy:2021ehe, Moshkin:SC_proceeding}.

\subsection{Analysis Train Framework}

The analysis of large volumes of simulated and future real data samples ($\sim$ 10 PB) requires a coordinated effort on the part of the \MPD Collaboration, which led to the implementation of the Analysis Train Framework (hereafter  referred to as Train). Train users interested in running over a particular data set sign up for a pass over the data with their analysis modules. Analysis codes are checked into 
 the \MPD code management system (Git). 
%and checked by users for memory leaks and realistic runtime. 
The required input files are read-out once by the Train manager and all analysis modules are sequentially run through the data. This approach reduces the number of input/output (I/O) operations and simplifies the storage architecture. The output files contain the required histograms and NTuples of small size and are stored on the local disks for further analysis.

The first modules in the Train are used to provide global information for all other physics analysis modules, such as event centrality and event plane orientation. In addition, special modules parametrize variables of common interest for each reconstructed track in terms of standard deviations, including the track matching to the primary vertex and outer detectors such as \TOF and \ECAL and the deviation of particle identification signals measured in the \TPC and \TOF from those expected for electrons, pions, kaons, protons and light ions. The use of centralized parametrizations minimizes the amount of work required to start a new analysis and ensures a consistent approach  throughout the MPD Collaboration. The Train architecture also simplifies storage and sharing of analysis codes and methods.
Most of the time, we are able to process the largest simulated datasets (50 M events) in 12 hours by running a Train with $\sim$ 15 modules. A thousand jobs, each processing 50,000 events, are submitted with a total equivalent consumption of one year of CPU time. The number of events per job should not be too small to correctly fill the mixing pools for invariant mass analyses. The first run of the Train took place in September 2023, with regular on-request runs since then.

\section{Global event categorization}
\label{sec:global}

\begin{figure*}[b]
\centering
\includegraphics[width=0.472\textwidth]{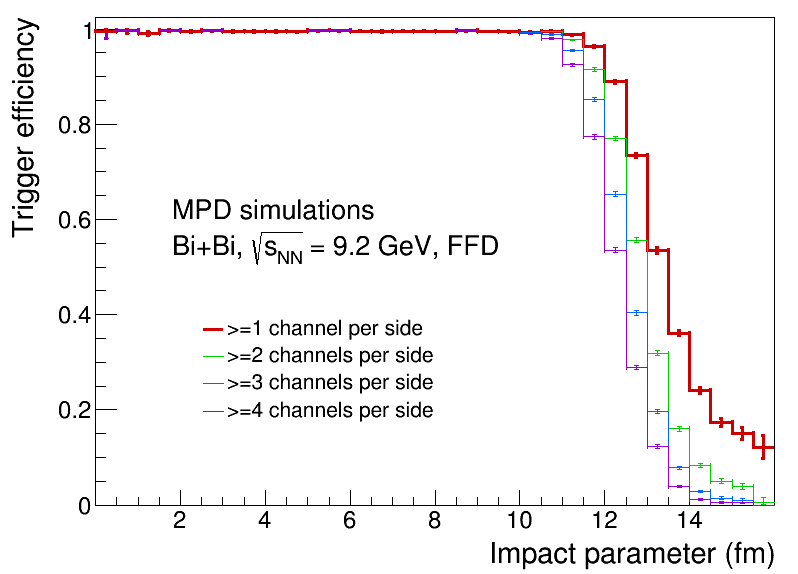}
\hfill
\includegraphics[width=0.472\textwidth]{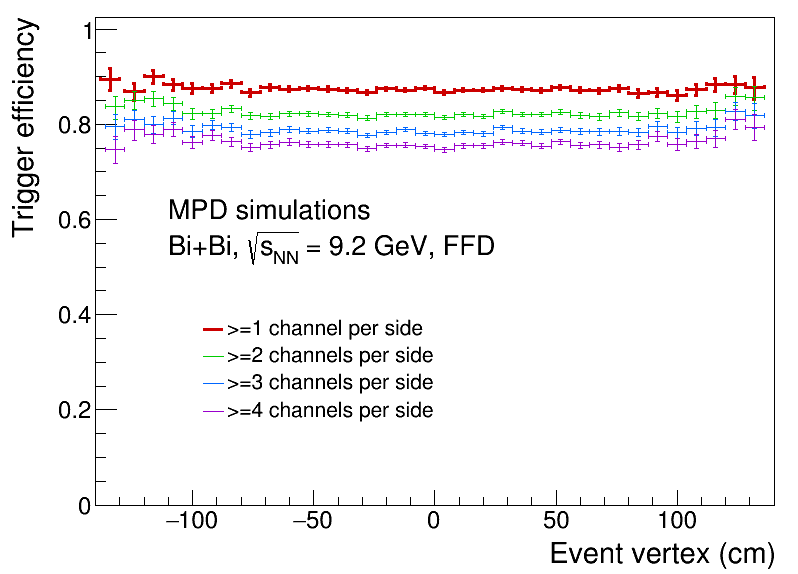}
\\
\includegraphics[width=0.472\textwidth]{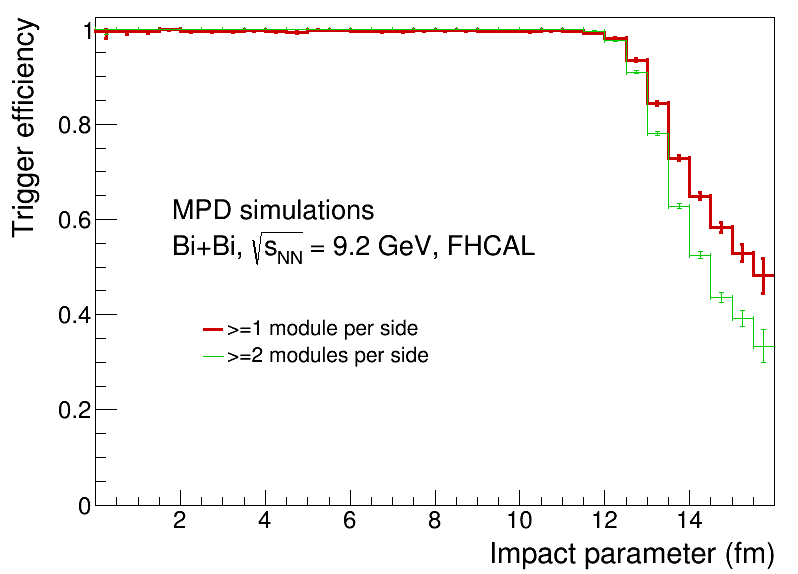}
\hfill
\includegraphics[width=0.472\textwidth]{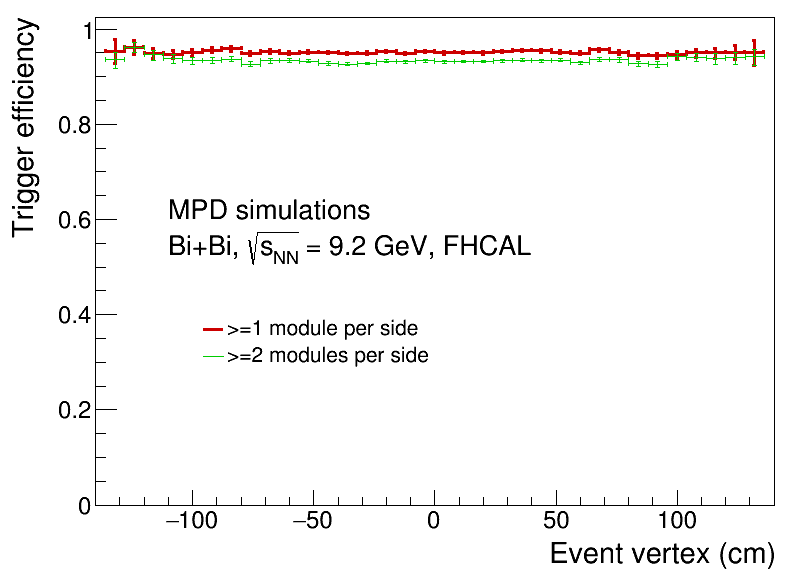}
\\
\includegraphics[width=0.472\textwidth]{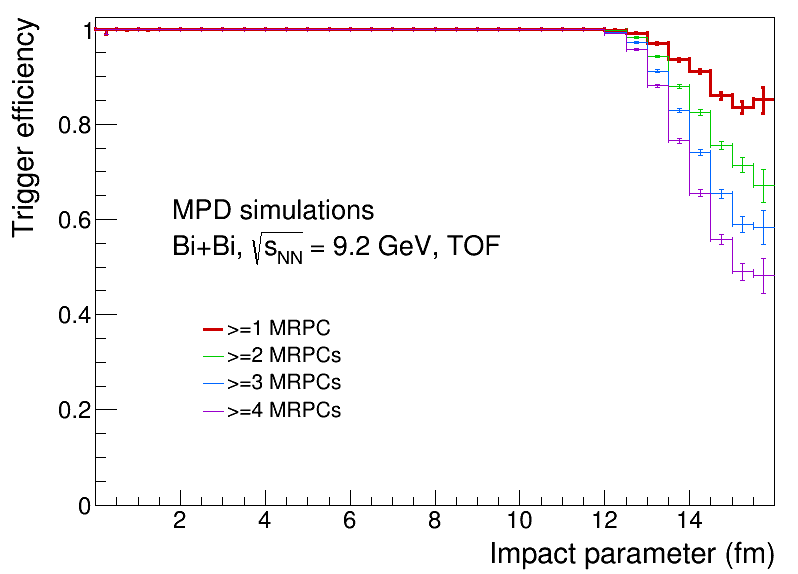}
\hfill
\includegraphics[width=0.472\textwidth]{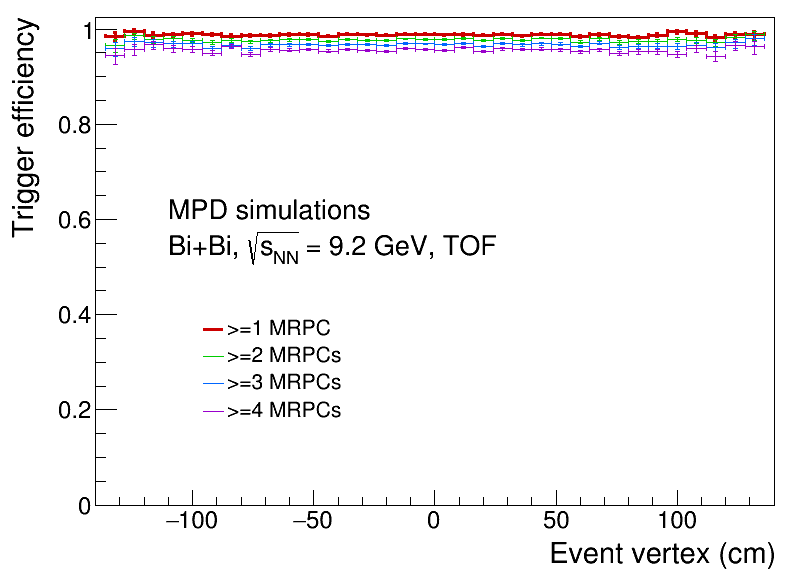}
\caption{Trigger efficiency of the \FFD (top), \FHCAL (middle) and \TOF (bottom) detectors estimated as a function of impact parameter with no $\zvtx$ selections (left) and event $\zvtx$ with no centrality selection (right) for \BiBi collisions at $\snn = 9.2$~GeV.\label{fig:treff}}
\end{figure*} 

The global event quantities discussed in this section are the event centrality and event plane, which characterize the geometry of heavy-ion collisions. These two observables provide basic information for more focused physics studies related to the onset of quark confinement, chiral symmetry restoration and for the search of the CEP in the QCD phase diagram.

\subsection{Trigger system and efficiency}
\label{sec:trigEff}
The trigger system of the \MPD experiment uses signals from three subsystems: \FFD, \FHCAL and \TOF. The performance of the trigger system was studied using centralized productions numbers 2, see Table~\ref{EvGens}.

The main trigger detector is the \FFD. The trigger requires a signal at least in one channel on each side of the detector. The high precision of the online vertex measurements with the \FFD allows an effective suppression of background events from beam-gas and beam-pipe collisions and the selection of events close to the center of the interaction region. 

The \FHCAL produces fast signals from the energy deposition in the 44 modules per side, which can also be used for a trigger decision. In spite of its modest time resolution of $\sim$ 1 ns, which results in a primary vertex resolution from $\sim$ 10 to $\sim$ 30 cm from central to peripheral events, the FHCAL is still useful for background rejection.

The \TOF subsystem generates a fast trigger signal for each of the 280 MRPCs that is hit by at least one particle. The \TOF detects particles produced at central rapidity and is sensitive even to events with small multiplicity. The actual threshold for the number of fired MRPCs in an event to make a trigger decision will depend on the noise conditions of the detector. The \TOF will not be able to provide online information on collision time or vertex position.

Figure~\ref{fig:treff} shows the trigger efficiencies estimated for \FFD, \FHCAL and \TOF as functions of the impact parameter and the event vertex for \BiBi collisions at $\snn = 9.2$~\GeV. Since the background situation is not yet known, the efficiencies are shown for a different number of fired channels for each subsystem.  All three subsystems show an efficiency of $\sim$ 100$\%$ in central and semi-central \BiBi collisions, which decreases rapidly in peripheral collisions. The \FHCAL and \TOF subsystems show higher trigger efficiencies compared with those of the \FFD. The trigger efficiency is not dependent on the vertex position over a wide range $|\zvtx| < 140$~cm, making it possible to collect data in a wide range of vertices with the same efficiency.

The simulated response of the trigger system is not realistic for most of the productions from Table~\ref{EvGens} because event generators such as UrQMD and PHSD do not simulate fragment production at forward rapidity. Therefore, the trigger efficiency estimates obtained in this section for \FHCAL using the DCM-QGSM-SMM event generator were used as a benchmark for the performance of the \MPD trigger system
\begin{figure}[H]
\centering
\includegraphics[width=\linewidth]{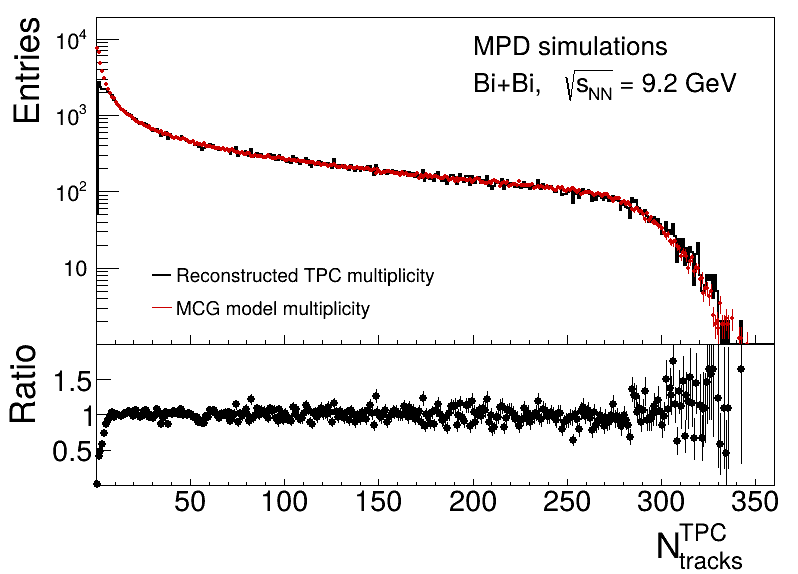}
\caption{The reconstructed \TPC (black) and MCG modeled (red) multiplicity distributions for \BiBi collisions at $\snn = 9.2$~GeV. The bottom part of the figure shows the ratio of the reconstructed and MCG modeled multiplicity distributions.\label{fig:mcg}}
\end{figure}
\noindent in all productions. The inefficiency of the trigger system was emulated for all productions by discarding peripheral events according to the estimated dependence of the trigger efficiency on event track multiplicity, providing an overall efficiency of 91\% for inelastic \BiBi collisions.

\subsection{Event centrality}
\label{sec:Centrality}

In heavy-ion collisions, the centrality of a collision is characterized by the impact parameter, which is the distance between the centers of the nuclei in the plane perpendicular to the beam axis. The impact parameter determines the overlap region of the nuclei.

In a nuclear collision event, the value of the impact parameter is not accessible experimentally. Therefore, the events are usually classified in centrality classes using some measurable quantity like multiplicity, transverse energy measured in a predefined pseudorapidity interval, or the energy of fragments registered in a hadronic calorimeter. Each class corresponds to a percentile of the total inelastic nucleus-nucleus cross section and an average impact parameter that is obtained from some model, usually a Monte Carlo Glauber (MCG) model.

%The value of the impact parameter is not accessible experimentally and must be obtained from the data  by using the measurable quantities approximately proportional to the nuclear overlap region. The centrality of an event can be characterized by the charged particle multiplicity or transverse energy measured in a predefined pseudorapidity interval, or by the energy of fragments registered in the forward direction. The default approach is the centrality determined from the charged-particle multiplicity measured in the \TPC at mid-rapidity, though alternative procedures can also be considered \cite{MPD:2022qhn}. 

In this study, we used centrality determined from the charged-particle multiplicity measured in the \TPC at mid-rapidity, though alternative procedures can also be considered \cite{MPD:2022qhn}. We consider such procedure to be sufficient in the initial stage of MPD at NICA. However, in order to avoid possible autocorrelation effects, future centrality determination using the charged-particle multiplicity measured in the TPC will be performed similarly to the procedure developed by STAR \cite{Chatterjee:2019fey,Luo:2013bmi}, i.e. by selecting centralities from a region different from the one used in the data analysis. The centrality was evaluated for events with a reconstructed vertex within $|\zvtx| < 130$~cm. 
As shown in Sec.~\ref{sec:trigEff}, the trigger efficiency remains constant in this range. A wider range would include collisions with vertices close to the \FFD. Rather loose selection criteria were used for the reconstructed tracks: number of \TPC hits $N_\mathrm{hits}^{\TPC} > 10$, transverse momentum $\pT > 0.1$~\GeVc, track matching to the primary vertex $< 2$~cm, and track pseudorapidity $|\eta| < 0.5$. Each track is corrected for the TPC reconstruction efficiency estimated as a function of the event $\zvtx$ and track pseudorapidity $\eta$. A typical  multiplicity distribution is shown in Figure ~\ref{fig:mcg}. 

The event centrality is estimated as a percentile of the total multiplicity with the maximum value of 91$\%$. By definition, the reconstructed centrality distribution is flat between 0 and 91~\%. 

\begin{figure*}[b]
\centering
\includegraphics[width=0.49\textwidth]{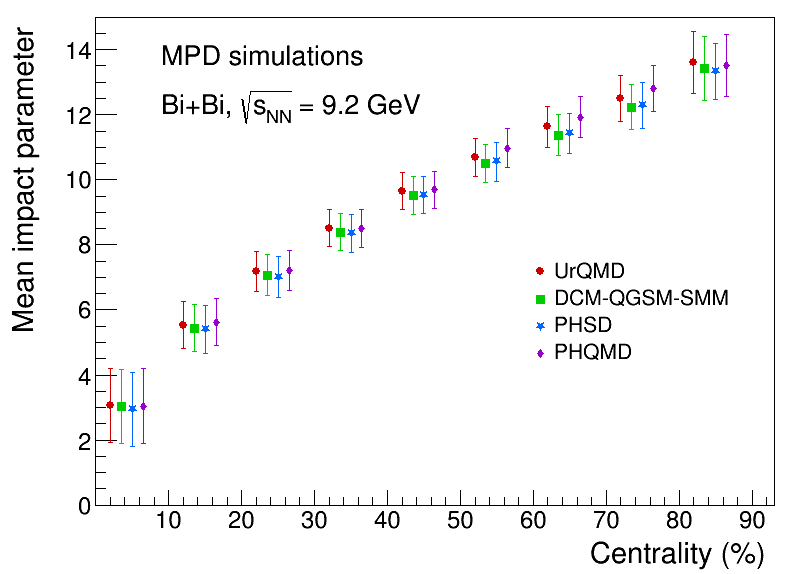}
\hfill
\includegraphics[width=0.49\textwidth]{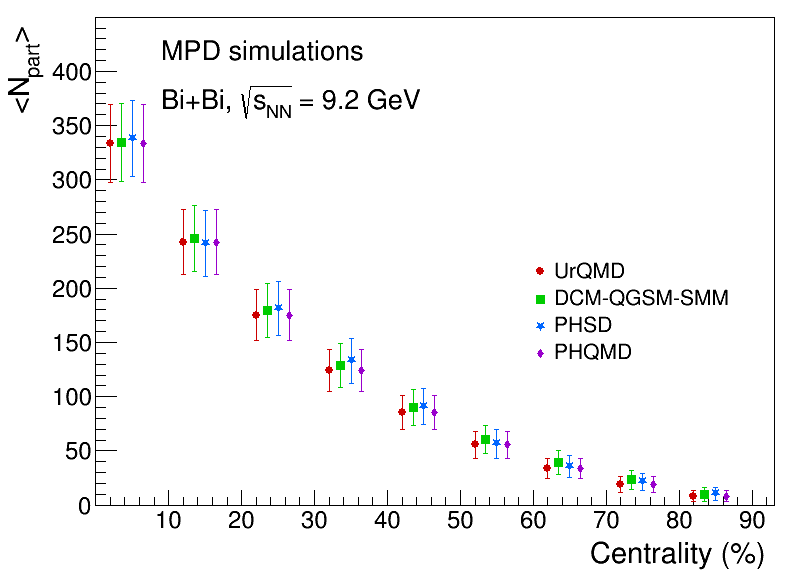}
\caption{Left: the value of the mean impact parameter for 10$\%$ centrality intervals estimated for \BiBi collisions at $\snn = 9.2$~\GeV, modeled with the UrQMD, DCM-QGSM-SMM, PHSD, and PHQMD event generators. Right: the same for the mean number of participants $\langle\Npart\rangle$. The symbols are slightly shifted horizontally for better visibility.\label{fig:mcg_bNpNc}}
\end{figure*} 

The standard MCG model~\cite{Loizides:2014vua} was used to parametrize the reconstructed multiplicity distribution and estimate the geometrical parameters of the collisions. The impact parameter distribution of the MCG model was reweighted to reproduce the distributions modeled in the event generators listed in Table~\ref{EvGens}.

Within the MCG model the particle multiplicity distribution is modeled as the sum of particles produced from a set of independent emitting sources ($N_\mathrm{a}$), each producing particles according to a negative binominal distribution NBD$(\mu, \mathrm{k})$. The number of emitting sources is parametrized as 
\begin{equation}
N_\mathrm{a} = f\Npart + (1-f)\Ncoll,
\end{equation} 
where $\Npart$ and $\Ncoll$ are the number of participating nucleons and the number of inelastic binary nucleon-nucleon collisions, respectively. The parameters $\mu$, k and $f$ are varied to minimize the $\chi^{2}$/NDF of the description of the measured multiplicity distribution in the range $N_\mathrm{tracks}^{\TPC} > 10$. This range can be varied for a systematic study and, by default, it was set to the minimal value corresponding to the saturation of the trigger efficiency.

The distribution, represented by the red markers in Figure~\ref{fig:mcg}, shows the result of this procedure. A good agreement between the measured and MCG-simulated multiplicity distributions in the overlap region can be observed. The ratio of the reconstructed and MCG multiplicity distributions is shown in the bottom part of the figure as an estimate of the trigger efficiency as a function of the event multiplicity.
%, $\varepsilon_\mathrm{MCG}$($N_\mathrm{tracks}^{\TPC}$). 
The weighted average efficiency estimated from the ratio is $\sim$ 90$\%$, which is very close to the expected value of 91~\%.

The MCG model is then used to estimate the initial geometry of the centrality classes. The values of the impact parameter, $\Npart$ and $\Ncoll$ for 10$\%$ centrality intervals are evaluated for the UrQMD, DCM-QGSM-SMM, PHSD, and PHQMD event generators. Figure~\ref{fig:mcg_bNpNc} shows the mean and RMS values with markers and error bars, respectively, evaluated for impact parameter and $\Npart$. The symbols for different event generators are shifted for visibility. A good agreement is found for the extracted values of the model parameters.

\subsection{Event plane}
\label{sec:evplane}

The event plane method correlates the azimuthal angle $\phi$ of each particle
with the azimuthal angle $\Psi_n$ of the event plane determined from the anisotropic
flow itself \cite{flowmpd1,flowmpd2}.  The event flow vector $Q_n=(Q_{n,x}, Q_{n,y})$  in the transverse $(x,y)$ plane  and the azimuthal angle of the event plane
$\Psi_n$ can be defined for each harmonic, $n$, of the Fourier expansion by
\begin{eqnarray}
  Q_{n,x}  = \sum_\mathrm{k=1}^\mathrm{M} w_\mathrm{k} \cos(\varphi_\mathrm{k}), Q_{n,y}  = \sum_\mathrm{k=1}^\mathrm{M} w_\mathrm{k} \sin(\varphi_\mathrm{k}), \quad  \notag \\
   \Psi_n = \frac{1}{n} \tan^{-1} \left( \frac{Q_{n,y}}{Q_{n,x}} \right),
\end{eqnarray}
where $\rm{M}$ is the multiplicity of the  particles $\rm{k}$ used in the event plane calculation, and $\varphi_\mathrm{k}$  and  $w_\mathrm{k}$ are the
laboratory azimuthal angle
and the weight for the particle $\rm{k}$,  which is used either to correct for
the azimuthal anisotropy of the detector or to account for the multiplicity of hadrons stopped in a particular cell of the segmented detector.
The details of $w_\mathrm{k}$
estimation can be found in Ref. \cite{flowmpd1,flowmpd2,flowmpd3}.
The reconstructed $\Psi_{n}$ values can be used to measure  the differential $v_n$ flow coefficients
of   particles detected in the \TPC  ($|\eta|<$ 1.5), 
\begin{equation}
v_\mathrm{n}(p_{\rm T},y)=\frac{\langle\cos(\mathrm{n}(\phi - \Psi_{n}))\rangle}{R (\Psi_{n})}, 
\end{equation}
where  ${R (\Psi_{n})}$ represents the event plane resolution factor and brackets denote the
average over the particles and events.
The 2-subevent method with extrapolation algorithm is used to estimate the ${R (\Psi_{n})}$
factors~\cite{posk}.

\begin{figure}[H]
\centering
\includegraphics[width=0.99\linewidth]{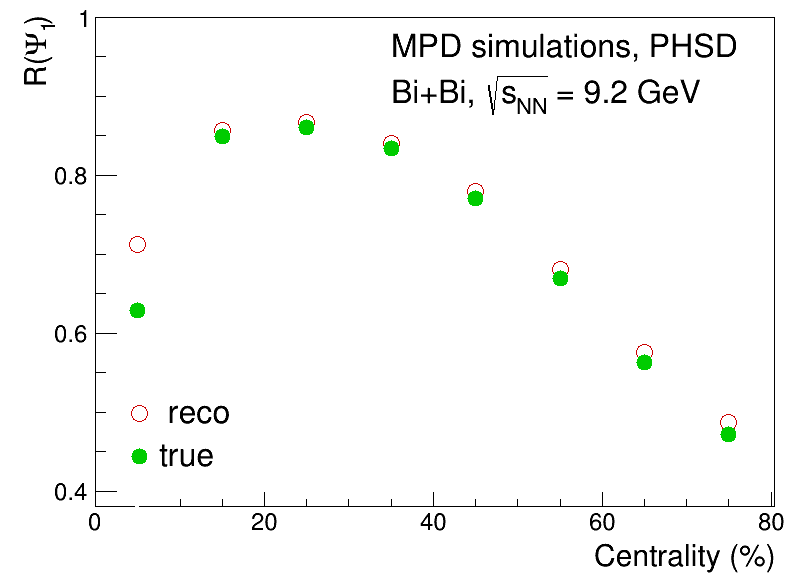}
\caption{Centrality dependence of the event plane resolution factor $R(\Psi_{1})$ for $v_1$  and $P_\Lambda$  measurements in 
\BiBi collisions at $\snn = 9.2$~\GeV, production 4 in Table~\ref{EvGens}. \label{fig:ResPol_cent}}
\end{figure}

Figure~\ref{fig:ResPol_cent} shows the
centrality dependence of the event plane resolution factor $R (\Psi_{1})$ for directed $v_1$ flow measurements for \BiBi collisions at $\snn$ = 9.2~\GeV simulated in production number 4 in Table~\ref{EvGens}. Here, the $\Psi_{1}=\Psi_{1,\FHCAL}$  
determined from the
directed flow ($n=$ 1) of particles detected in the FHCal (2 $<|\eta|<$ 5).

The open symbols correspond to the $R(\Psi_{1})$ values from the analysis of the fully reconstructed events “reco” and closed symbols
to the results from generated “true” PHSD events. 
For the mid-central  events,  the  resolution factor $R(\Psi_{1})$  is as
large as 0.85 for $v_1$ and  the global polarization $P_\Lambda$ of $\Lambda$ hyperon \cite{global1} measurements.

\begin{figure}[H]
\centering
\includegraphics[width=0.99\linewidth]{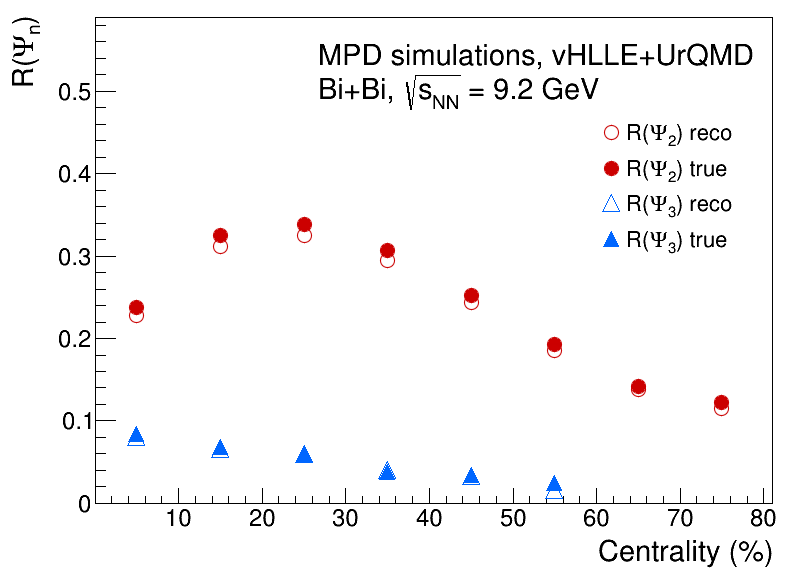}
\caption{Centrality dependence of the event plane resolution factor $R(\Psi_\mathrm{n})$ for the second $n=2$ (circles) and third $n=3$ (triangles) order event planes constructed from the tracks of charged particles in the \TPC.
Open markers correspond to the reconstructed data, closed markers to the generated vHLLE+UrQMD model events.\label{fig:Res2_Res3_cent}}
\end{figure}

Figure~\ref{fig:Res2_Res3_cent} shows the centrality dependence of the event plane resolution factor $R(\Psi_\mathrm{n})$ for elliptic ($v_2$) 
and triangular ($v_3$) flow measurements  for \BiBi collisions at $\snn$ = 9.2~\GeV.
Here, the flow vectors $\Qn = Q_{n,\TPC}$ and the azimuthal angle of the event plane
$\Psi_n=\Psi_{n,\TPC}$ are  constructed from the charged particle tracks reconstructed in the
\TPC ($|\eta|<$ 1.5)\cite{flowmpd3}.
The open markers correspond to $R(\Psi_\mathrm{n})$ values from the analysis of the fully reconstructed  vHLLE+UrQMD events (production 5 in Table~\ref{EvGens}) and the 
\noindent closed markers to results from the  generated  events. The difference in the
resolution factors for different flow harmonics reflect the observed ordering at NICA energies: $v_1 > v_2 > v_3$. The details of the extraction of collective flow parameters of different species are discussed in Sec.~\ref{sec:physics}. 

\section{ Physics performance studies}
\label{sec:physics}

%=================================================================

In this section, we present selected results of physics feasibility studies for the MPD experiment in \BiBi collisions at $\snn = 9.2$~\GeV  with emphasis on the measurements expected for the first years of MPD operation.

\subsection{Light flavor hadron production}
Light-flavored hadrons are copiously produced and play an important role in understanding the physics of relativistic heavy-ion collisions. Experimental studies of charged pion, kaon and (anti)proton spectra and yields are used to determine the properties of the hot and dense baryonic matter at the moment of its decay into final-state hadrons, allow testing of thermal and chemical equilibrium in the system,
%at this moment, % VK 03/12 "actually, these are two moments not one"
and provide insight into the underlying reaction dynamics by addressing the collective effects in the longitudinal and transverse expansion of the fireball. 
The shapes of particle $\pT$ distributions and $\langle \pT\rangle$ probe the reaction dynamics and are sensitive to particle production mechanisms in different kinematic regions, and to the interplay of the radial flow and parton recombination at intermediate transverse momenta. Measurements of hadrons containing strange quarks allow to study the strangeness enhancement in heavy-ion collisions. Studying  strangeness enhancement of  particles with open and hidden strangeness, provides much more details of the strangeness production mechanisms. The production of short-lived resonances with lifetimes comparable to  the fireball lifetime is measured to study the rescattering and regeneration processes in a dense hadronic medium.

\subsubsection{Yields of charged pions, kaons and (anti)protons}
\label{sec:hadrons}

The present analysis of charged hadron yields uses data of production number 3 from
Table~\ref{EvGens}. To select events, we apply a primary vertex position cut of $|\zvtx|<100$~cm. To minimize the contamination of secondary tracks, the Distance of Closest Approach (DCA) from the track to the collision vertex is taken to be less than 
3~cm. To select tracks with good momentum and \dEdx resolution and to reject split tracks, the number of TPC points associated with the track is required to be larger than 20. The center-of-mass rapidity and transverse momentum windows to perform the analysis are $|y|<1.1$ and $0.05<\pT<2.5$~\GeVc. Two different approaches have been used for identification of charged hadrons.

\begin{figure}[H]
\centering
\includegraphics[width=0.46\textwidth]{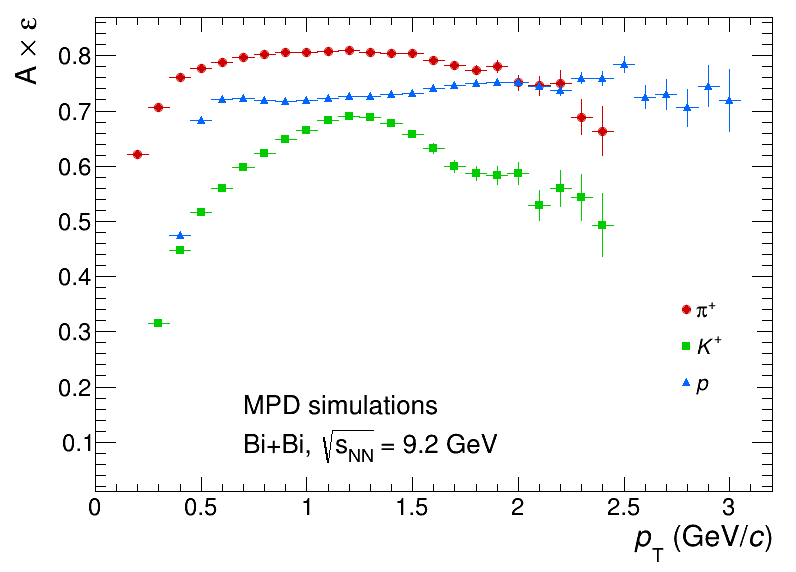}
\hfill
\includegraphics[width=0.46\textwidth]{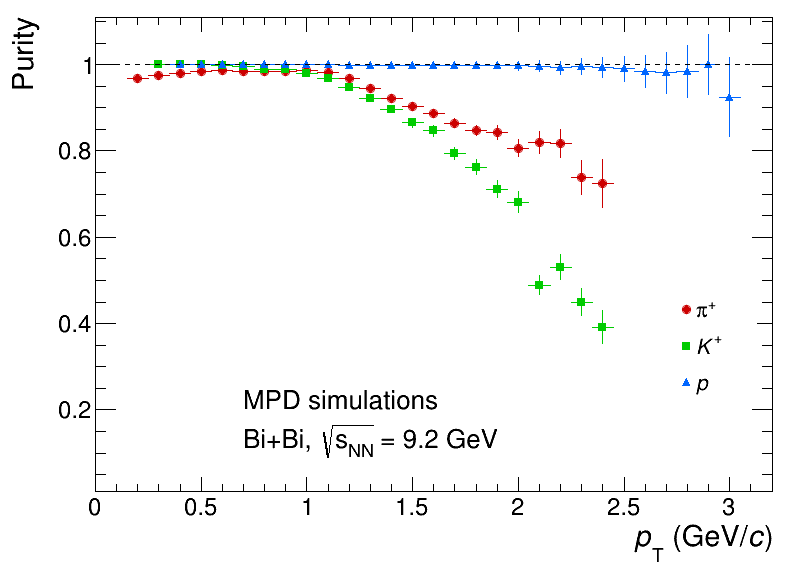}
\caption{Top: Overall efficiency for positively charged hadrons as a function of \pT in Approach 1. Bottom: Purity for positively charged hadrons as a function of \pT. \label{hadrons_effic}}
\end{figure}

\begin{figure*}[t]
\centering
\includegraphics[width=0.32\textwidth]{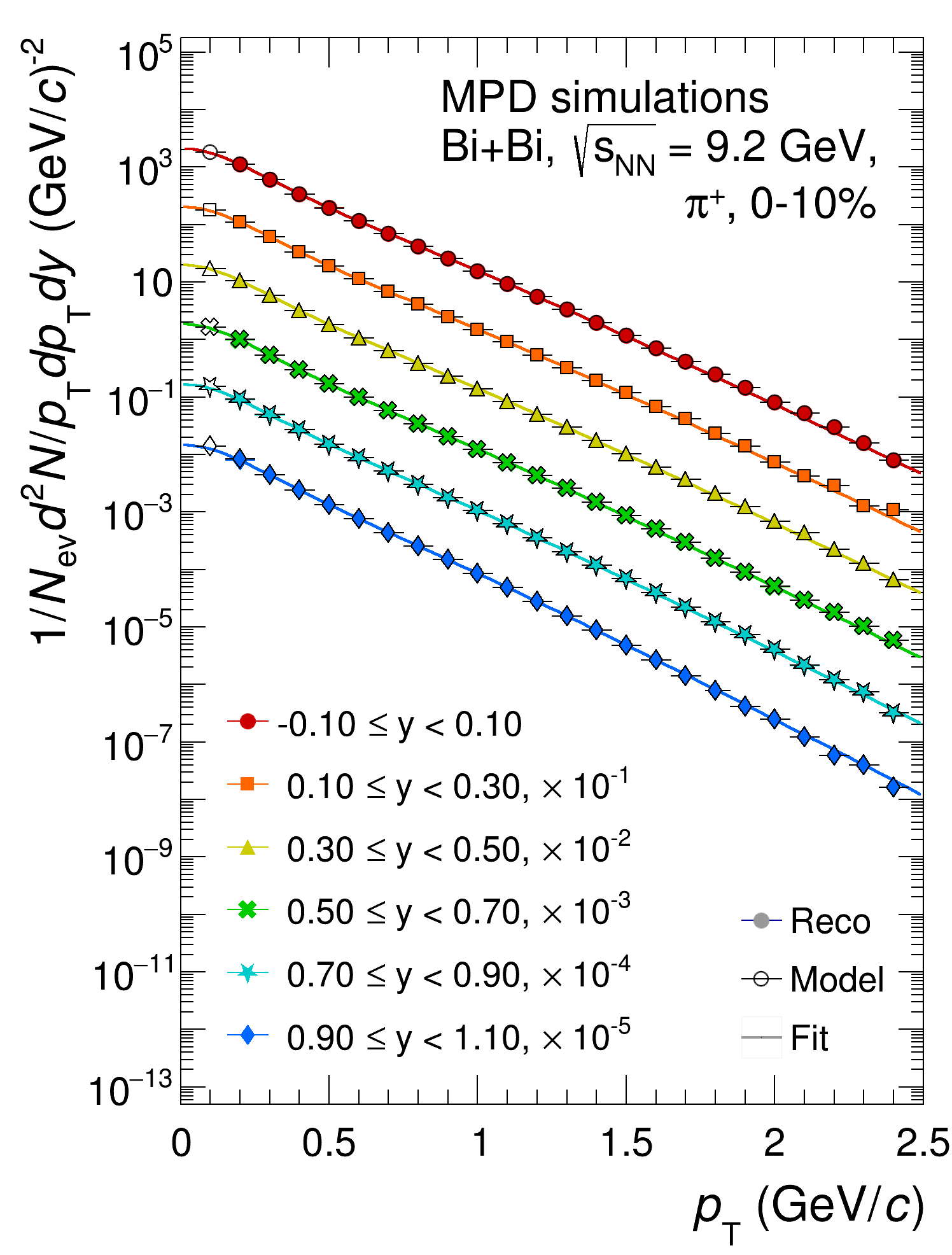}
\hfill
\includegraphics[width=0.32\textwidth]{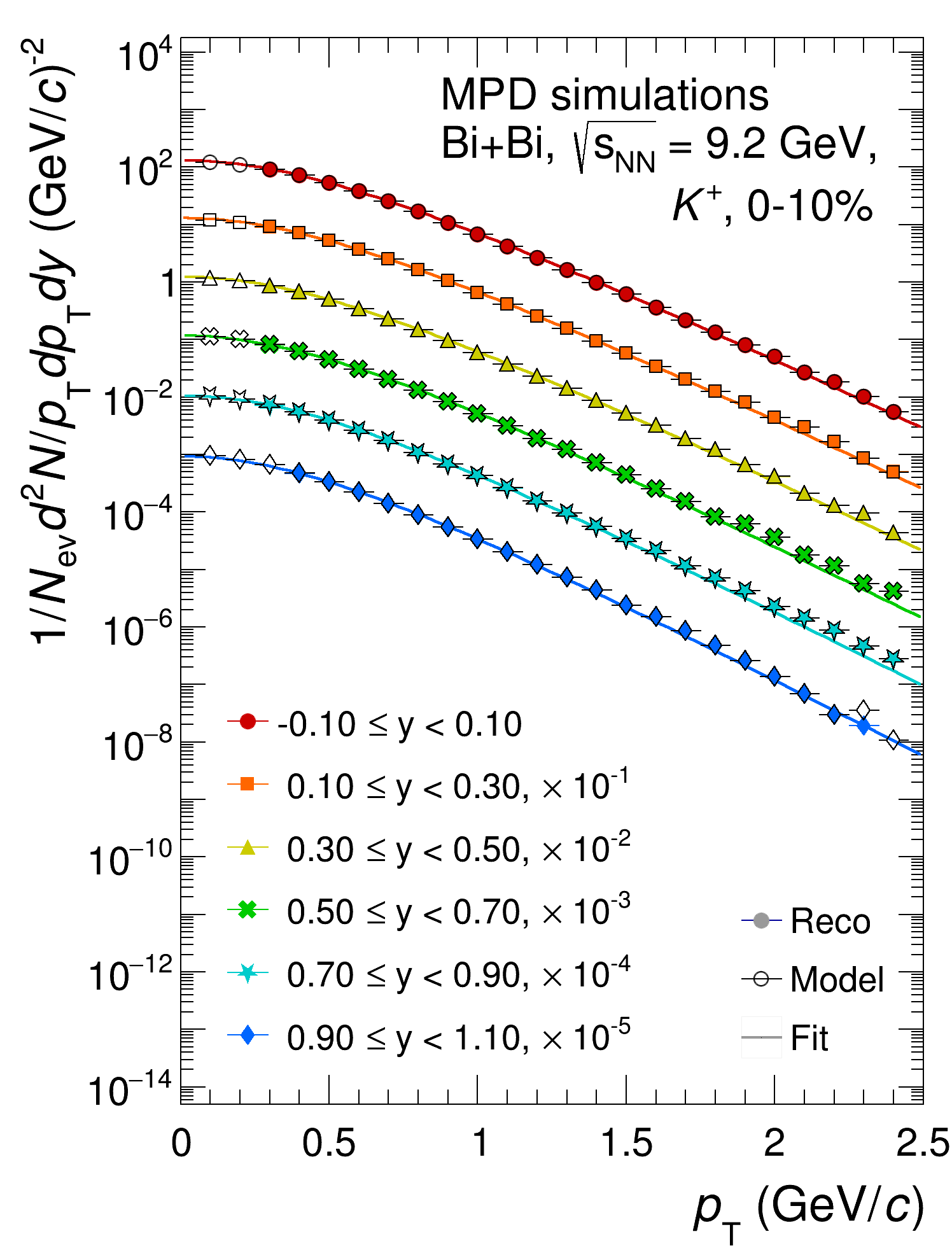}
\hfill
\includegraphics[width=0.32\textwidth]{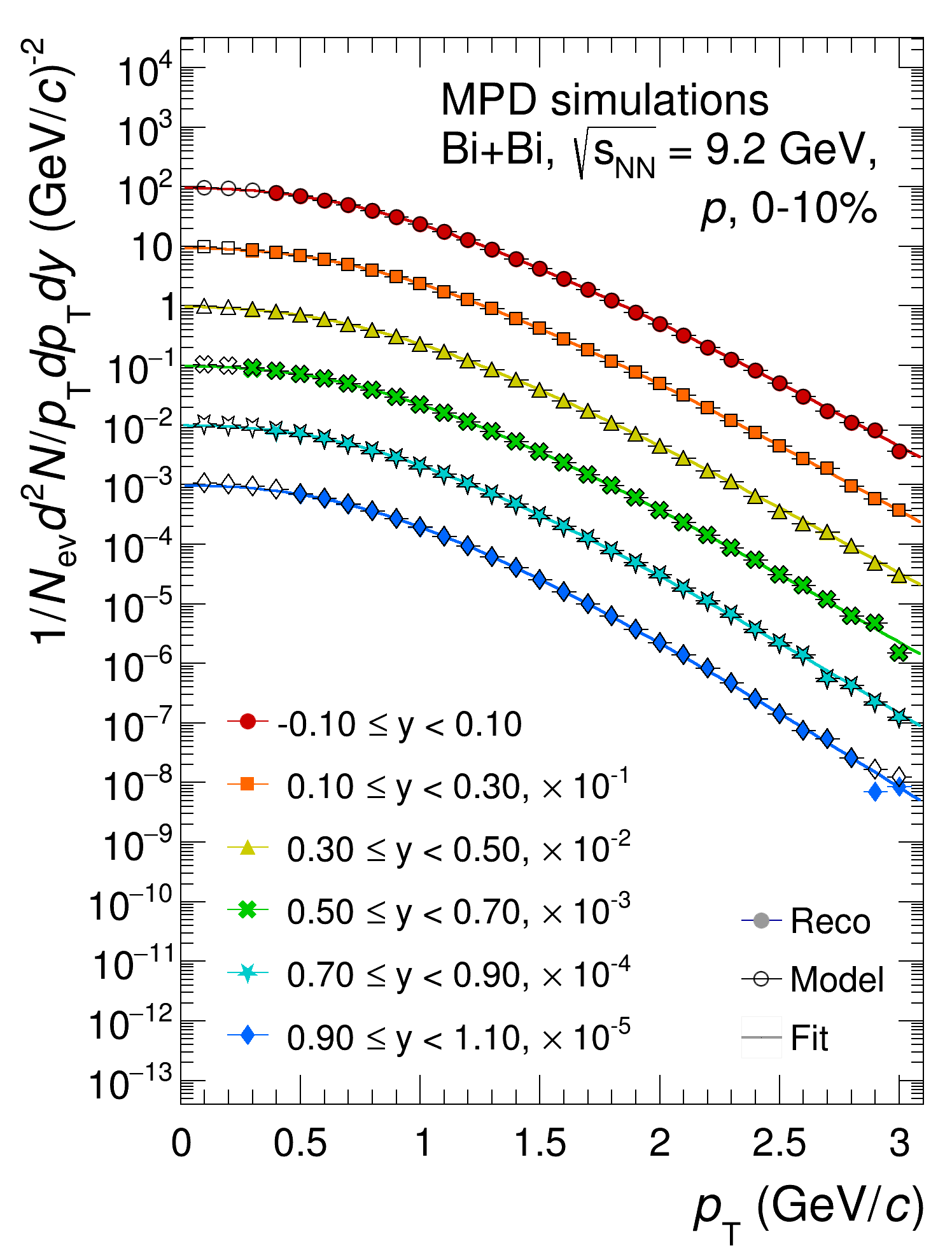}
\caption{Invariant \pT-spectra of \pip (left), \kap (center) and $p$ (right) in several rapidity intervals for 0-10\% central \BiBi collisions. The reconstructed data are shown by filled symbols while the model data are depicted by open symbols. Fits to invariant spectra are shown by lines (see text for details). \label{hadrons_pt}}
\end{figure*}

{\bf Approach 1} Signals in the \TPC and \TOF are required for each charged particle track to be accepted and particle identification is achieved by a combination of energy loss \dEdx and time-of-flight measurements. Such an approach provides the best purity of the measured signals (see Figure~\ref{hadrons_effic}, bottom panel), but limits the measurement ranges at low \pT due to limited \TOF acceptance, see Sec.\ref{sec:detector}.  The measured raw yields of hadrons are corrected for reconstruction efficiency (see Figure~\ref{hadrons_effic}, top panel), which accounts for hadron misidentification, reconstruction losses, geometrical acceptance, and contamination from secondary interactions in the detector material and from 
weak decays of hyperons (relevant for pions and protons). The yields of charged hadrons 
are divided in centrality classes (0-10\%, 10-20\%,
20-30\%, 30-40\%, 40-80\%) and in several rapidity intervals. As an example, 
Figure~\ref{hadrons_pt} shows the comparison of transverse momentum spectra of positively charged pions (left panel), kaons (central panel) and protons (right panel) reconstructed in  0-10\% central \BiBi collisions to the generated ones. 
The comparison of spectra is shown in rapidity intervals of $\Delta y=0.2$, where spectra are
scaled down relative to the data  at midrapidity by successive orders of ten for clarity. 
We found good agreement between reconstructed and generated spectra in all cases.

\begin{figure*}[b]
\centering
\includegraphics[width=0.32\textwidth]{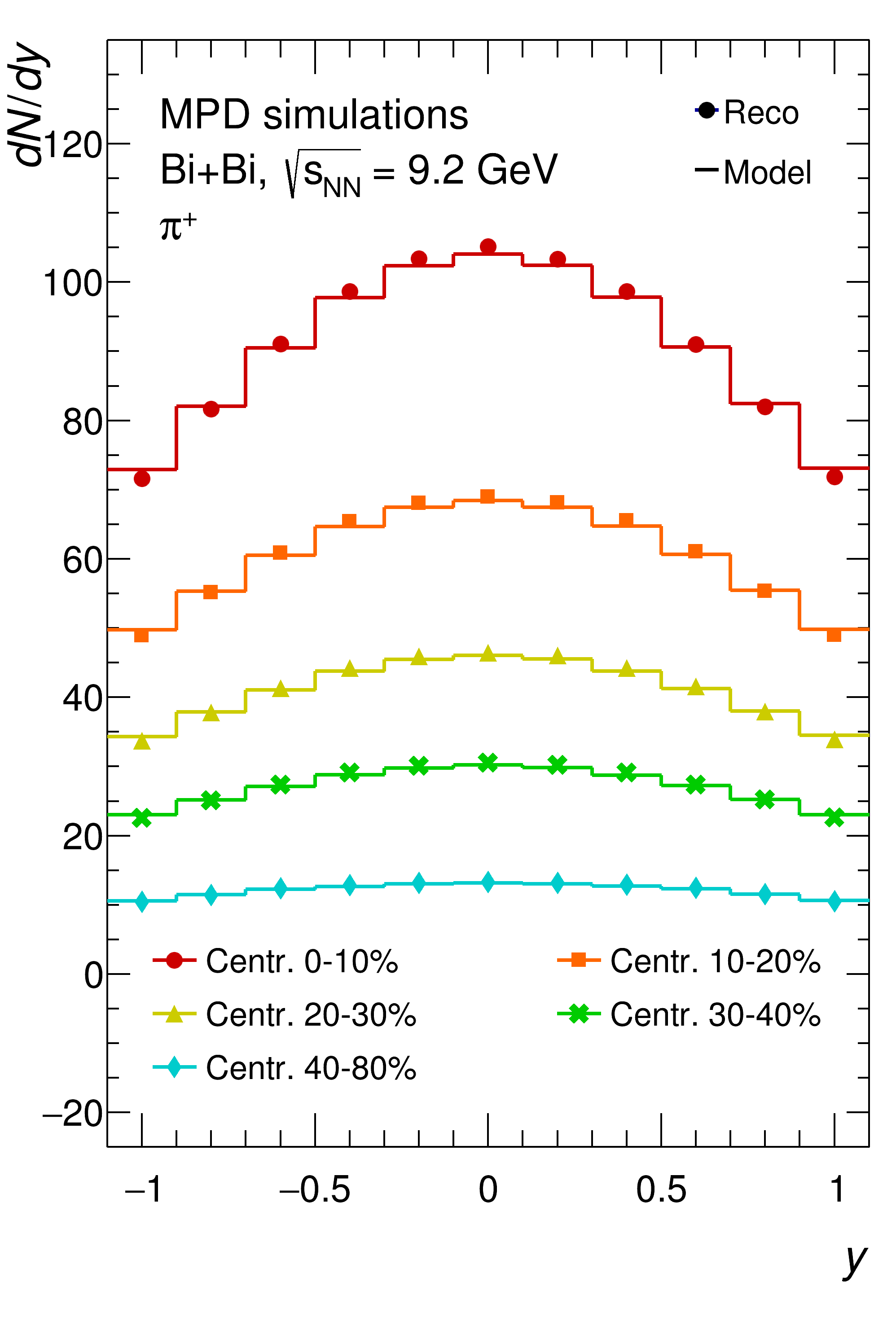}
\hfill
\includegraphics[width=0.32\textwidth]{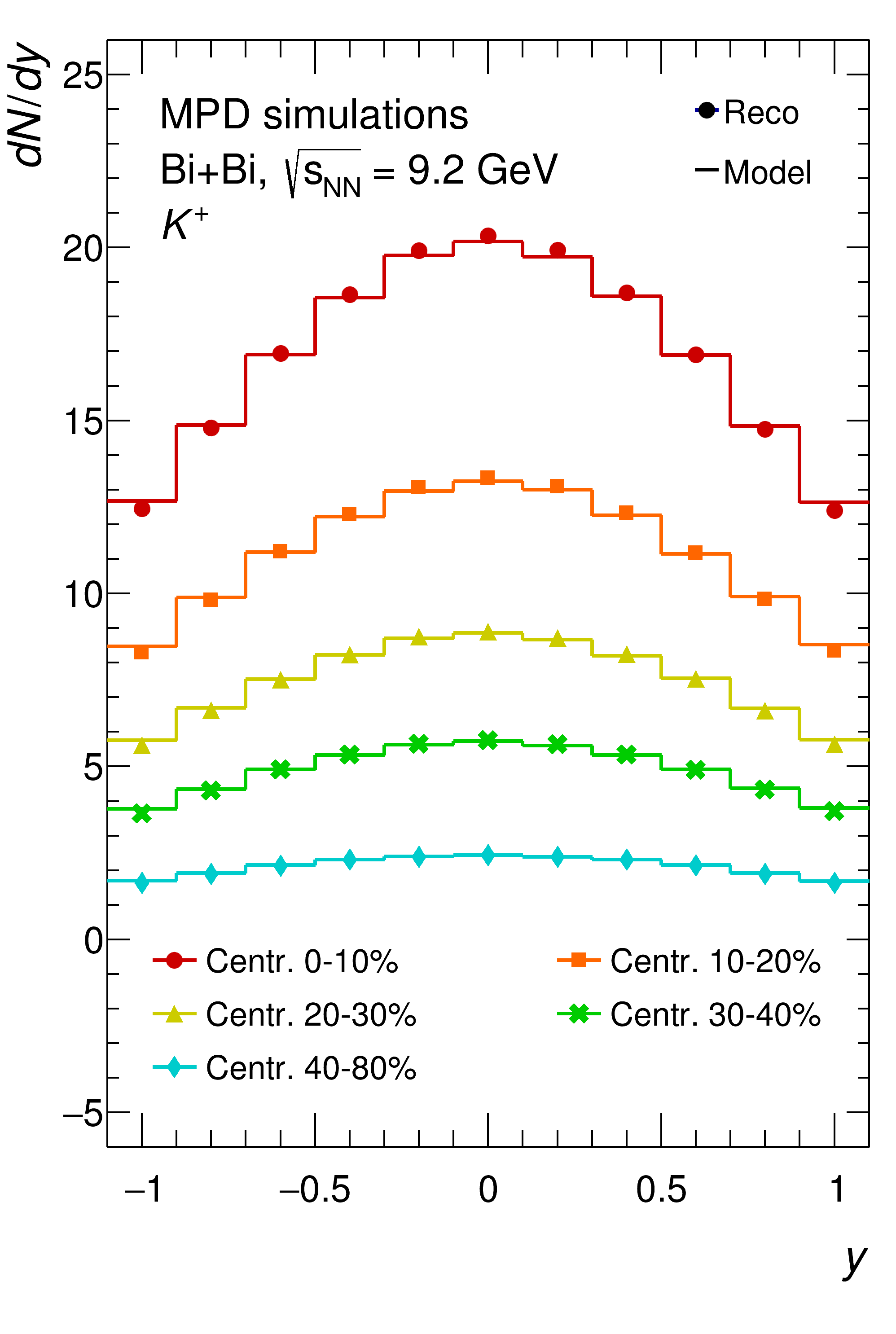}
\hfill
\includegraphics[width=0.32\textwidth]{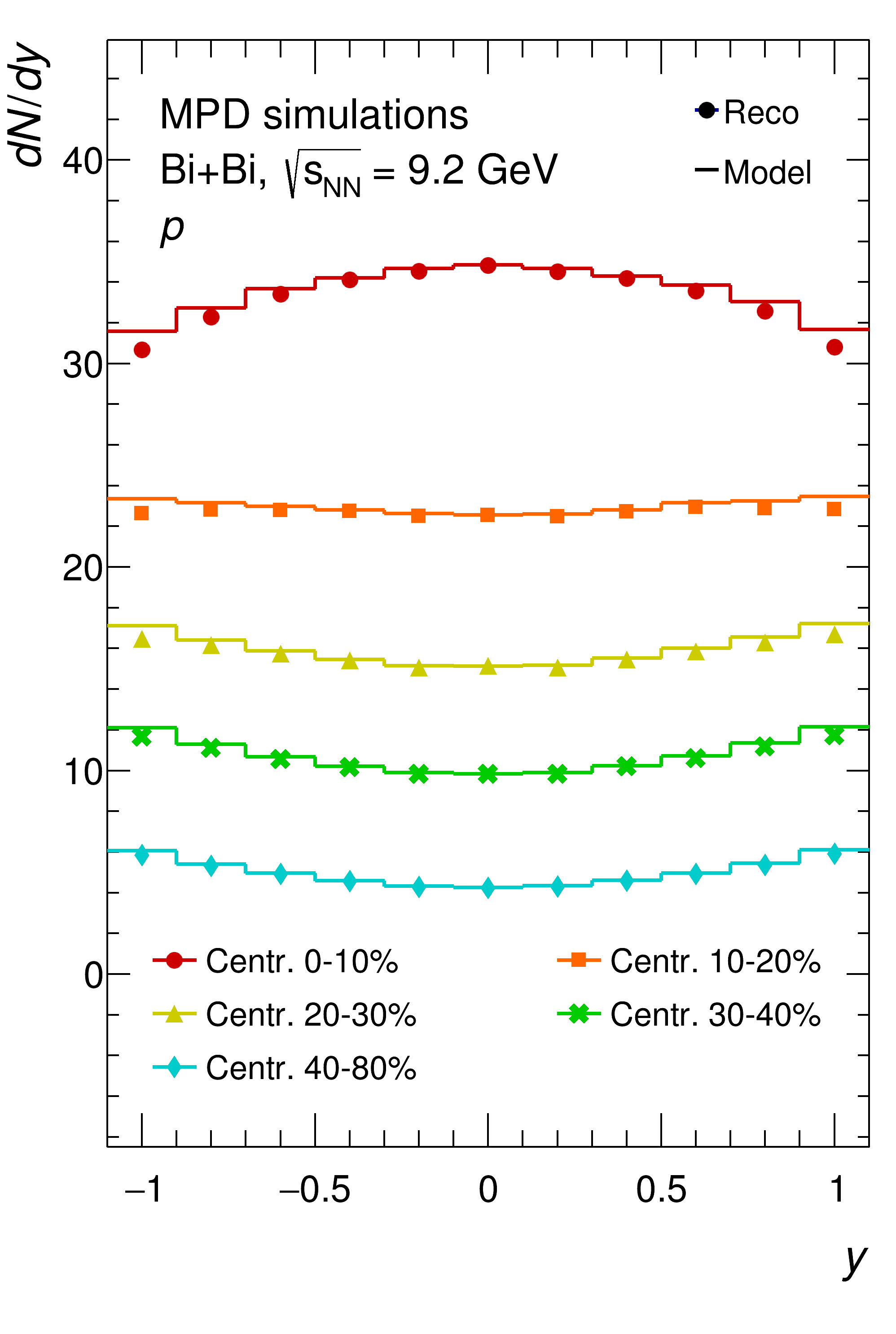}
\caption{Rapidity distributions of \pip (left), \kap (center) and $p$ (right) in \BiBi collisions in different centrality classes. The reconstructed data are shown by symbols while the spectra at generator level are depicted by lines. \label{hadrons_dndy}}
\end{figure*}

In this approach, the MPD has limited \pT coverage at low transverse momenta and to calculate the integrated yields one 
\noindent has to extrapolate spectra to an unexplored \pT-range. 
To do this, the spectra are approximated with appropriate functional forms. The yield of pions
is enhanced at low-\pT due to a contribution from resonance decays, thus a sum 
of two exponentials in \mT (thermal function) is used. 
The kaon distributions are well 
described with a thermal function, while for protons a Blast-Wave motivated function~\cite{bw} is
used. 
The contribution of the extrapolation region varies for different particle species, but it does not exceed 5\%,
10\%, and 15\% for pions, kaons, and protons, respectively.   
The rapidity density distributions (\dndy) of positively charged hadrons (\pip, 
\kap, $p$), obtained by integrating the transverse momentum spectra in 
Figure~\ref{hadrons_pt}, are shown in Figure~\ref{hadrons_dndy}, where 
the reconstructed data are shown with symbols, while spectra at generator level are shown 
by lines. 
The measurements for pions and kaons cover approximately 65\% of the total 
phase-space and the rapidity distributions can be approximated by a Gaussian. Thus, an 
integrated mean total multiplicity of $\pi, K$ can be obtained with $\sim10$\% 
uncertainty. The situation for protons is more difficult because the shape of their 
rapidity distributions changes with centrality. The \MPD phase-space coverage for protons
is not sufficient to reconstruct the total (4$\pi$) yield of protons without model 
assumptions that can accurately predict the proton yields near the beam rapidity.

{\bf Approach 2} In this case, hadron spectra are measured separately using the particle identification capabilities of the \TPC or \TOF, and then combined by switching from one to another at a given \pT value. The spectra based on \TPC identification ("TPC-spectra") consist of  particles that are: 1) identified in the \TPC within two standard deviations $2\sigma_{\TPC}(\pT)$ and not consistent with signals expected for other species within $3\sigma_{\TPC}(\pT)$; 2) identified in the \TOF within two standard deviations $2\sigma_{\TOF}(\pT)$ if the track is matched to \TOF. Similarly, spectra based on \TOF identification ("TOF-spectra") consist of particles that are: 1) identified in \TOF within two standard deviations $2\sigma_{\TOF}(\pT)$ and not consistent with signals expected for other species within $3\sigma_{\TOF}(\pT)$; 2) identified in \TPC within two standard deviations $2\sigma_{\TPC}(\pT)$. The spectra are reconstructed in the momentum ranges where signal purity exceeds 95\%. The main advantage of this approach is that it provides access to measurements of identified hadrons down to as low transverse momenta as is possible with the existing track reconstruction algorithms in MpdRoot~\cite{mpdroot}: $\pT> 100$, 150 and 200~\MeVc for \pipm, \kapm and $p(\overline{p})$, respectively. The disadvantage is limited coverage at higher momenta because of the imposed strict requirements of high signal purity.

For charged pions, the veto requirement for other species keeps the signal purity close to $\sim 100\%$ over the entire momentum range, but limits the measurement range to $\pT < 1$~\GeVc in both \TPC and \TOF. For charged kaons, the requirement for high signal purity limits the measurements to $\pT < 0.45 (1.5)$~\GeVc with \TPC (\TOF). At higher momenta, the purity of kaons decreases rapidly due to admixture of pions. The proton measurements with \TPC or \TOF are limited by veto requirement to $\pT < 1.0$~\GeVc and $\pT < 4.0$~\GeVc, respectively. The situation is more complicated for antiprotons because of the high baryon asymmetry at NICA energies. The main contamination of the sample of identified antiprotons comes from the back-scattered protons, which are misidentified as antiprotons due to an incorrectly determined momentum direction. The purity requirement limits \TPC measurements for antiprotons to transverse momenta from 0.2~\GeVc to 0.9~\GeVc. Since the \TOF has no acceptance for low-\pT protons, the measurements with the \TOF are not affected by proton contamination. However, the admixture of kaons limits \TOF measurements to $\pT < 1.2$~\GeVc.

\begin{figure*}[t]
\centering
\includegraphics[width=0.48\textwidth]{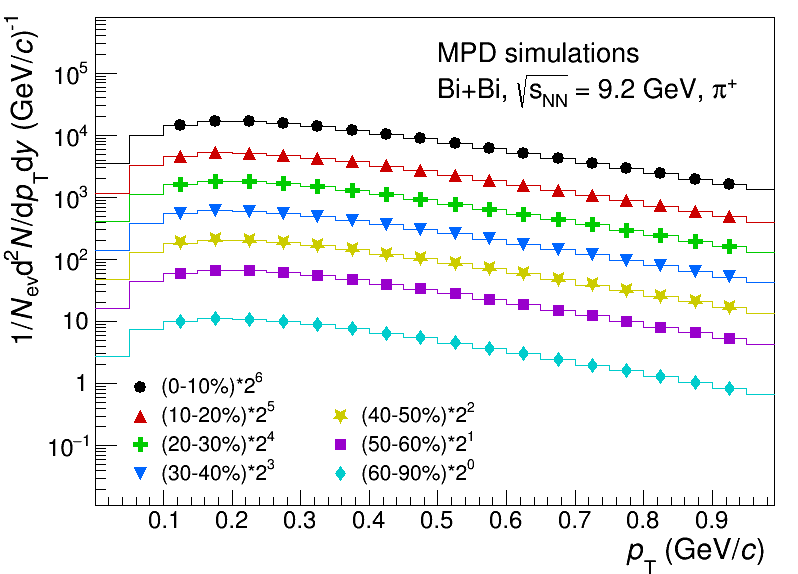}
\hfill
\includegraphics[width=0.48\textwidth]{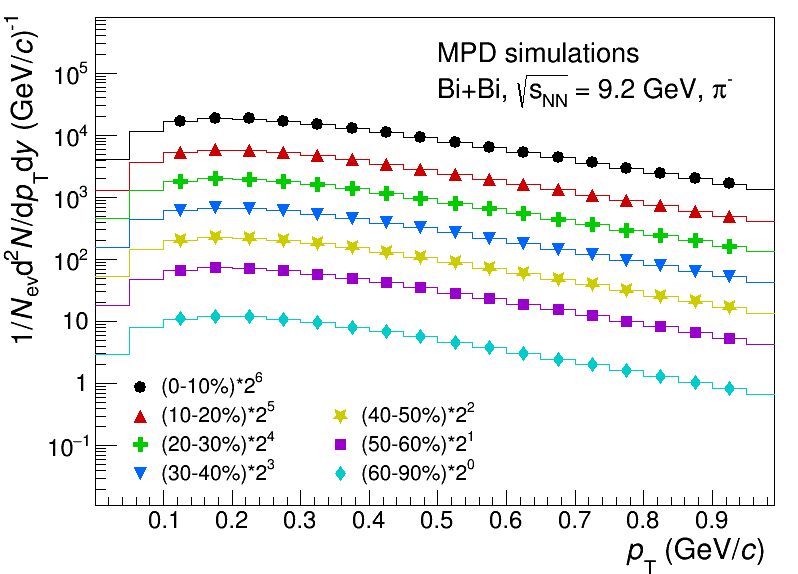}
\includegraphics[width=0.48\textwidth]{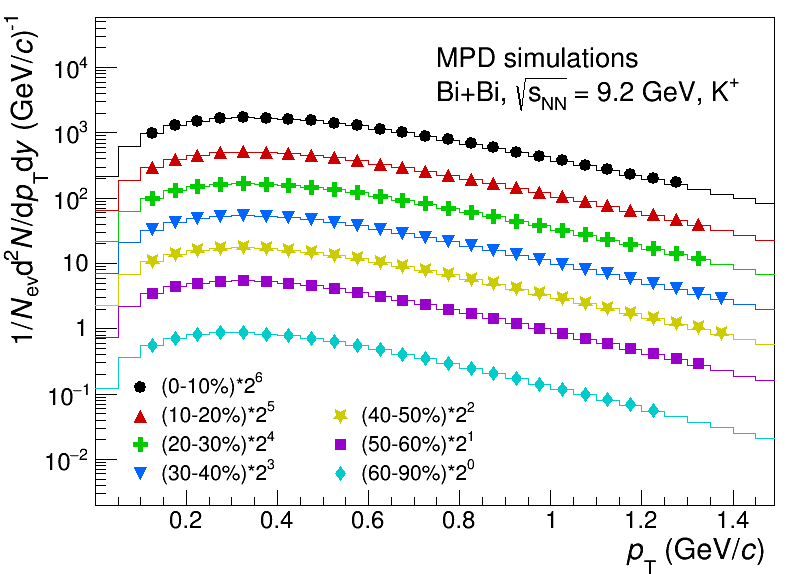}
\hfill
\includegraphics[width=0.48\textwidth]{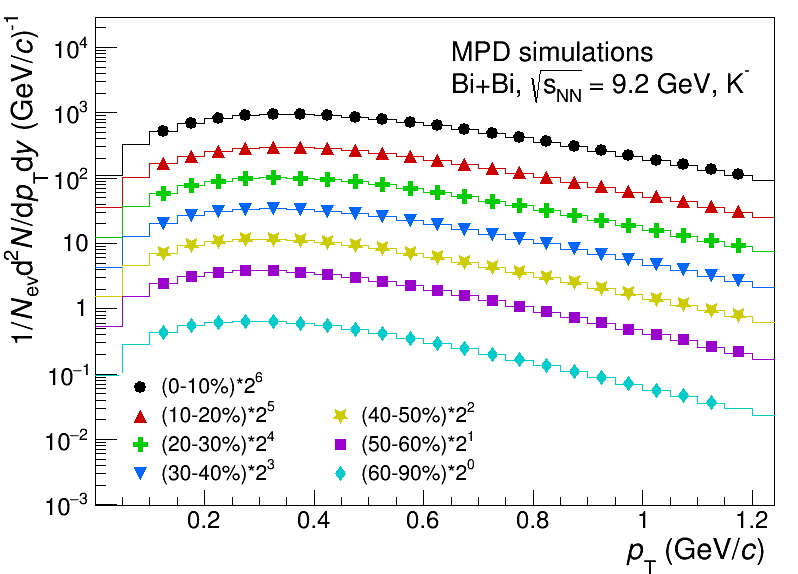}
\includegraphics[width=0.48\textwidth]{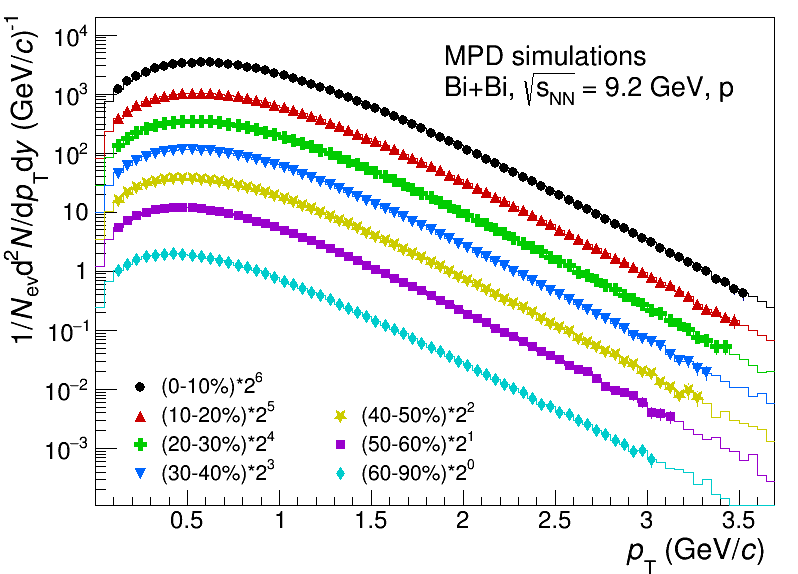}
\hfill
\includegraphics[width=0.48\textwidth]{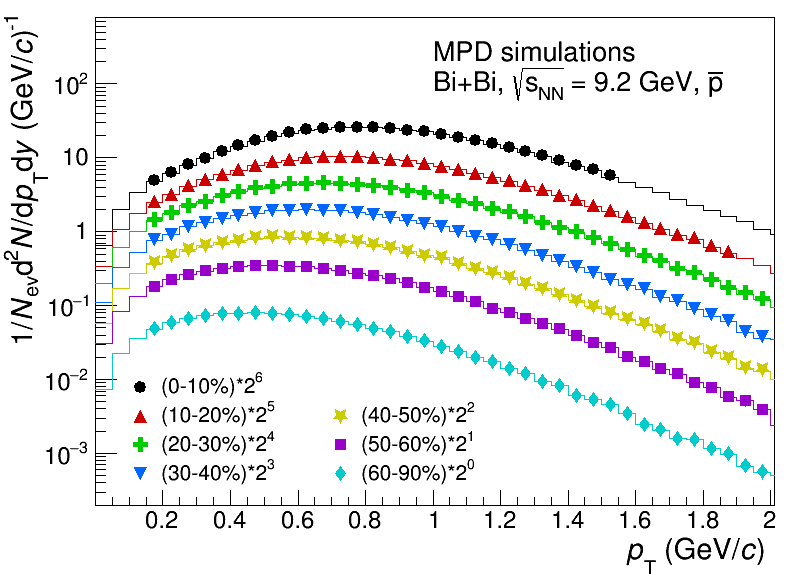}
\caption{The reconstructed (markers) and generated (histograms) transverse momentum spectra for $\pi^{+}$, $\pi^{-}$, $\kap$, $\kam$, $p$ and $\overline{p}$ for midcentral ($|y|<0.5$) Bi+Bi collisions at $\snn = 9.2$~GeV in different centrality intervals. \label{fig:PiKP_ap2}}
\end{figure*} 

The raw-yields ($N_\mathrm{raw}^\mathrm{a}$), obtained for particles of type a (a stands for 
charged pions, kaons or (anti)protons) in different intervals of transverse momentum, are 
corrected for the reconstruction efficiencies, estimated as a product of acceptance ($A$) and 
detector efficiency ($\epsilon$),
$A \times \epsilon = N_\mathrm{raw}^\mathrm{a}/N_\mathrm{gen}^\mathrm{a}$,
where $N_\mathrm{gen}^\mathrm{a}$ is the
number of  generated primary particles of a given type. The evaluated reconstruction efficiencies 
for the \TPC and \TOF depend on the particle transverse momentum and are on average a few tenths of 
percent. The transition points from \TPC-spectra to \TOF-spectra are chosen based on an analysis 
of statistical uncertainties and are set equal to $\pT=0.95$, 0.4 and 0.7 \GeVc for charged pions, 
kaons and (anti)protons, respectively.

The feed-down contributions from decays of heavier hadrons do not exceed $5(10)\%$ for \pip(\pim) at $\pT < 0.2$~\GeVc and are negligible for charged kaons for all momenta. The corresponding contributions for $p(\overline{p})$ vary from $40\%$ to $10\%$ with transverse momentum, with $\Lambda$-hyperon decays giving the main contribution. The reconstructed proton yield is also significantly contaminated by protons produced in collisions with the beam pipe at $\pT < 0.2$~\GeVc. As a result, the measurements of the $p$ and $\overline{p}$ spectra are limited to the momentum range $\pT > 0.2$~\GeVc.

The fully corrected \pT spectra of charged pions, kaons and (anti)protons, reconstructed with approach 2 are shown in Figure~\ref{fig:PiKP_ap2} for different centrality intervals. Within the measurement ranges, MPD samples $91\%$ of the charged pion production with $4\%$ and $5\%$ of the total yield in the unmeasured regions at low and high \pT, respectively. The situation is similar for charged kaons, for which MPD samples $>93\%$ of the total yield with $1\%$ and $<7\%$ of the total yield unmeasured at low and high \pT. The best coverage is provided for protons for which more than $98\%$ of the total yield is sampled in the detector with $2\%$ of the remaining yield in the unmeasured region at low \pT. For antiprotons, MPD samples $> 92\%$ of the total yield with $2\%$ and $<6\%$ of the unmeasured yield at low and high \pT, respectively. The unmeasured yields can be recovered by extrapolating the fits to the measured spectra, similar to that described for the  approach 1, with smaller uncertainties due to a wider coverage at low momentum. 

The two approaches produce fully consistent results. The first approach provides \pT measurements in a wider momentum range, but relies on purity corrections that are model dependent and can be quite significant at higher \pT. When analyzing real data, the corrections should be carefully evaluated in an iterative process by reweighting the particle differential yields in the event generators to the measured ones. However, this is the only possible approach to study the production of charged pions, kaons and (anti)protons at intermediate and high \pT. The second approach limits \pT measurements to ranges where particle purity exceeds 95\%, leaving little room for purity corrections and corresponding uncertainties, making the whole analysis more straightforward. Such measurements have better coverage at low \pT and are best suited to measure particle integrated yields.
%As one can conclude from Figs.~\ref{hadrons_pt} and~\ref{fig:PiKP_ap2}, both spectra agree with spectra at generator level and thus with each other. 
%In analysis of real data two approaches can be used to provide crosscheck and alternative estimate of the systematic uncertainties.

%\textcolor{red}{Comparison of approach1/approach2? Comparison of TPC/TOF spectra?}

%\subsubsection{$\Lambda$, $\Xi$ and $\Omega$  hyperons}
\subsubsection{Hyperon reconstruction}
\label{sec:hyperons}

Since the energy threshold for strangeness production in the QGP phase is smaller than in the hadron gas phase, an enhanced production of strange particles (kaons and hyperons) was proposed as a signature of the transition to QGP~\cite{Rafelski:1982pu}. Relative strangeness production, tested via the $K/\pi$ ratio, was observed to be enhanced in central heavy-ion collision at CERN SPS energies~\cite{NA49:2007stj}.  For hyperons, the increase of the production rate with respect to elementary $p p$ reactions was observed in a broad energy range~\cite{WA85:1999iwr,STAR:2011fbd,ALICE:2013xmt}, stronger for particles with larger strangeness content. However, there are other possible  explanations for the observed strangeness enhancement such as multi-mesonic reactions in dense nuclear matter~\cite{Greiner:2000tu}, partial chiral symmetry restoration~\cite{Palmese:2016rtq}, vanishing of the canonical suppression with increasing multiplicity~\cite{Cleymans:1990mn} or calculations within the core-corona approach~\cite{Becattini:2008ya}. 
In addition to the yields, the $\pT$ distributions of hyperons provide important information on the reaction dynamics. 

Due to their small hadronic reaction cross sections, multi-strange hadrons cannot effectively pick up collective flow during the fireball evolution.  Therefore, the transverse momentum spectra of cascades reflect the initial conditions of a collision. Investigation of strange particle production as a function of beam energy and system size remains an essential part of the NICA research program.

The hyperon analysis is performed using the UrQMD event generator (first production in 
Table~\ref{EvGens}).  
All events with the reconstructed vertex position within $|\zvtx|<130$~cm are used.
The reconstruction of \lmb (\almb) is carried out using the V0
decay mode $\lmb\rightarrow p + \pim$ ($\almb\rightarrow \overline{p} + \pip$).
For a given event, all possible pairs of (anti)protons and charged pions,
having $N_\mathrm{hits}^{\TPC} > 20$ per track, are identified.
For each pair, the point of closest approach of particle trajectories (i.e. a potential decay vertex) is then determined by extrapolating
tracks back to the beam axis. 

\begin{figure}[H]
\centering
\includegraphics[width=0.9\linewidth]{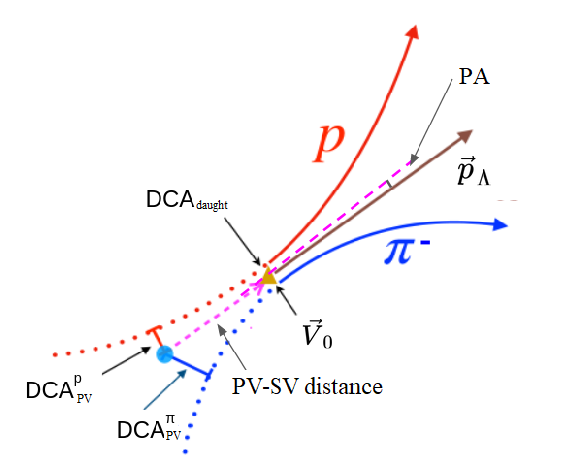}
\caption{Topology of V0 decay shown for the case of $\lmb\rightarrow p + \pim$. \label{fig:V0_Geo}}
\end{figure}

% \begin{figure}[H]
% \centering
% \caption{Invariant mass spectra of ($\bar{p}$,\pip) pairs in transverse momentum interval 1.25$<$\pT$<$1.5~\GeVc. Reconstructed data are plotted by symbols, the results of fits by a Gaussian plus a polynomial function is shown by the line. \label{antilambda_invar}}
% \end{figure} 
%%%% Do not remove, otherwise Figure13 will go next page

\begin{figure*}[t]
\includegraphics[width=0.49\textwidth]{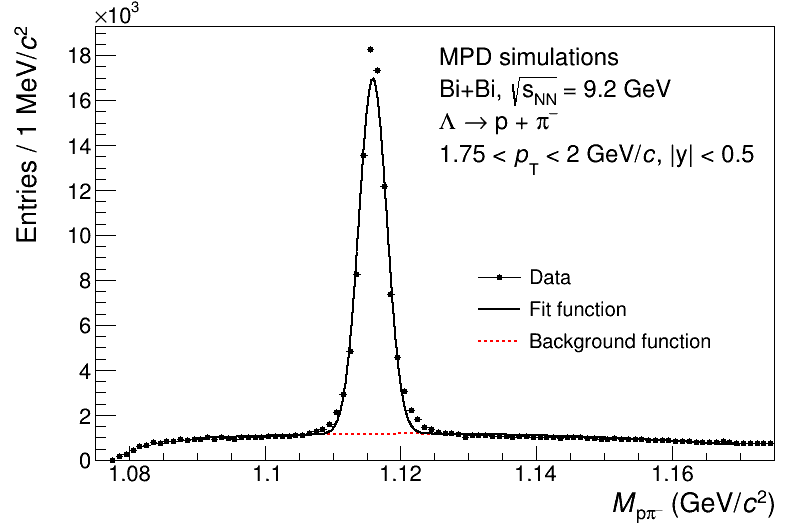}
\hfill
\includegraphics[width=0.49\textwidth]{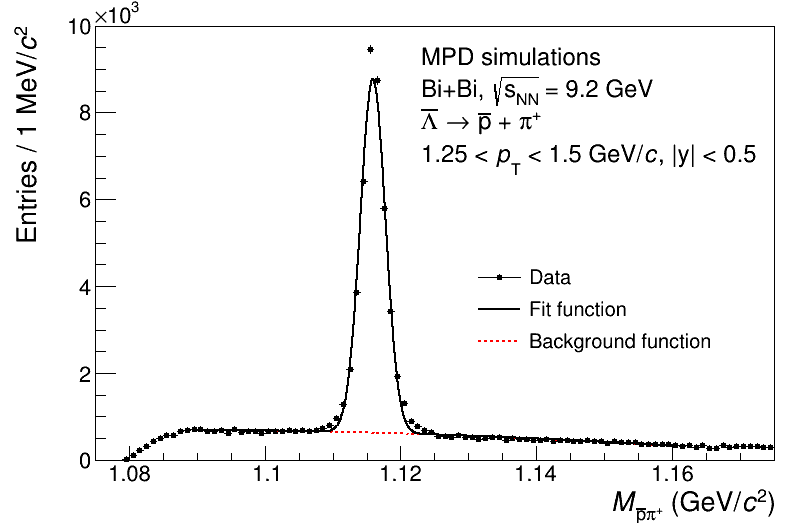}
\caption{Left: Invariant mass spectra of ($p$,\pim) pairs in the transverse momentum interval $1.75<\pT<2$~\GeVc. Reconstructed data are plotted by symbols, the result of a fit using a Gaussian plus a polynomial function of second order is shown by the line. Right: invariant mass spectra of ($\overline{p}$,\pip) pairs in the transverse momentum interval 1.25$<$\pT$<$1.5~\GeVc. Reconstructed data are plotted by symbols, the result of a fit using a Gaussian plus a polynomial function is shown by the line. \label{lambda_invar}}
\end{figure*}

In order to reduce the background from random track
crossings (combinatorial background), several cuts are imposed as explained in the text below and illustrated in Figure~\ref{fig:V0_Geo}. These cuts
include:  a) DCA of decay daughter particles to the primary vertex (DCA$_{\rm PV}$) - this cut is imposed in the $\chi^2$-space, i.e., after
normalization to respective parameter errors; b) quality of the secondary vertex reconstruction ($\chi_\mathrm{vertex}^{2}/$NDF); c) DCA between the daughters in the secondary vertex (DCA$_{\rm daught}$); d) the distance between the primary and secondary vertices (PV-SV Distance); e) the value of the pointing angle (PA), defined as the angle between the reconstructed parent particle momentum vector and the line connecting the primary and secondary vertices. The selection criteria have been optimized to achieve the best significance, defined as $S$/($S$+$B$), where $S$ is the hyperon signal and $B$ is 
the background under the signal peak. 
The actual values of the topological cut parameters for $\Lambda$($\bar{\Lambda}$) are given in Table~\ref{table_cuts_lambda}.

\begin{figure}[H]
\centering
\includegraphics[width=0.9\linewidth]{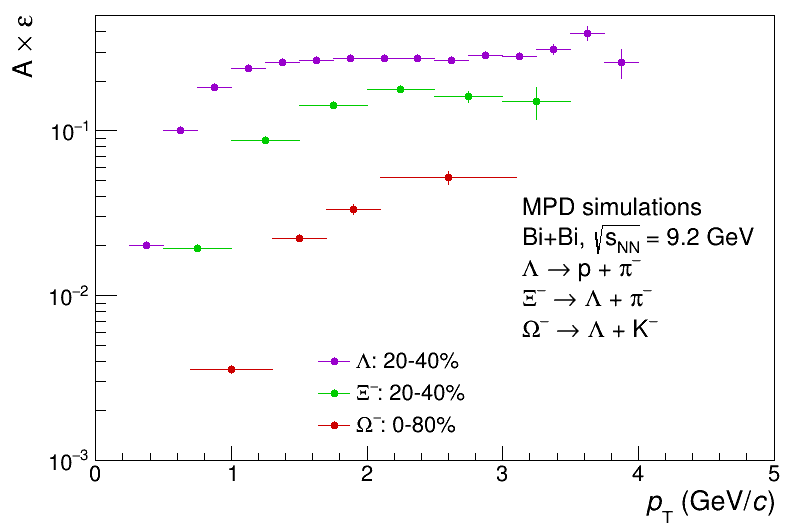}
\caption{The reconstruction efficiency ($A \times \epsilon$) for $\Lambda$, $\Xi$, and $\Omega$ as functions of \pT in centrality selected \BiBi collisions. \label{hyper_eff}}
\end{figure}

For each selected pair of daughter particles, the invariant mass of the
parent hyperon is then calculated. Figure~\ref{lambda_invar} shows 
invariant mass distribution for $p\pim$ (left panel) and $\overline{p}\pip$
 (right panel) pairs. In order to extract the raw signal, the background under the peak 
region has to be estimated. 
For this purpose, a combined fit of a Gaussian 
for the signal and a second order
polynomial function for the background is applied. The raw yields of hyperons are determined by bin counting in the
$\pm 5\sigma$ interval around the measured peak position with subsequent subtraction of the polynomial function integral estimated for \noindent
the same invariant mass range. 
The resulting hyperon yield is then corrected for the reconstruction efficiency ($A \times \epsilon$, see Figure~\ref{hyper_eff}), which accounts for signal
losses due to the finite detector acceptance, track reconstruction efficiency,
and the applied cuts

\begin{equation}
\frac{d^{2}N}{dy d\pT} = \frac{1}{\Nev}\frac{N_\mathrm{raw}}{\Delta \pT \Delta y }\frac{1}{A \times \epsilon} \frac{1}{\mathrm{BR}}, 
\label{eq:CorrYield}
\end{equation}
where $N_\mathrm{raw}$ is the number of reconstructed particles from the invariant mass distributions, $\Delta \pT$ and $\Delta y$ are the intervals in $\pT$ and rapidity, $A \times \epsilon$ is the reconstruction efficiency, BR is the decay branching ratio and \Nev is the number of analyzed events in a given centrality interval. 

\begin{table}[H]
\centering
\caption{Selection criteria for $\Lambda$ and $\bar{\Lambda}$ reconstruction. \label{table_cuts_lambda}}
\vspace{1mm}
\begin{tabular}{c|c|c}
\hline
\textbf{Selection} & \boldmath{$\Lambda$} & \boldmath{$\overline{\Lambda}$} \\
\hline
DCA$_{\rm PV}$ (cm) & $>4.0(\pim)$ & $>4.0(\pip)$ \\
 &  $>2.5(p)$ &  $>1.5(\overline{p})$ \\
$\chi_\mathrm{vertex}^{2}/$NDF & $<1.75$ & $<1.75$\\
DCA$_{\rm daught}$ (cm) & $<3.0$ & $<2.8$ \\
PV-SV Distance (cm)  & $>2.0$ & $>2.0$ \\
PA (radians) & $<0.08$ & $<0.14$ \\
\end{tabular}
\end{table}

\begin{figure*}[t!]
\centering
\includegraphics[width=0.47\textwidth]{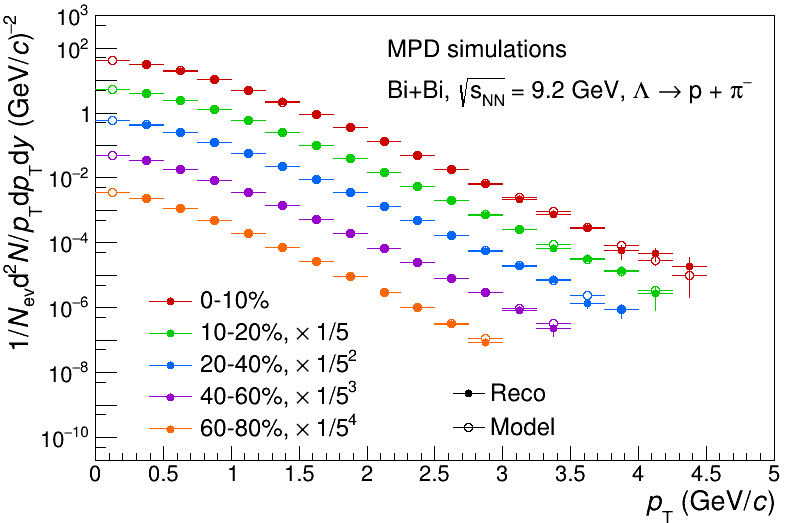}
\hfill
\includegraphics[width=0.47\textwidth]{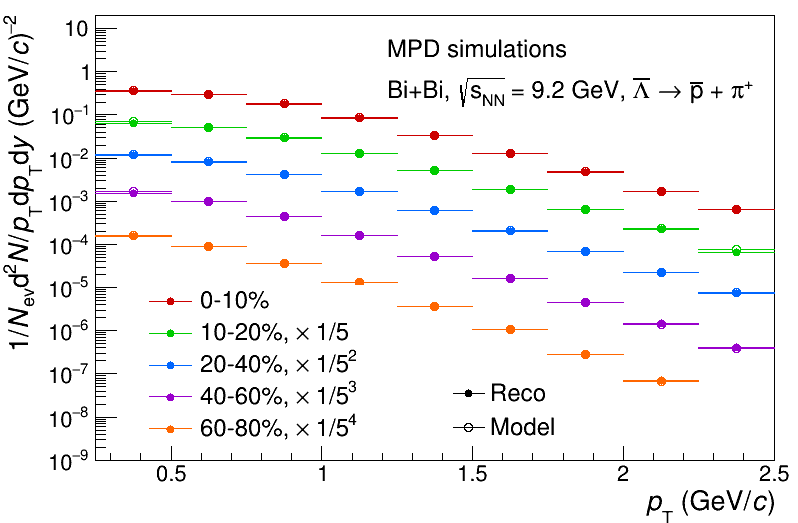}
\caption{Left: Midrapidity transverse momentum spectra of $\Lambda$ in centrality selected \BiBi collisions. Reconstructed distributions are shown with solid symbols; empty symbols show the initially generated distributions from the model. Right: The same for $\bar{\Lambda}.$ \label{lambda_ptspec}}
\end{figure*}

The reconstruction efficiency for \lmb
%, $\Xi$, and $\Omega$
as a function of \pT is shown in Figure~\ref{hyper_eff}.
We found small variations in $A \times \epsilon$ with 
the collision centrality. The reconstructed invariant transverse momentum spectra of \lmb and 
\almb in centrality selected \BiBi collisions are shown in 
Figure~\ref{lambda_ptspec}. The distributions, reconstructed within the rapidity range $|y|<0.5$, are shown with solid symbols, while corresponding distributions, calculated at generator level, are shown with empty symbols. Both spectra agree within the  uncertainties. Due\ \ to \ the very short  time scale of the electromagnetic
decay $\Sigma^0\rightarrow \lmb + \gamma$ ($\sim 10^{-19}$ s), the \lmb-hyperons originated in $\Sigma^0$ decays are experimentally indistinguishable from the primary \lmb-hyperons.  Therefore, 
the results for the yield of $\lmb$ and $\almb$-hyperon represent the summed contribution from \lmb and  ($\Sigma^0\rightarrow \lmb + \gamma$),  \almb  and  ($\bar{\Sigma^0}\rightarrow \almb + \gamma$), respectively.

\begin{figure*}[b]
\centering
\includegraphics[width=0.47\textwidth]{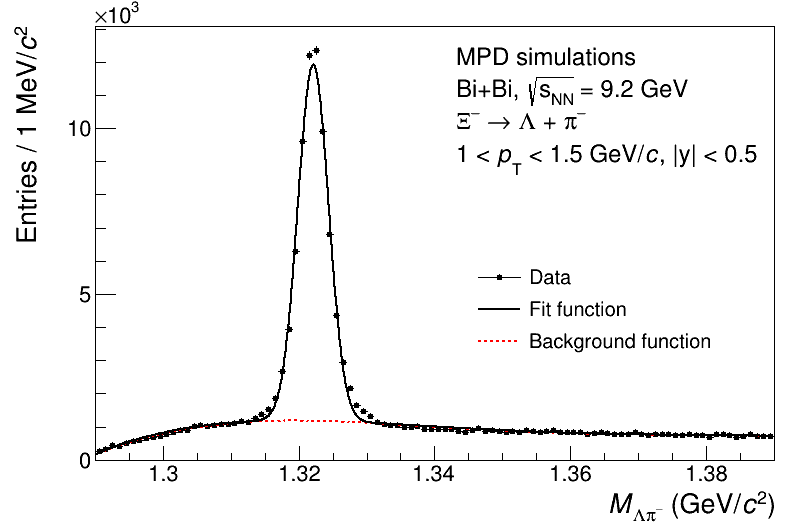}
\hfill
\includegraphics[width=0.47\textwidth]{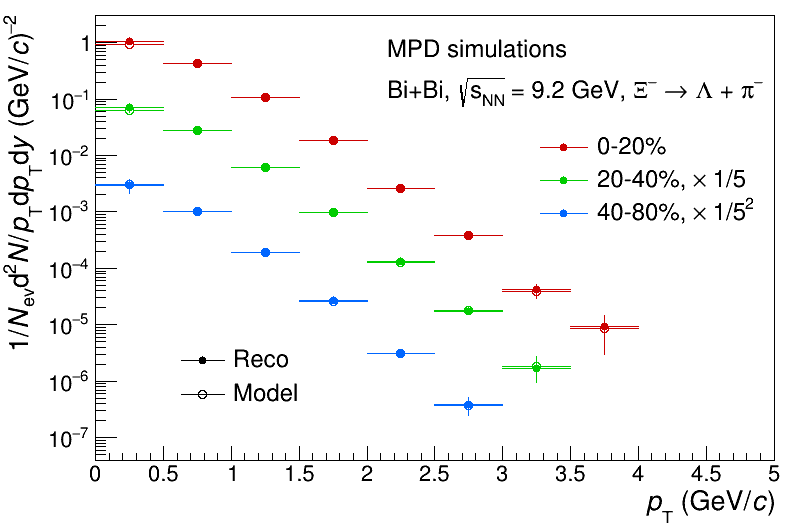}
\caption{Left: Invariant mass distribution for $\Lambda \pi^-$ pairs at $1.0<p_T<1.5$~\GeVc. Right: Midrapidity transverse momentum spectra of $\Xi^-$ in centrality selected \BiBi collisions. Reconstructed distributions are shown with solid symbols; empty symbols show the initially generated distributions of the model. \label{ksi_ptspec}}
\end{figure*} 
\noindent 
The capability of the MPD detector to reconstruct \lmb(\almb), $\Xi^-$($\bar{\Xi}^+$) and $\Omega^-$($\bar{\Omega}^+$) hyperons in central Au + Au collisions at $\snn = 9$~GeV was investigated previously in \cite{Ilieva:2015qwa}  showing reasonable yields of these particles in 10 weeks of data taking with the expected operational luminosity. However, the yield of multi-strange anti-hyperons is very low at NICA energies decreasing systematically with an increasing number of strange quarks. Therefore, in what follows for the 1st stage of heavy-ion collisions at NICA, we perform only an analysis of multi-strange hyperons.
%The yield of multi-strange anti-hyperons is very low at NICA energies  decreasing systematically with increasing number of strange quarks (for example, $\bar{\Xi}^+$/$\Xi^-$\,$\approx$\,0.05 in Bi+Bi collisions at $\snn$\,=\,9~GeV in the UrQMD model).
%Therefore in what follows, we perform only an analysis of multi-strange hyperons. 
Once \lmb-hyperons are reconstructed, the cascade hyperons are reconstructed as well using the decay mode $\Xi^-\rightarrow\lmb+\pim$. The
\lmb candidate for pairing with $\pim$ is determined requiring the invariant
mass to be within $\pm\ 5\sigma$ relative to the nominal value. To improve the signal purity, the topological selection criteria similar to (a)–(e) described above are 
applied (see Table~\ref{table_cuts_ksi_omega}). For example, Figure~\ref{ksi_ptspec} (left panel) shows an invariant 
mass distribution for $\lmb\pim$ pairs in the transverse momentum interval 
$1.0<\pT<1.5$~\GeVc. The reconstruction efficiency for $\Xi^-$ as a function of \pT is shown in Figure~\ref{hyper_eff}. The right panel of Figure~\ref{ksi_ptspec} shows the 
reconstructed invariant \pT-spectra of $\Xi^-$-hyperons in centrality selected 
\BiBi collisions. The difference between the reconstructed and the generator level spectra is small.

\begin{figure*}[t]
\centering
\includegraphics[width=0.49\textwidth]{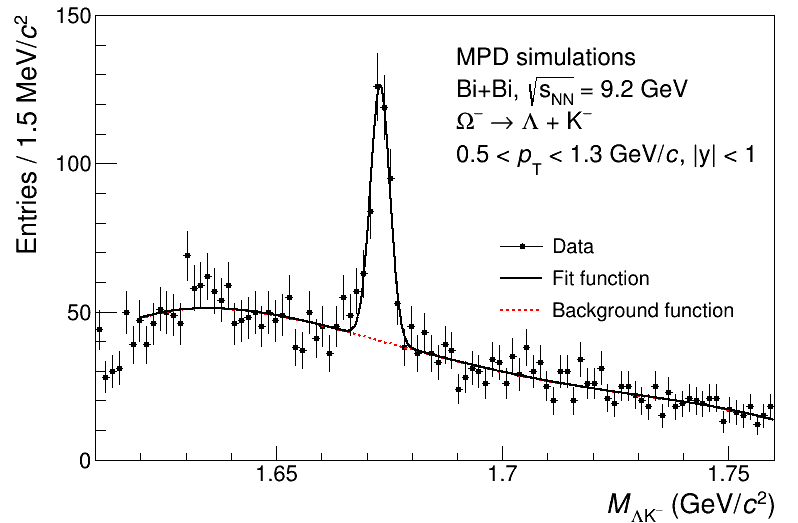}
\hfill
\includegraphics[width=0.49\textwidth]{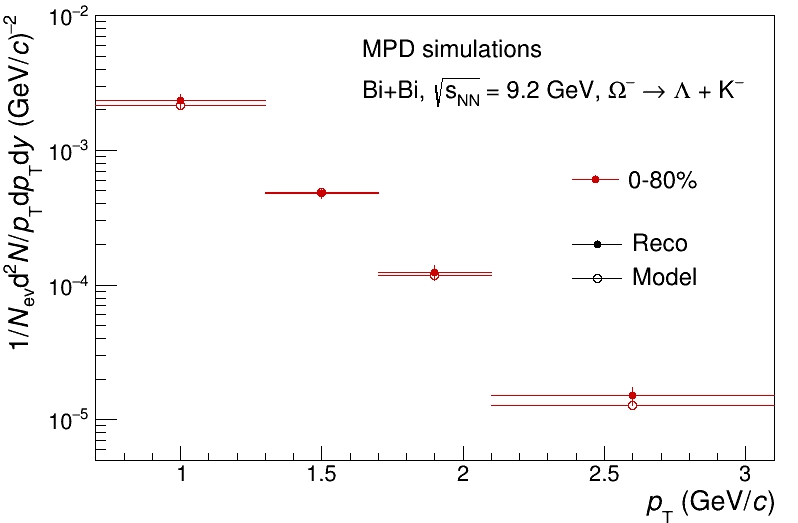}
\caption{Left: Invariant mass distribution for ($\lmb,\kam$) pairs at $0.5<\pT<1.3$~\GeVc. Right: Midrapidity transverse momentum spectrum of $\Omega^-$ in 0-80\% central \BiBi collisions. Reconstructed distributions are shown with solid symbols, empty symbols show the initially generated distributions of the model. \label{omega_ptspec}}
\end{figure*} 

The yield of $\Omega$-hyperons in heavy-ion collisions is small, thus, the analysis was performed in a larger rapidity interval ($|y|<1$) and for a wider centrality selection 0-80\%. The selection criteria applied for $\Omega$ are given in Table~\ref{table_cuts_ksi_omega}, the efficiency $p_T$-dependence is plotted in Figure~\ref{hyper_eff}. The left panel of Figure~\ref{omega_ptspec}  shows the invariant mass distribution for ($\lmb,\kam$) pairs in the \pT-interval $0.5<\pT<1.3$~\GeVc, while the right panel shows a good agreement of the reconstructed \pT-spectrum of $\Omega^-$ in \BiBi interactions with the spectrum obtained at generator level.

\begin{table}[H]
\caption{Selection criteria used for $\Xi^-$ and $\Omega^-$.
\label{table_cuts_ksi_omega}}
\vspace{1mm}
\begin{tabular}{c|c|c}
\hline
Selection & $\Xi^-$ & $\Omega^-$ \\
\hline
DCA$_{\rm PV}$ (cm) & $>8.0(\pim)$ & $>7.5(\kam)$\\ 
 & $>2.5(\Lambda)$ & $>4.0(\Lambda)$ \\
DCA$_{\rm daught}$ (cm) & $<0.8$ & $<0.5$ \\
PV-SV Distance (cm) & $>1.0$ & $>1.0$ \\
PA (radians) & $<0.06$ & $<0.06$ \\
%\bottomrule
\end{tabular}
\end{table}

The hyperon feasibility study shows that measurements of $\Lambda$($\bar{\Lambda}$) and $\Xi$ are
possible with a data set of several million events.
%with MPD during the first period of data taking.
Much larger data sets are needed to measure the production and centrality dependence of
(multi)strange hyperons at NICA energies.
%However, a data set of $10^7$ \BiBi events is not sufficient to obtain enough signal statistics
%for $\Omega$. In order to measure the system-size dependence of (multi)strange hyperon production
%at NICA energies, much larger data sets are needed.     

\subsubsection{Short-lived hadronic resonances}

Measurements of short-lived hadronic resonances such as $\rho(770)^{0}$, K$^{*}$(892), $\phi$(1020), $\Sigma$(1385)$^{\pm}$ and $\Lambda$(1520) at RHIC~\cite{STAR:2004bgh,STAR:2010avo,STAR:2008twt,Yamamoto:2001ph,STAR:2004yym,STAR:2006vhb,STAR:2008lcm} and the LHC~\cite{ALICE:2014jbq,ALICE:2016sak,ALICE:2017ban,ALICE:2017pgw,ALICE:2019smg,ALICE:2018ewo,ALICE:2018qdv,ALICE:2019aid} have been used to study enhanced strangeness production, dominant hadronization mechanisms and vector meson spin alignment. However, resonances are most useful in studying the lifetime and properties of the late hadronic phase~\cite{Riabov:2019hfe,ALICE:2019xyr}, which may distort signals of 
the crossover or the chiral symmetry restoration phase transition.
Measurements of resonance properties in heavy-ion collisions at $\snn = 7.7-5020$~GeV revealed that  production of resonances with lifetimes $\tau < 20$~fm/$c$ is
suppressed in central collisions, while production of longer-lived resonances like $\phi$(1020) remains almost unchanged from peripheral to central collisions. The observed modifications show a smooth evolution with the final state charge particle multiplicity in different collision systems. The suppression of resonance yields in central heavy-ion collisions is explained by rescattering of daughter particles in the hadronic phase. The modifications  occur at multiplicities expected in (semi)central heavy-ion collisions at NICA energies~\cite{PHENIX:2015tbb}. The yield modifications are also predicted by cascade model calculations at NICA energies~\cite{Riabov:2022mzg,Ivanishchev:2021ujg,Ivanishchev:2021gai}. This provides a strong incentive to study resonances in heavy-ion collisions at intermediate energies with the ultimate goal of achieving a comprehensive understanding of the hadronic phase.

\begin{figure*}[t]
\centering
\includegraphics[width=0.48\textwidth]{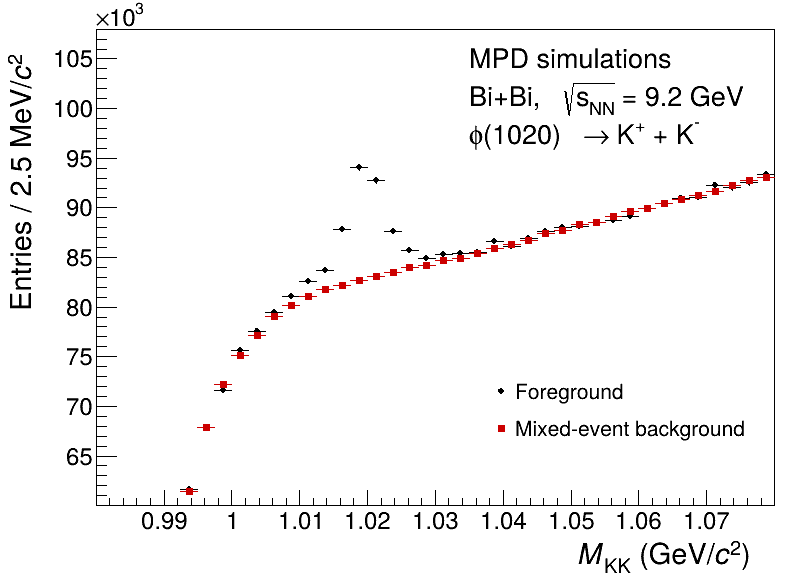}
\hfill
\includegraphics[width=0.48\textwidth]{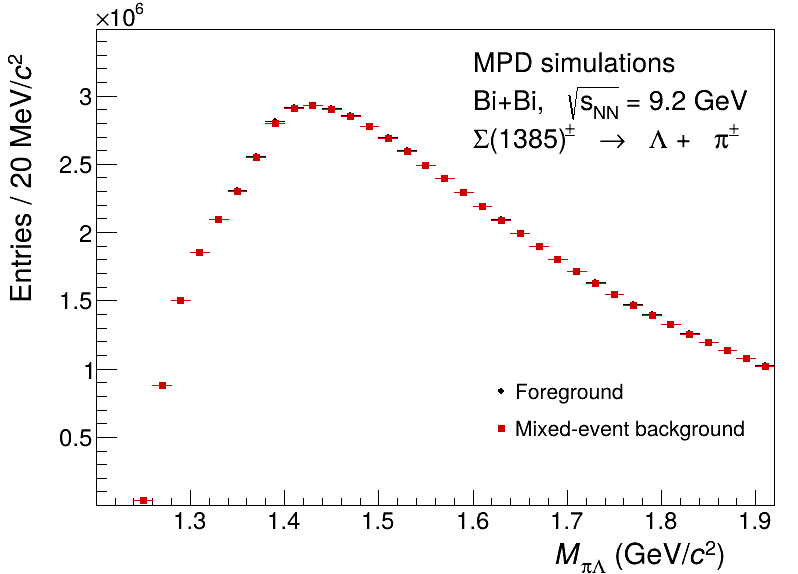}
\includegraphics[width=0.48\textwidth]{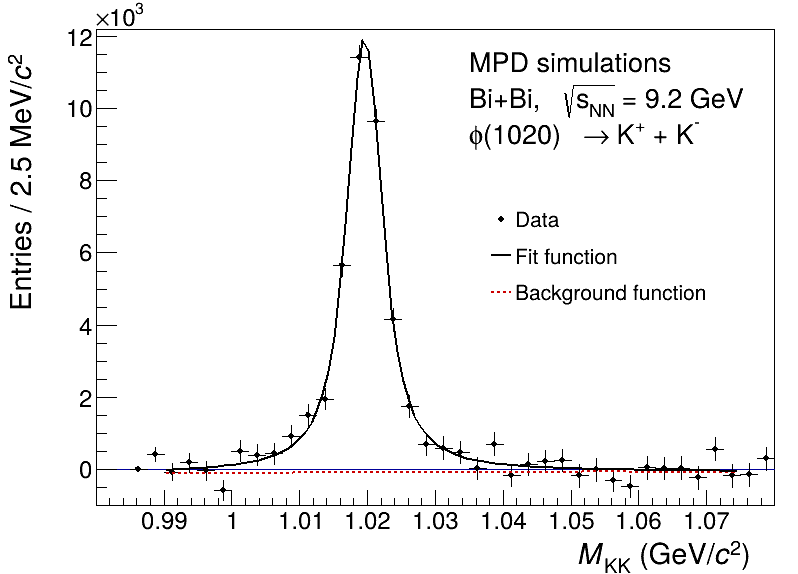}
\hfill
\includegraphics[width=0.48\textwidth]{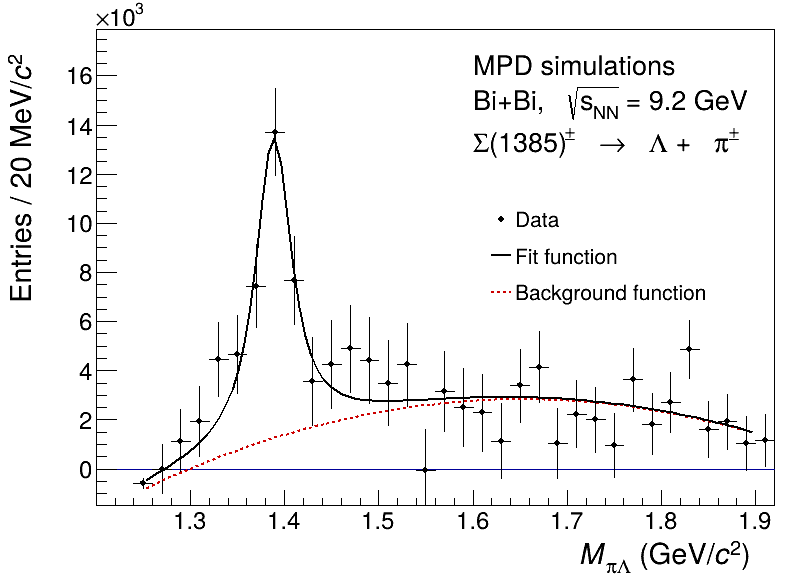}
\caption{The invariant mass distributions for $\kap\kam$ (left) and $\pi^{\pm}\Lambda$ (right) pairs accumulated for the same and the mixed events in \BiBi collision at $\snn = 9.2$~GeV. The bottom panels show the distributions after subtraction of the mixed-event background. The resulting distributions are fit to a combination of a second-order polynomial and  the Voitian function. Examples are shown for 0-10\% central \BiBi collisions at $\snn = 9.2$~GeV in the transverse momentum interval 0.2-0.4 (0.4-0.6)~\GeVc for $\kap\kam$ ($\pi^{\pm}\Lambda$) pairs.\label{fig:MinvReso}}
\end{figure*}

Production 1 from Table~\ref{EvGens} was used to study MPD capabilities to reconstruct short-lived resonances in \BiBi collisions at $\snn = 9.2$~GeV. The reconstructed vertex had to be within |$\zvtx$| < 130 cm, and only events with reconstructed centrality in the range 0-91$\%$ were accepted in the analysis. Charged daughter particles from resonance decays were treated as primary particles because the vertices of resonance decays are indistinguishable from the primary vertex. Such particles had to have at least 24 hits (out of a maximum 53) reconstructed in the TPC and to match to the primary vertex within 3$\sigma$. Secondary particles from K$_{s}^{0}$ and $\Lambda$ decays were required to have at least 10 hits in the TPC.  Only tracks with \pT > 0.1 \GeVc were accepted. Charged hadrons were identified by a 2$\sigma$ cut on the value of $\langle \dEdx \rangle$ measured in the \TPC. If the track was matched to TOF, the track was additionally required to be identified by a 2$\sigma$ cut on the measured value of particle velocity $\beta$.

The weakly decaying daughter particles (K$_{s}^{0} \rightarrow \pi^{+} + \pi^{-}$ and $\Lambda \rightarrow\ p  + \pi^{-}$) were reconstructed by using topological selections described in Sec.\ref{sec:hyperons} and summarized in Table~\ref{KsLamCuts}. The values were optimized to increase the significance of the reconstructed resonance signals. The $\pi^{+}\pi^{-}$ and $p\ \pi^{-}$ pairs were selected as K$_{s}^{0}$ and $\Lambda$ candidates if their reconstructed invariant masses were within 2$\sigma$ of the expected values, where   $\sigma$ was parametrized as a function of particle transverse momentum. The PDG~\cite{ParticleDataGroup:2020ssz} masses of daughter particles and the reconstructed momenta were used for the measurement of parent resonances. 

\begin{table}[H]
\centering
\caption{The topological selection used to reconstruct weak decays of K$_{s}^{0}$ and $\Lambda$.}
\begin{tabular}{c|c|c}
Selection & K$_{s}^{0}$ & $\Lambda$  \\
\hline
$\chi_\mathrm{vertex}^{2}/$NDF & 3.0 & 3.0 \\
DCA$_{\rm daught}$ (cm) & 1.0 & 1.0 \\
PV-SV Distance (cm) & 0.5 & 0.5 \\
PA (radians) & 0.1 & 0.1 \\
DCA$_{\rm PV}^{\pi}$ (cm) & 7 & 7 \\
DCA$_{\rm PV}^{p}$  (cm) & - & 3 \\
\end{tabular}\label{KsLamCuts}
\end{table}

The daughter particle candidates are paired to accumulate $\kap\kam$, $\pi^{+}\pi^{-}$, $\pi^{+}\kam$, $p\kam$, $\pi^{\pm}\rm{K}_{s}^{0}$ and $\pi^{\pm}\Lambda$ invariant mass distributions  for different centrality intervals 0-10$\%$, 10-20$\%$, 20-30$\%$, 30-40$\%$, 40-50$\%$, 50-60$\%$ and 60-90$\%$ at
\begin{figure*}[t]
\centering
\includegraphics[width=0.49\textwidth]{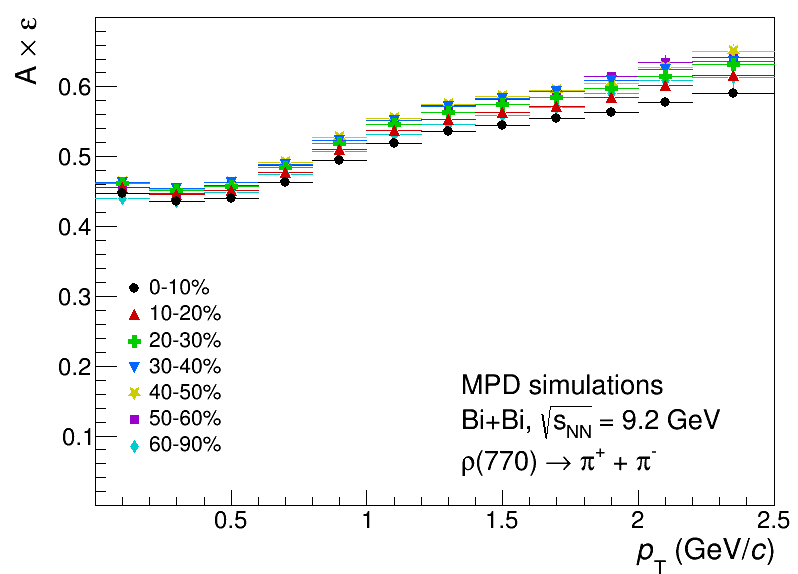}
\hfill
\includegraphics[width=0.49\textwidth]{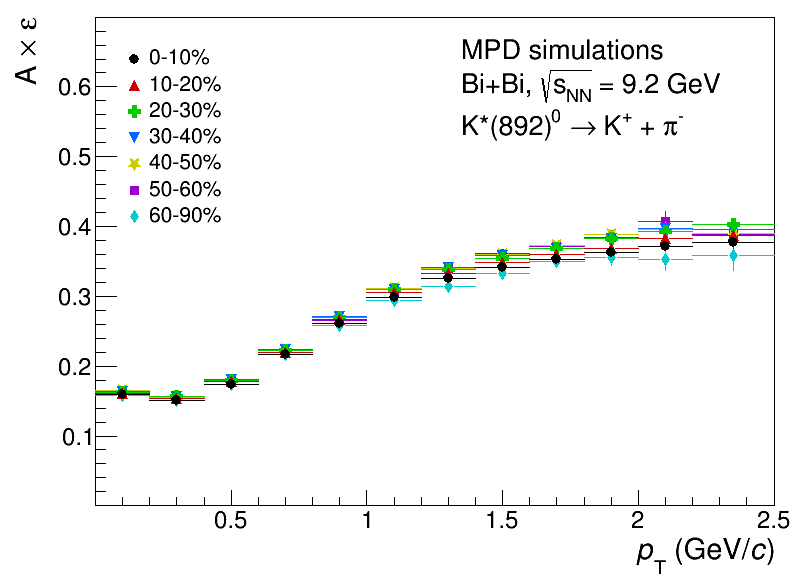}
\includegraphics[width=0.49\textwidth]{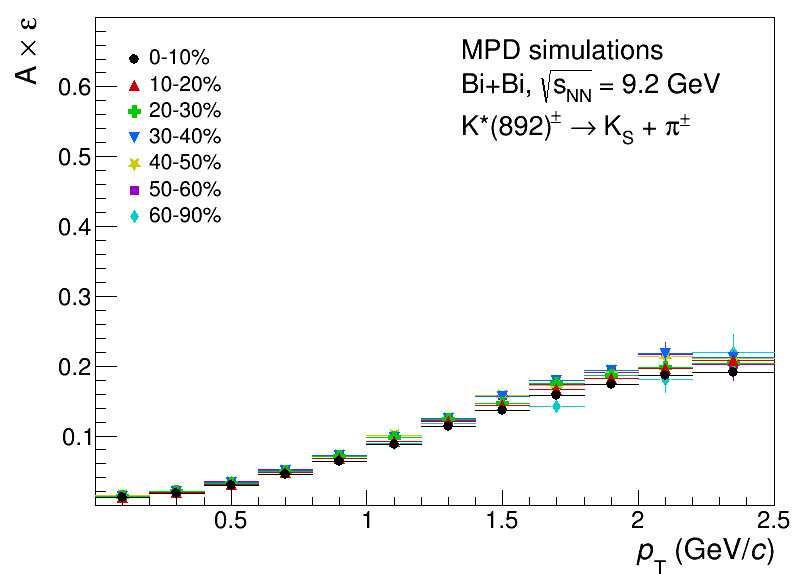}
\hfill
\includegraphics[width=0.49\textwidth]{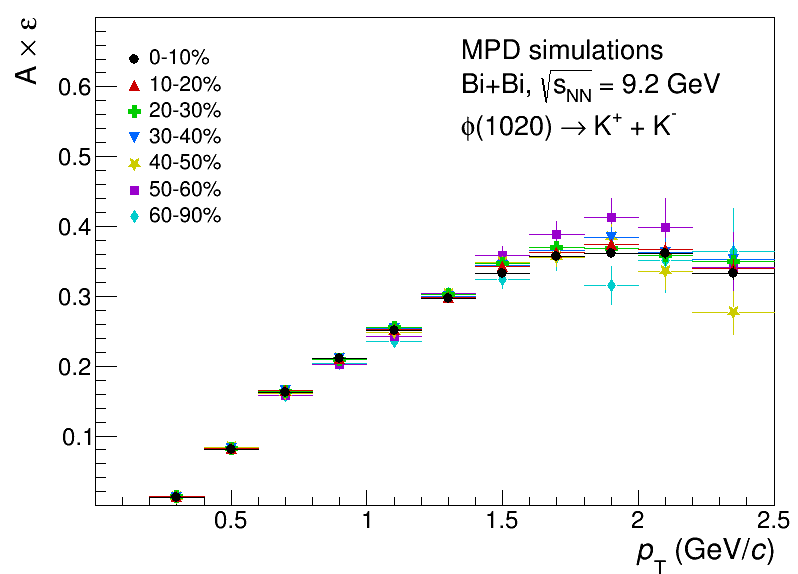}
\includegraphics[width=0.49\textwidth]{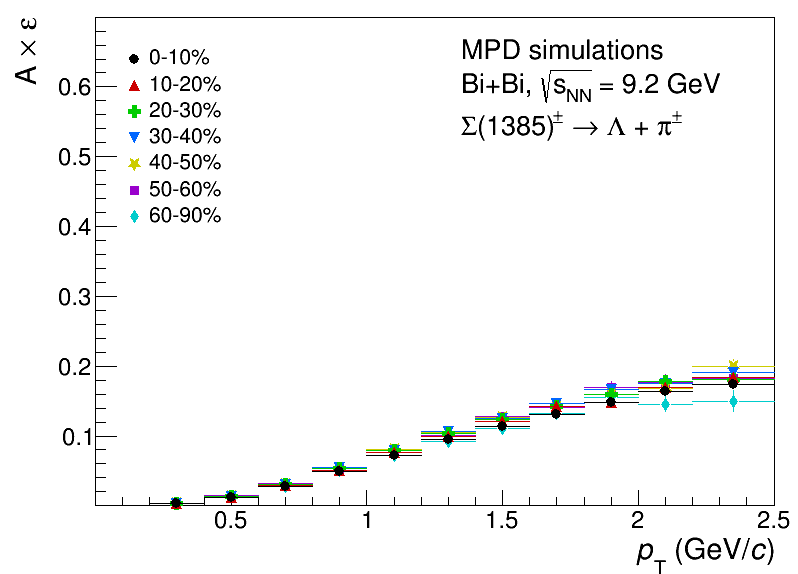}
\hfill
\includegraphics[width=0.49\textwidth]{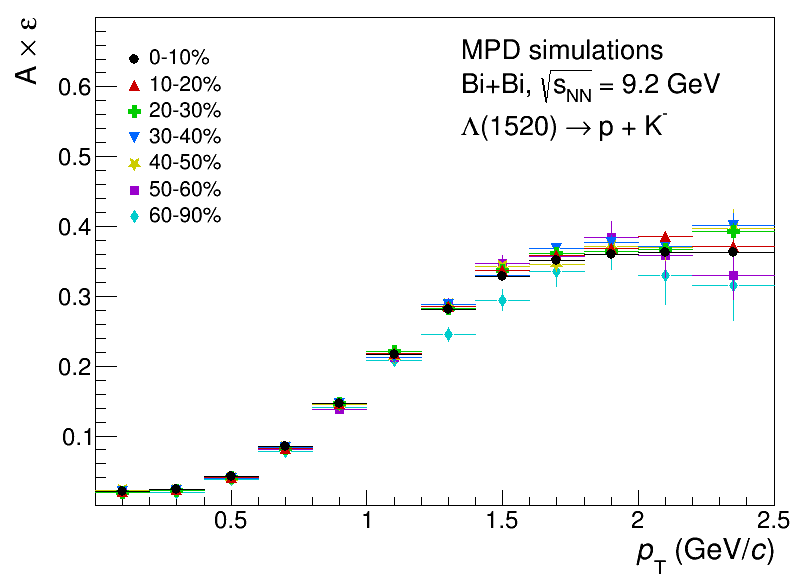}
\caption{Reconstruction efficiencies evaluated for $\rho(770)^{0}$, $\rm{K}^{*}(892)^{0}$, $\rm{K}^{*}(892)^{\pm}$, $\phi$(1020), $\Sigma$(1385)$^{\pm}$ and $\Lambda$(1520) resonances as a function of transverse momentum in different centrality \BiBi collisions at $\snn = 9.2$~GeV.\label{fig:RecEffReso}}
\end{figure*} 
\noindent midrapidity $|y| < 0.5$.  Examples of $\kap\kam$ and $\pi^{\pm}\Lambda$ invariant mass distributions accumulated in 0-20\% central \BiBi collisions at $\snn = 9.2$~GeV are shown in the upper panels of Figure~\ref{fig:MinvReso} by black symbols. 

\begin{figure*}[t]
\centering
\includegraphics[width=0.49\textwidth]{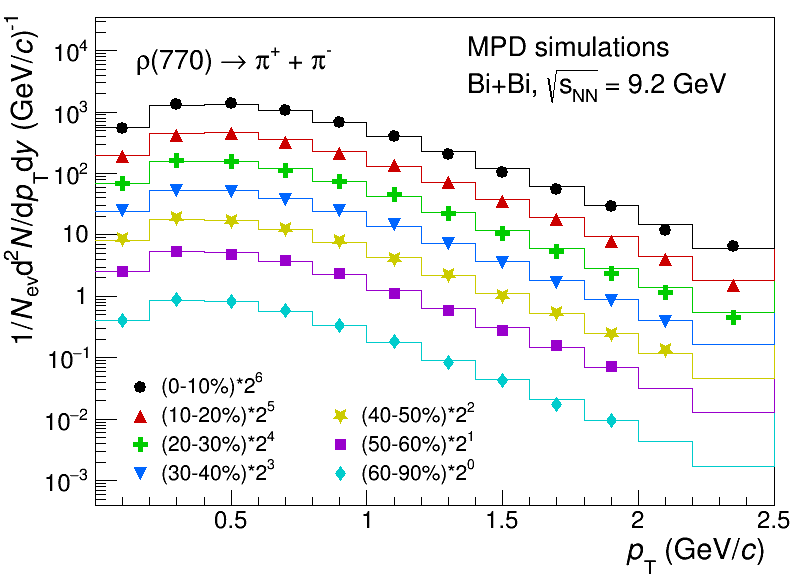}
\hfill
\includegraphics[width=0.49\textwidth]{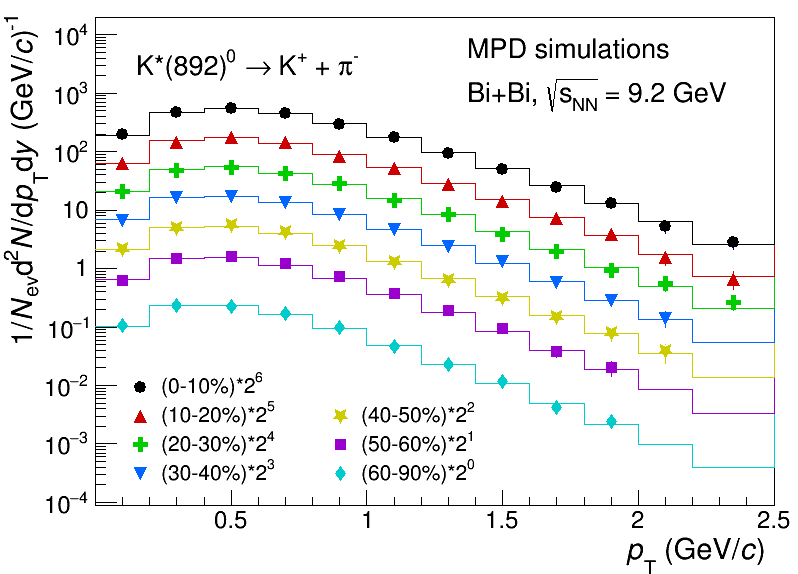}
\\
\includegraphics[width=0.49\textwidth]{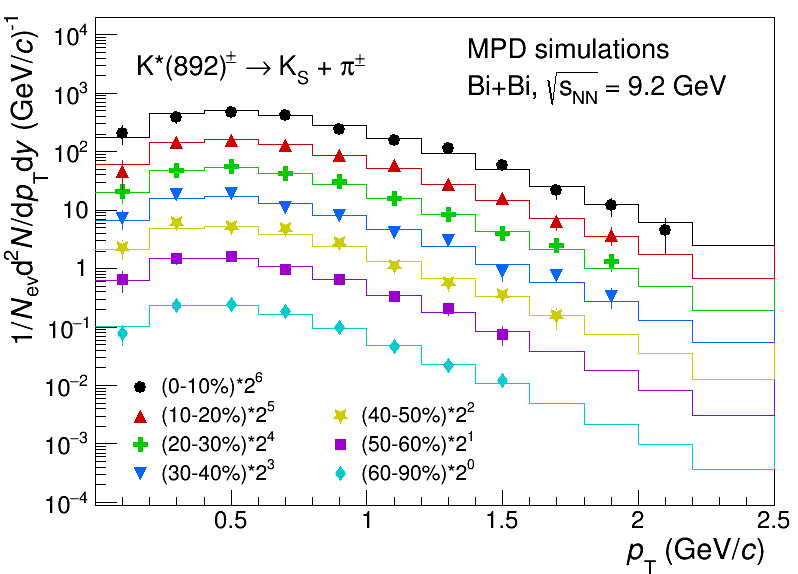}
\hfill
\includegraphics[width=0.49\textwidth]{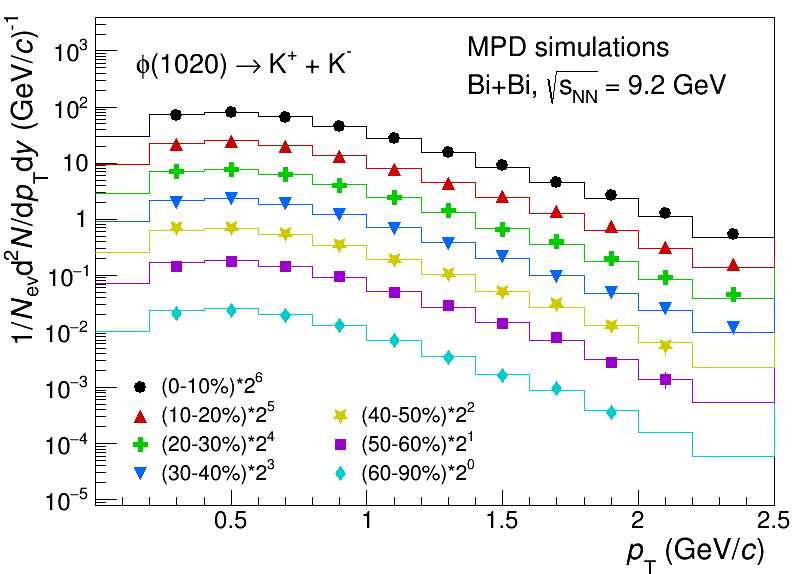}
\\
\includegraphics[width=0.49\textwidth]{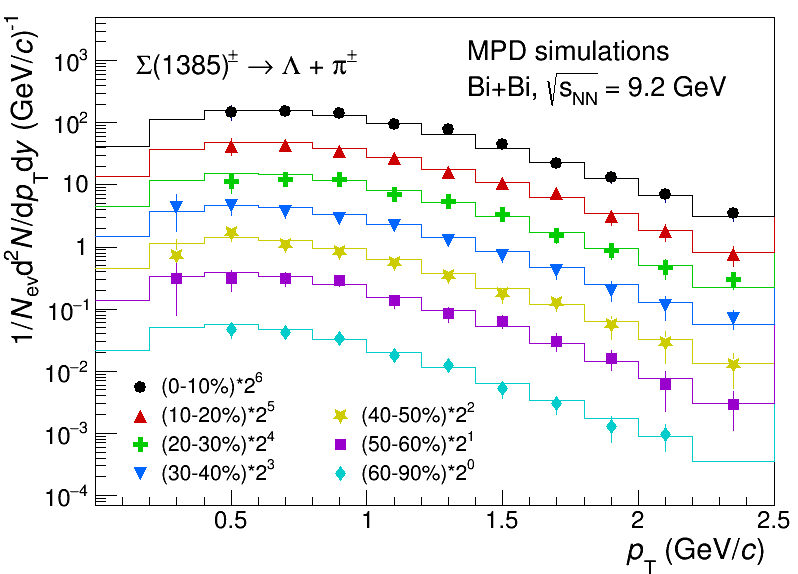}
\hfill
\includegraphics[width=0.49\textwidth]{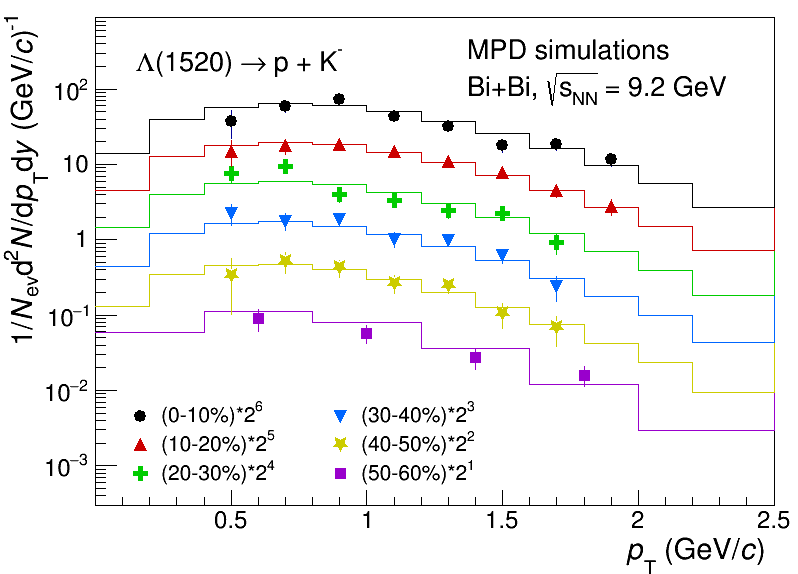}
\caption{The reconstructed (markers) and generated (histograms) transverse momentum spectra for $\rho(770)^{0}$, $\rm{K}^{*}(892)^{0}$, $\rm{K}^{*}(892)^{\pm}$, $\phi$(1020), $\Sigma$(1385)$^{\pm}$ and $\Lambda$(1520) resonances for \BiBi collisions at $\snn = 9.2$~GeV in different centrality intervals.\label{fig:SpectraReso}}
\end{figure*} 

The accumulated invariant mass distributions contain signals from resonance decays and combinatorial background. The uncorrelated combinatorial background is estimated using a mixed-event approach, where one of the daughter particles is taken for the same event and the other from another event with similar multiplicity, $\zvtx$ and event plane. The invariant mass distributions of the mixed events are then scaled to the invariant mass distributions of the same events at higher masses, and then subtracted. The invariant mass distributions of the mixed events are shown by the red symbols in Figure~\ref{fig:MinvReso}. The distributions remaining after subtraction contain peaks from the resonance decays and some remaining correlated background from jets and misreconstructed decays of heavier particles as shown in the lower panels of Figure~\ref{fig:MinvReso}. The remaining background was found to be a smooth function of the mass in the neighborhood of the resonance peaks and can be described with a polynomial. To extract the resonance raw yields, the invariant mass distributions are fitted to a combination of a second-order polynomial, to describe the remaining background, and a Voitian function (the Breit-Wigner function convolved with a Gaussian to account for the finite mass resolution of the detector) for the signal. Examples of the fits are shown in the same plots. The mass resolution of the detector was estimated as a function of transverse momentum and collision centrality for each decay mode studied as the width of a Gaussian fit to the distribution with the difference between the generated and reconstructed resonance masses.

The efficiency of resonance reconstruction at midrapidity  in the MPD setup was estimated as $A \times \epsilon = N_{\rm rec}/N_{\rm gen}$, where $N_{\rm rec}$ and $N_{\rm gen}$ are the number of reconstructed and generated resonances. The number of reconstructed resonances is determined after all event and track selection cuts, while the number of generated resonances accounts for the branching ratios of particular decay channels. The evaluated reconstruction efficiencies for $\rho(770)^{0} \rightarrow \pi^{+} + \pi^{-}$, $\rm{K}^{*}(892)^{0} \rightarrow \kap + \pi^{-}$, $\rm{K}^{*}(892)^{\pm} \rightarrow \pi^{\pm} + K_{s}^{0}$, $\phi(1020) \rightarrow \kap + \kam$, $\Sigma(1385)^{\pm} \rightarrow \pi^{\pm} + \Lambda$ and $\Lambda(1520) \rightarrow p + \kam$ resonances are shown in Figure~\ref{fig:RecEffReso} as functions of transverse momentum and centrality in \BiBi collisions at $\snn = 9.2$~GeV.  The estimated efficiencies are much smaller for resonance decays with weakly decaying daughters because more particles need to be reconstructed. The efficiencies decrease at low momentum, but most resonances can be measured from zero transverse momentum. The efficiencies show a modest dependence on event centrality, they are smaller in central collisions because of the higher detector occupancy.

The fully corrected transverse momentum spectra of $\rho(770)^{0}$, $\rm{K}^{*}(892)^{0}$, $\rm{K}^{*}(892)^{\pm}$, $\phi$(1020), $\Sigma$(1385)$^{\pm}$ and $\Lambda$(1520) resonances are calculated according to Eq.~(\ref{eq:CorrYield}) and are shown with markers of different colors in Figure~\ref{fig:SpectraReso} for different centrality intervals. The obtained spectra are compared to the  generated ones shown by histograms in the same plots. The reconstructed spectra are consistent with the generated ones within statistical uncertainties, which confirms the consistency of the analysis chain. To study resonance production, as a function of centrality, a sample of about $10^{8}$  \BiBi collisions at $\snn = 9.2$~GeV will be required. 
Most resonances, with the exception of $\phi$(1020), can be measured starting from $\pT=0$, which is important to minimize systematic uncertainties in the integrated yield measurements needed for physics studies.

\subsubsection{Light nuclei production}
\begin{figure*}[t]
\centering
\includegraphics[width=0.49\textwidth]{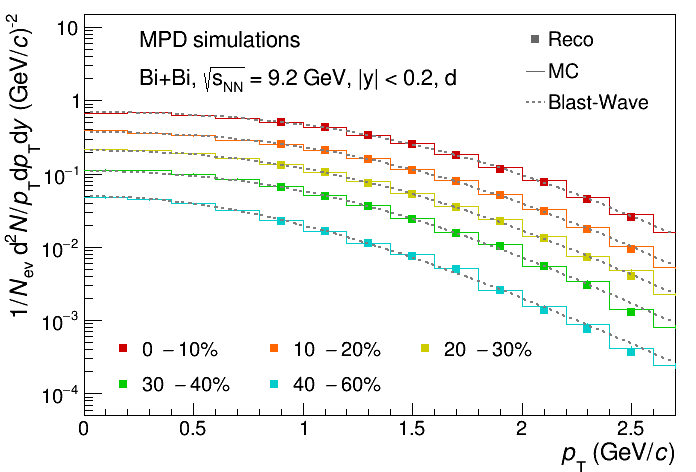}
\hfill
\includegraphics[width=0.49\textwidth]{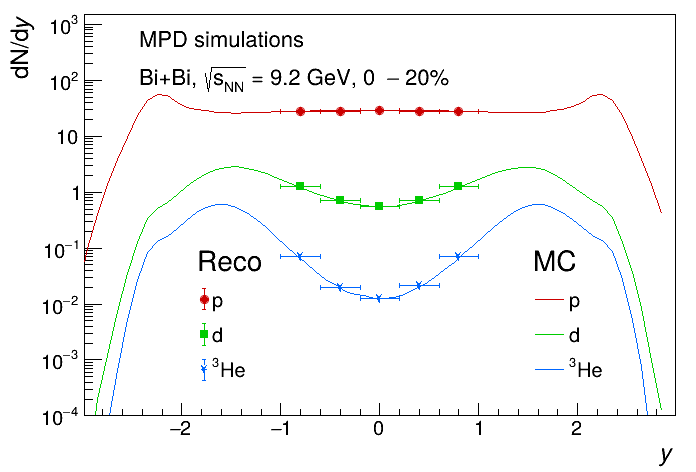}
\caption{Left: Invariant \pT-spectra of $d$ in centrality selected \BiBi collisions. Right: Rapidity distributions of $p$, $d$ and $^3$He in 0-20\% central \BiBi collisions. The reconstructed data are shown by symbols while the model data are drawn by lines. \label{fig:light_nuclei}}
\end{figure*} 

\begin{figure*}[b]
\centering
\includegraphics[width=0.49\textwidth]{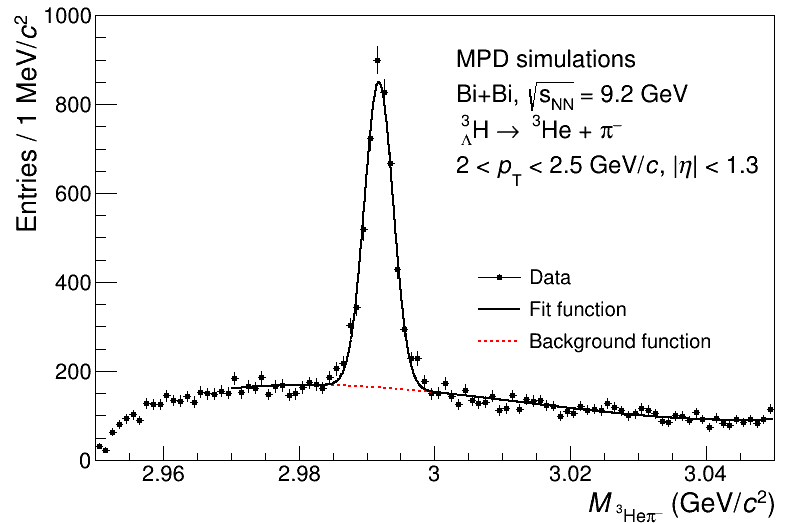}
\hfill
\includegraphics[width=0.49\textwidth]{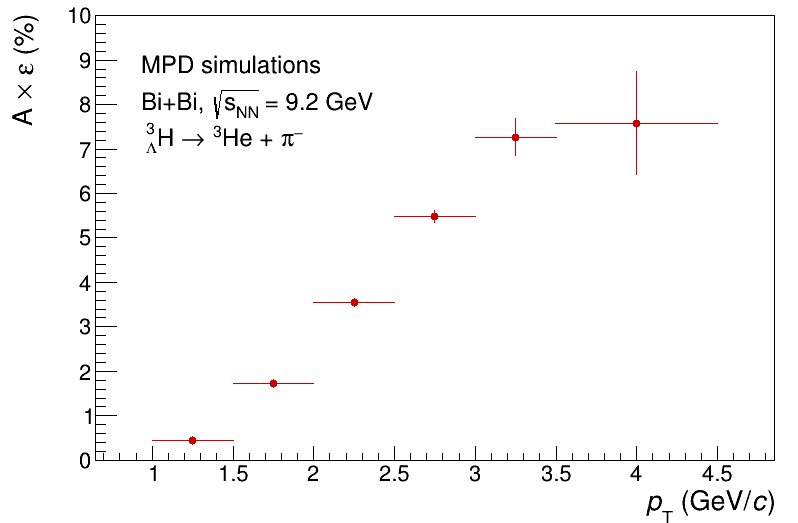}
\caption{Left: Invariant mass distribution for $^3$He$\pi^-$ pairs at $2.0<p_T<2.5$~\GeVc. Reconstructed data are shown by symbols, the solid line indicates a fit to a Gaussian and a third order polynomial. Right: The overall reconstruction efficiency for hypertritons in \BiBi collisions.\label{just_label}} 
\end{figure*}

\label{sec:light_nuclei}
The study of the production of light nuclei is of particular interest in view of the puzzling fact
that weakly bound objects are abundantly produced inside hot and dense hadronic matter. 
Light nuclei at near midrapidity can be formed as a result of the
fusion reaction of secondary nucleons located close to each other in space and having
small relative momentum. Thus, the process of cluster formation is sensitive not only
to the nucleon density in phase space, but also to spatial-momentum correlations that
appear in the collective velocity field during the fireball evolution. In order to obtain
detailed information on the structure of the particle source, detailed measurements
of the transverse momentum and rapidity distributions for clusters of different masses
at several collision energies and centralities are necessary.

The MPD performance for light nuclei measurements was studied using mass production 3 from Table~\ref{EvGens}.
Particle identification was achieved by combining information about particle energy losses
measured in the \TPC and time-of-flight measured in the \TOF. The
overall efficiency correction procedure is similar to that used in the analysis of hadrons (see Sect.~\ref{sec:hadrons} for details). The left panel of Figure~\ref{fig:light_nuclei} shows the invariant \pT-spectra of deuterons in centrality
selected \BiBi collisions. Reconstructed data are shown by symbols, model distributions 
are depicted by histograms. Extrapolations to the unmeasured regions of transverse momentum
are based on the Blast-Wave fit function (shown by dashed lines). 

Figure~\ref{fig:light_nuclei} (right panel) shows the rapidity distributions of reconstructed
protons and light nuclei ($d,^3$He). As one can see, the MPD acceptance
allows the measurements of cluster yields over the rapidity range $|y|<1$.

\subsubsection{ Hypernuclei}
\label{sec:hypernuclei}

Hypernuclei are bound nuclear systems consisting of nucleons and hyperons. 
Therefore, the process of their formation in heavy-ion collisions is determined 
by hyperon-nucleon correlations in the phase space of the reaction and the 
magnitude of the nucleon-hyperon potential~\cite{hypernucl_qgp}. The latter is of 
fundamental importance for astrophysics, since the appearance of hyperon degrees 
of freedom is expected in the interior of neutron stars~\cite{neutr_star}. New 
experimental data on the yields, binding energies, and lifetimes of hypernuclei 
can provide important information on the nature of the interaction between nucleons 
and hyperons in dense baryon matter. The NICA energy range is very well suited 
for such studies because the maximum in the freezeout baryon density and in the 
strangeness-to-entropy ratio is achieved in the NICA energy range~\cite{nica_predictions}.

\begin{figure*}[b]
\centering
\includegraphics[width=0.49\textwidth]{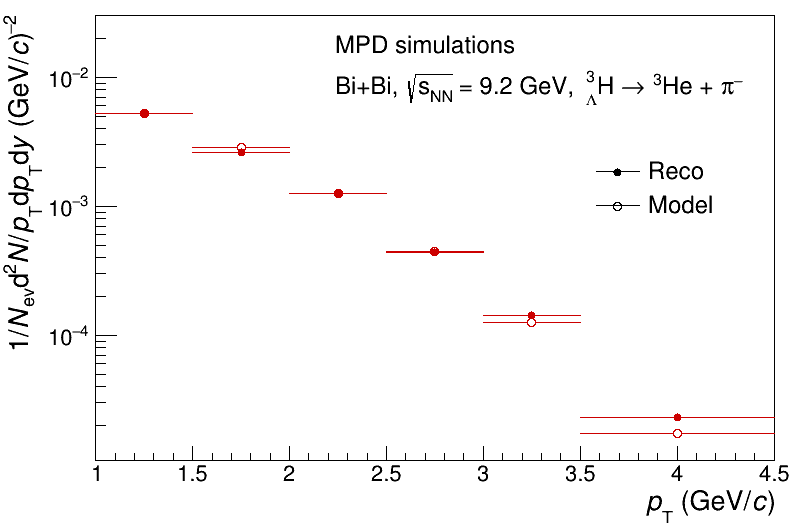}
\hfill
\includegraphics[width=0.49\textwidth]{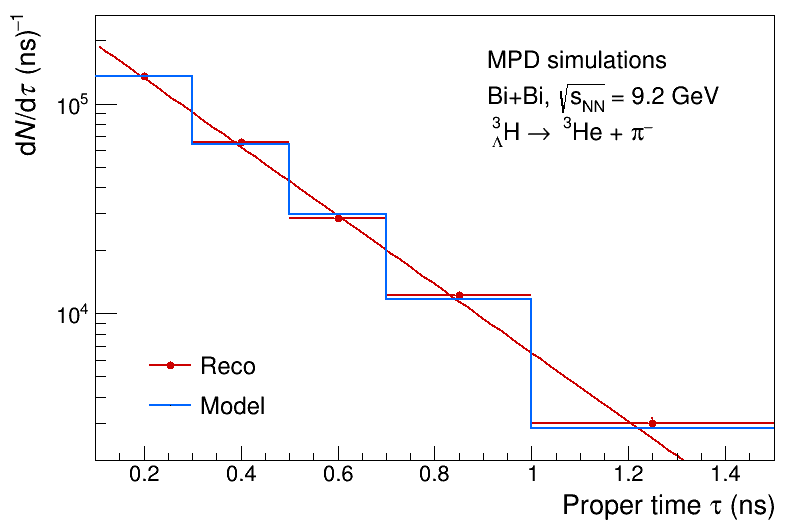}
\caption{Left: Invariant yield distribution for hypertritons. Reconstructed and generated data are shown with triangles and rectangles, respectively. Right: Distribution of the number of  hypertritons in intervals of proper time $\tau$. The blue and red histograms represent the generated and reconstructed distributions, respectively, the line shows the fit according to 
Eq.~(\ref{eq2}).\label{hypertrit_ptspec}}
\end{figure*}

To study the MPD characteristics for the reconstruction of hypernuclei, data from mass production 3 from Table~\ref{EvGens} were used.  The reconstruction of hypertritons
was carried out using the
$_{\lmb}^3$H$\rightarrow$$^3$He + $\pim$ decay mode.
The daughter particles were identified
using the information about the ionization energy loss in the \TPC gas and the mass 
squared from the \TOF. The particle species is considered to be determined if the values 
of \dEdx and $M^2$ lie within $\pm 3\sigma$ of the values expected for true protons and pions. To reduce the combinatorial 
background, topological selections were applied to the reconstructed pairs, similar to those
used in the reconstruction of hyperons in Sec.~\ref{sec:hyperons}. The invariant mass spectrum 
of $^3$He \pim pairs, that passed through each of the selection criteria, is displayed 
in Figure~\ref{just_label} (left panel).
The distribution was fitted by the sum of a
Gaussian distribution for the signal and a third order polynomial for the background. The signal was 
determined by histogram bin counting within a $\pm 5\sigma$ window of the  Gaussian peak
position and subtracting the integral of the background function in the same mass range. 
The raw yield of hypertritons is then corrected for the reconstruction efficiency, which 
includes the detector acceptance and signal losses due to the selection criteria and 
particle identification. The \pT dependence of the evaluated efficiency is shown in 
Figure~\ref{just_label} (right panel).

Figure~\ref{hypertrit_ptspec} (left panel) shows the invariant \pT-spectrum of hypertritions from \BiBi collisions as evaluated using Eq.~(\ref{eq:CorrYield}). The spectra are obtained for the rapidity interval $|y|<0.5$ without selection on the collision centrality. The reconstructed distribution is shown with solid symbols, while the initially generated distribution of the model is shown with empty symbols. As can be seen from the figure, the agreement between the reconstructed spectra is good for all \pT intervals.

According to the standard method of determining the life-time, the yield of
unstable particles in intervals of proper time $\tau$ decreases exponentially, 
\begin{equation} 
N(\tau) = N(0)\exp\left(- \frac{\tau}{\tau_0}\right) = N(0)\exp\left(-\frac{ML}{cp\tau_0}\right), \label{eq2}
\end{equation}
where the slope parameter $\tau_0$ is the particle lifetime
and $\tau=t/\gamma$ is the proper time, $\gamma= 1/\sqrt{1-(v/c)^2}$, with $v$ 
the velocity, $L$ the decay length, $p$ the particle momentum and
$M = 2.991$~GeV/$c^2$ the hypertriton rest mass~\cite{ParticleDataGroup:2020ssz}.
The hypertriton yield was analyzed in several $\tau$ intervals in the range \mbox{[0.1--1.5]}~ns.
%As an example, Figure~\ref{hypernucl} (left panel) shows the invariant mass distribution for 
%pairs ($^3$He,$\pi^-$) in the interval $\tau =[0.1,0.3]$~ns.
Figure~\ref{hypertrit_ptspec} (right panel) shows the fully
corrected hypertriton yields as a function of proper time $\tau$. A fit of the obtained distribution using Eq.~(\ref{eq2}) is shown as a line. The slope parameter (lifetime) of $265 \pm 4$~ps agrees well with the expected value of the lifetime used in the event generator, 263~ps.

According to simulation-based estimates of the MPD efficiency for
hypertritons and model predictions on (hyper)nuclei yields, about $10^3$ hypertritons can be registered in one week of data taking of \BiBi collisions at $\snn=9.2$~\GeV with luminosity $L\approx 10^{25}$~cm$^{-2}$s$^{-1}$.

\subsection{Hyperon global polarization}

Global spin polarization ($P_\lmb$) of \lmb  and \almb  hyperons  was found and measured in relativistic heavy-ion collisions over a broad collision energy range~\cite{nature2global,global2inv,global3inv}. The data indicate a trend of increasing $P_{\Lambda}$ with decreasing collision energy   from 1-2\% at $\snn$ = 200 GeV to 5-7\% at $\snn$ = 3 GeV.
Different scenarios for the global polarization mechanism are predicted by phenomenological~\cite{Ayala:2020soy,Ayala:2021xrn} and MC hydrodynamic and transport models, highlighting the importance of collecting new experimental data \cite{global2,global3}. Here we report on the MPD performance analysis of global polarization of
\lmb-hyperons. Data from mass production number 4 in Table~\ref{EvGens} served as the basis for this study, as the hyperon global polarization was included
in the PHSD model \cite{kol1,kol2}.  The procedure was developed in \cite{global1} to transfer the hyperon spin polarization signal from the transport code to the final moments distribution of particles after weak decays. This allowed us to investigate the reconstruction of the spin signal within the detector simulation. The global polarization observable $P_\lmb$ is defined as~\cite{nature2global,global2inv,global3inv,global4inv}
\begin{equation}
P_\lmb = \frac{8}{\pi \alpha_\lmb} \frac{\langle \sin(\Psi_{\rm 1} - \phi_{\rm p}^*)\rangle}{ R(\Psi_{1}) }.
\label{Eq:1}
\end{equation}
Here $\alpha_\lmb = 0.732 \pm 0.014$~\cite{ParticleDataGroup:2020ssz} is the \lmb decay parameter, $\Psi_{\rm 1}$ the first-order
event plane angle from \FHCAL, $\phi_{\rm p}^*$ the azimuthal angle of the proton in the \lmb rest frame, $R(\Psi_{1})$  the resolution of the first-order event plane angle and the brackets denote the average over all produced \lmb hyperons.

Protons and pions measured in the \TPC were used to reconstruct \lmb  hyperons, which decay via $\lmb\rightarrow p+\pim$ with a branching ratio
63.9\%. The \lmb candidates have been  reconstructed using the invariant mass technique. 
\begin{figure}[H]
\centering
\includegraphics[width=0.9\linewidth]{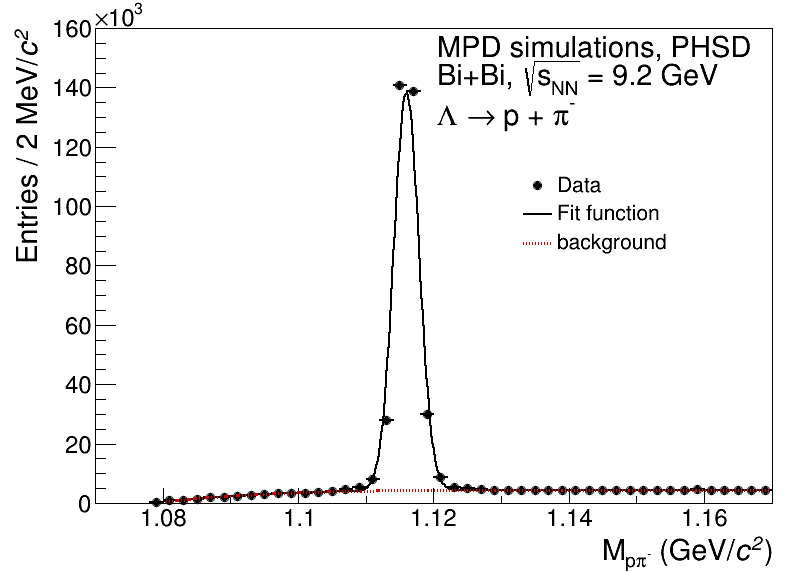}
\hfill
\includegraphics[width=0.9\linewidth]{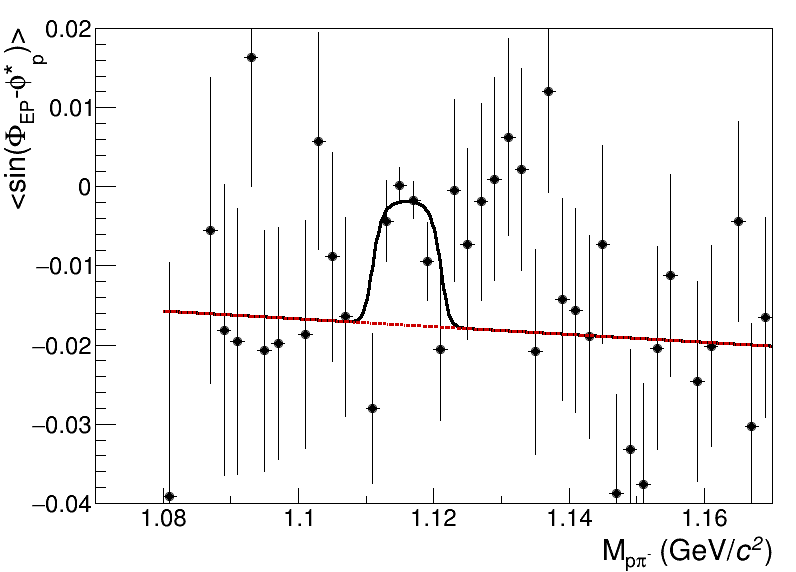}
\caption{Invariant mass distribution (top) and global polarization distribution $\langle \sin(\Psi_{\rm 1}-\phi_{\rm p}^*)\rangle(M_\mathrm{p\pi})$ (bottom) for $\Lambda$ particles at  $0.5 < \pT < 3$ \GeVc for 20-50\% central \BiBi collisions at $\snn = 9.2$~GeV. Reconstructed data are plotted by black symbols, the fit results are shown by the solid black line for the signal and red dotted  line for background.\label{fig:lambda_invar_polar}}
\end{figure}
The combinatorial background from uncorrelated particles has been reduced by the selection criteria based on the decay topology with quality assurance selections, such as the  primary and secondary decay vertex positions, the DCA of the daughter particles to the primary vertex, the DCA of the mother particle to the primary vertex, and the DCA between the daughter tracks, see details in Sec.~\ref{sec:hyperons}. As an example, the upper panel of Figure~\ref{fig:lambda_invar_polar} shows the invariant mass distribution for \lmb-particles with
$0.5 < \pT < 3$~\GeVc for 20-50\% central \BiBi collisions at $\snn = 9.2$~GeV.  The background region is fitted with a second-order polynomial while the signal is fitted with a Gaussian distribution. From these fits, a background $f^\mathrm{B}(M_\mathrm{p\pi})$  and signal $f^\mathrm{S}(M_\mathrm{p\pi})$ fractions, as functions of invariant mass, are extracted.  The  selected sample $P_{\lmb}^{\rm{all}}=\langle \sin(\Psi_{\rm 1}-\phi_{\rm p}^*)\rangle(M_\mathrm{p\pi})$ contains both the
signal $P^\mathrm{S}_{\lmb}=\langle \sin(\Psi_{\rm 1} - \phi_{\rm p}^*)\rangle^\mathrm{S}$ and the combinatorial background contribution
$P^\mathrm{B}_{\lmb}(M_\mathrm{p\pi})=\langle \sin(\Psi_{\rm 1} - \phi_{\rm p}^*)\rangle^\mathrm{B}(M_\mathrm{p\pi})$. The distribution $P_{\lmb}(M_\mathrm{p\pi})$ is fitted as a function of invariant mass $M_\mathrm{p\pi}$ (invariant mass fit method)\cite{nature2global,global2inv,global3inv,global4inv},  according to
\begin{equation}
\label{eq:InvMassMethod}
P_{\lmb}^{\rm{all}}(M_\mathrm{p\pi}) =f^\mathrm{B}(M_\mathrm{p\pi})P^\mathrm{B}_{\lmb}(M_\mathrm{p\pi})+f^\mathrm{S}(M_\mathrm{p\pi})P^{S}_{\lmb}~~
\end{equation}
\begin{figure}[H]
\centering
\includegraphics[width=0.83\linewidth]{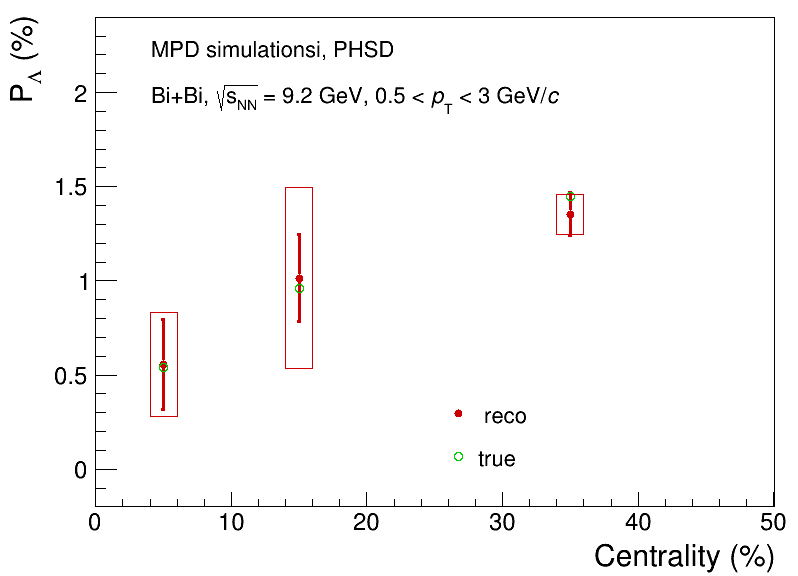}
\hfill
\includegraphics[width=0.83\linewidth]{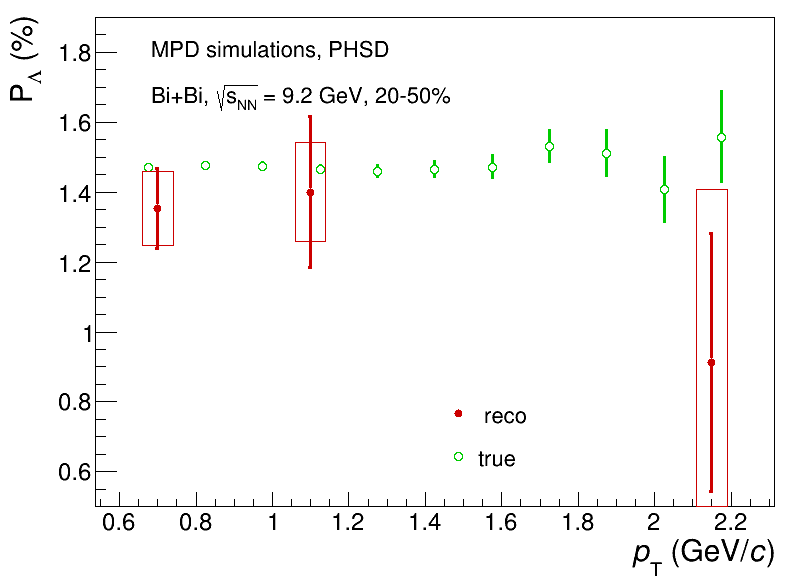}
\hfill
\includegraphics[width=0.83\linewidth]{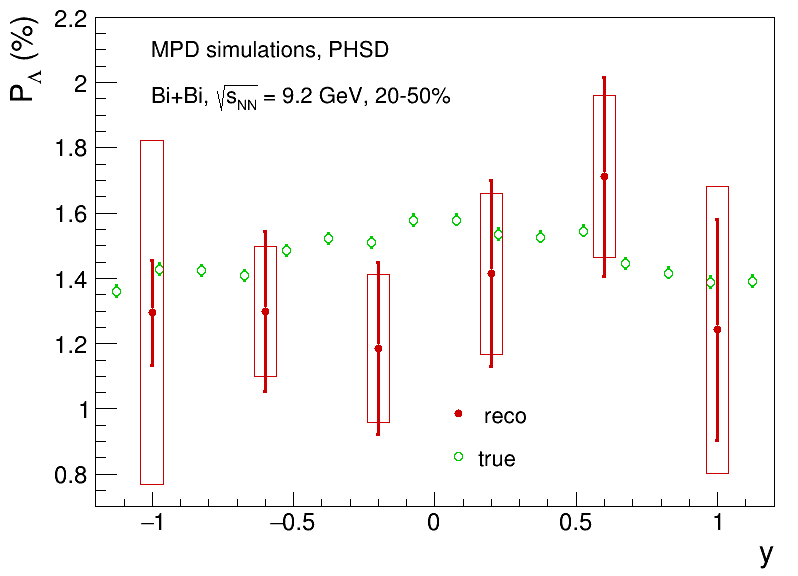}
\caption{Top: global polarization of \lmb as a function of centrality  in \BiBi collisions at $\snn = 9.2$~GeV. Middle:  the same as a function of \pT. Bottom: the same as a function of  rapidity $y$. Open and closed markers correspond to generated and reconstructed data, respectively. \label{fig:Polar_vs_pt_eta}}
\end{figure}

\noindent to extract the signal contribution $P^\mathrm{S}_{\lmb}$ to the measured  polarization signal, see the bottom panel of Figure~\ref{fig:lambda_invar_polar}.
That is, the background $P^\mathrm{B}_{\lmb}(M_\mathrm{p\pi})$    was parametrized as a linear  function of $M_\mathrm{p\pi}$
 and $P^\mathrm{S}_{\lmb}$ is taken as a fit parameter.
Figure ~\ref{fig:Polar_vs_pt_eta} presents the resulting values of the global polarization
$P_{\lmb}=P^\mathrm{S}_{\lmb}/R(\Psi_{1})$ as a function of centrality (upper panel) for \lmb particles at  $0.5 < \pT < 3$~\GeVc, as a function of  transverse momentum \pT (central panel) and  rapidity $y$ (lower panel) for 20-50\% central  \BiBi collisions at $\snn = 9.2$~GeV.  Good agreement is observed
between the $P_\lmb$ results obtained from the analysis of fully reconstructed data “Reco” and generated “MC” PHSD model events. The analyzed statistics of 15 M events allows to perform the differential measurements of \lmb global  polarization in mid-central \BiBi collisions only. The more detailed \pT-differential studies as a function of centrality and rapidity, 
as well as the measurements for \almb-hyperons,
will require a larger data sample of up to 200-300 M of minimum-bias events.

\subsection{Anisotropic flow}
The sensitivity of the azimuthal anisotropic collective flow to the equation of state
(EoS) and the transport properties of the strongly interacting matter makes it one of the promising observables in the relativistic heavy-ion experiments~\cite{posk,flow1,flow2,eos}. The collective flow (assuming a perfect event plane resolution) is usually quantified by the Fourier coefficients
$\vn$ in the expansion of the particle azimuthal
distribution relative to the collision symmetry plane
given by the angle $\Psi_\mathrm{n}$~\cite{posk,flow1}, see Sec.~\ref{sec:evplane} for details. In this section,
we discuss the anticipated performance of the MPD
detector for differential measurements of the directed ($v_1$), elliptic ($v_2$) and triangular ($v_3$) flow of identified hadrons in \BiBi collisions at $\snn = 9.2$~\GeV~\cite{MPD:2022qhn,flowmpd1,flowmpd2}. Although theoretical models can successfully describe flow observables at RHIC and LHC energies, none of them can quantitatively describe the existing $v_\mathrm{n}$
measurements in the NICA energy range  \snn = 4–11~\GeV ~\cite{MPD:2022qhn}. Therefore, we have used two models (productions number 1 and 5 listed in Table~\ref{EvGens}) to
simulate minimum bias \BiBi collisions: the
viscous hydro + hadronic cascade vHLLE+UrQMD hybrid model \cite{Karpenko:2013wva,Karpenko:2015xea} with QGP formation and the
cascade version of UrQMD \cite{Bleicher:1999xi,Bass:1998ca}, which is a purely 
hadronic transport model. We refer to the \vn results obtained from the flow analysis of the generated model events as “true”, 
whereas “reco” denotes the $v_n$ results derived from
the flow analysis of the fully reconstructed events.

\begin{figure}[H]
\centering
\includegraphics[width=0.85\linewidth]{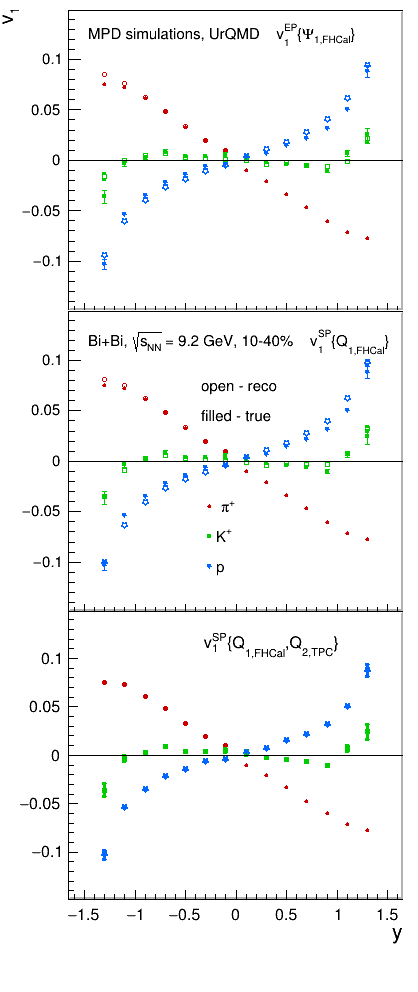}
\vspace{-0.2in}
\caption{ Directed flow $v_1(y)$  of identified charged hadrons as functions of rapidity in 10-40\% central \BiBi collisions at $\snn = 9.2$~\GeV for different methods of flow analysis
of fully reconstructed  events (filled markers) and generated UrQMD events (open markers). \label{fig:hadrons_v1}}
\end{figure}

\begin{figure}[H]
\centering
\includegraphics[width=0.83\linewidth]{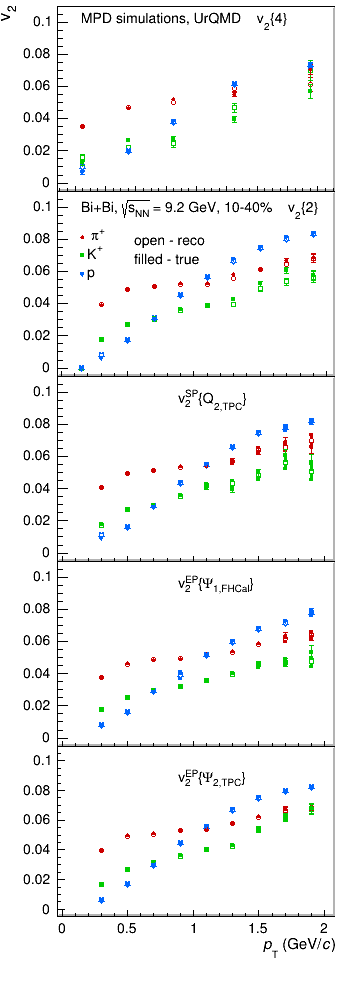}
\vspace{-0.2in}
\caption{Elliptic flow $v_2(\pT)$  of identified charged hadrons as a function of \pT
  in 10-40\% central \BiBi collisions at $\snn = 9.2$~\GeV for different methods of flow analysis
  of fully reconstructed  events (filled markers) and generated UrQMD events (open markers).\label{fig:hadrons_v2}}
\end{figure}

Figure~\ref{fig:hadrons_v1} shows the
rapidity dependence of directed $v_1(y)$ flow   of charged pions (triangles), kaons (boxes)
and protons (circles) for 10-40\% central \BiBi collisions at $\snn$ = 9.2~\GeV from the analysis of
UrQMD model  events.  A momentum dependent $\pm 2\sigma$ 
cut around each peak in the mass-squared mass$^2$ distribution was used to identify pions, kaons and protons. The figure shows results obtained with three different analysis methods with respect 
to the flow vector $Q_1=Q_{1,\FHCAL}$ of 
spectator fragments detected in FHCAL, the event plane method $v_{1}^\mathrm{EP}( \Psi_{1,\FHCAL})$ (upper panel), the scalar 
product method $v_{1}^\mathrm{SP}( Q_{1,\FHCAL})$ (middle panel) and the scalar product method using mixed harmonics $v_{1}^\mathrm{SP}( Q_{1,\FHCAL},Q_{2,\TPC})$ (lower panel).
For all particle species, the directed flow $v_1$ crosses zero at midrapidity and  the reconstructed
values “reco” of $v_1$ (open symbols) are fully consistent with the generated “true” values (filled symbols). 
Figure ~\ref{fig:hadrons_v2} shows the results of the \pT-differential elliptic flow $v_2$ measurements for charged pions (triangles), kaons (boxes) and protons (circles) in 10-40\% central \BiBi collisions.

The large and uniform acceptance of the \TPC allows us to use multiparticle methods, such as direct cumulants,
for elliptic flow measurements. The top panel of Fig.~\ref{fig:hadrons_v2} shows the four-particle  $v_2\{4\}$. The other panels 
show the two-particle methods: b) two particle cumulants $v_2\{2\}$, c) scalar product method using \TPC tracks
values for reconstructed and generated signals is observed for all particle species and flow analysis methods.
Different methods of flow measurements have different degrees of sensitivity to the flow fluctuations and to so-called non-flow correlations~\cite{flowmpd2,flowmpd4,flow1}. They include the transverse momentum
conservation, small azimuthal angle correlations due to final state interactions, resonance
decays, and quantum correlations due to the Hanbury Brown--Twiss (HBT) effect~\cite{flow1}. 

\begin{figure}[H]
\centering
\includegraphics[width=0.83\linewidth]{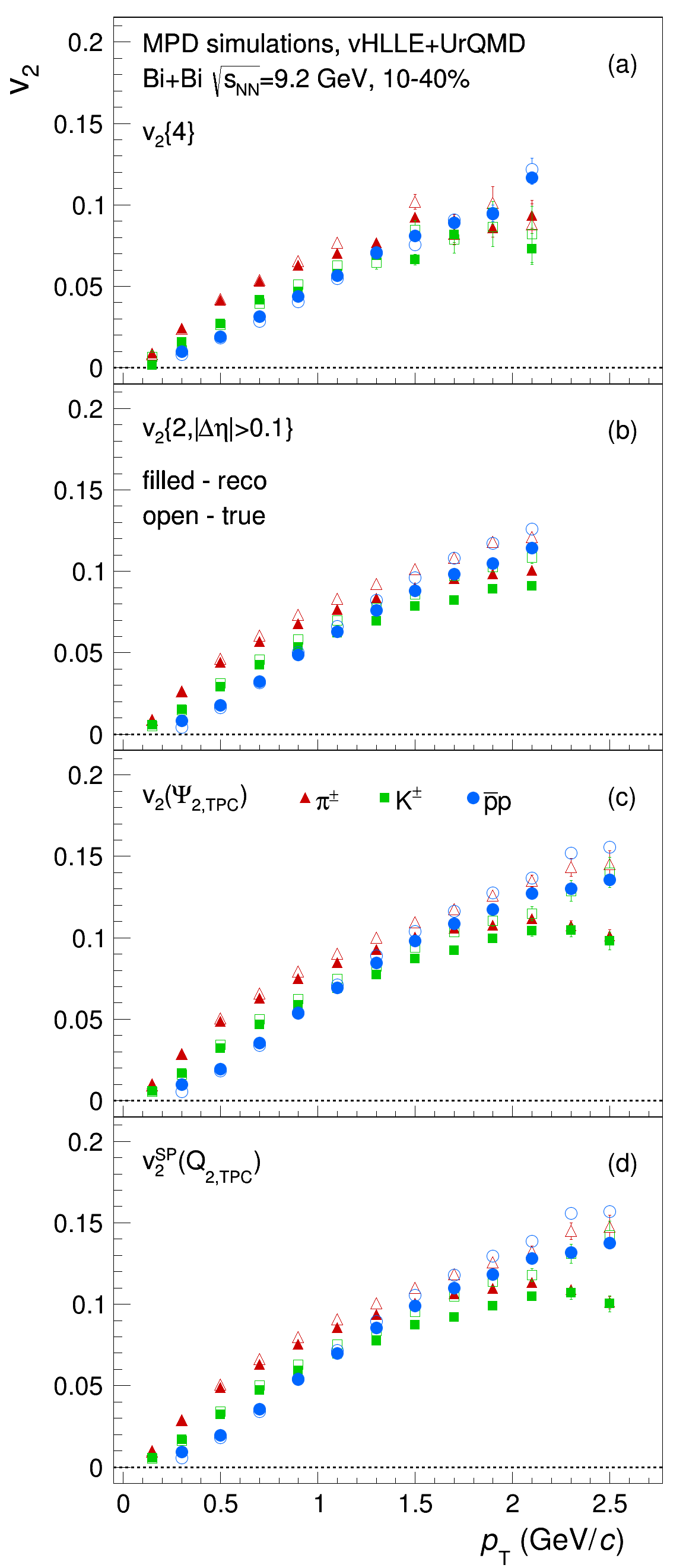}
\caption{Elliptic flow $v_2(\pT)$  of identified charged hadrons as a function of \pT
  in 10-40\% central \BiBi collisions at $\snn = 9.2$~\GeV for different methods of flow analysis
  of fully reconstructed  events (filled markers) and generated vHLLE+UrQMD model events (open markers).
\label{fig:pid_v2pt_req32}}
\end{figure}

The main cause of non-flow effects is few particle correlations, so estimates of the $v_2$ flow coefficients based on four-particle  cumulants
$v_2\{4\}$ have the benefit of great suppression of non-flow effects contribution. To suppress
the non-flow effects in two-particle methods, we have applied the  pseudo-rapidity gaps $\Delta\eta$ between sub-events: 
$|\Delta\eta|>0.1$ between the two \TPC sub-events for $v_2\{2\}$, $v_{2}^\mathrm{SP}( Q_{2,\TPC})$, $v_{2}^\mathrm{EP}( \Psi_{2,\TPC})$ and $|\Delta\eta|>0.5$ between
the \TPC and \FHCAL detectors for $v_{2}^\mathrm{EP}( \Psi_{1,\FHCAL})$.

\begin{figure}[H]
\centering
\includegraphics[width=0.88\linewidth]{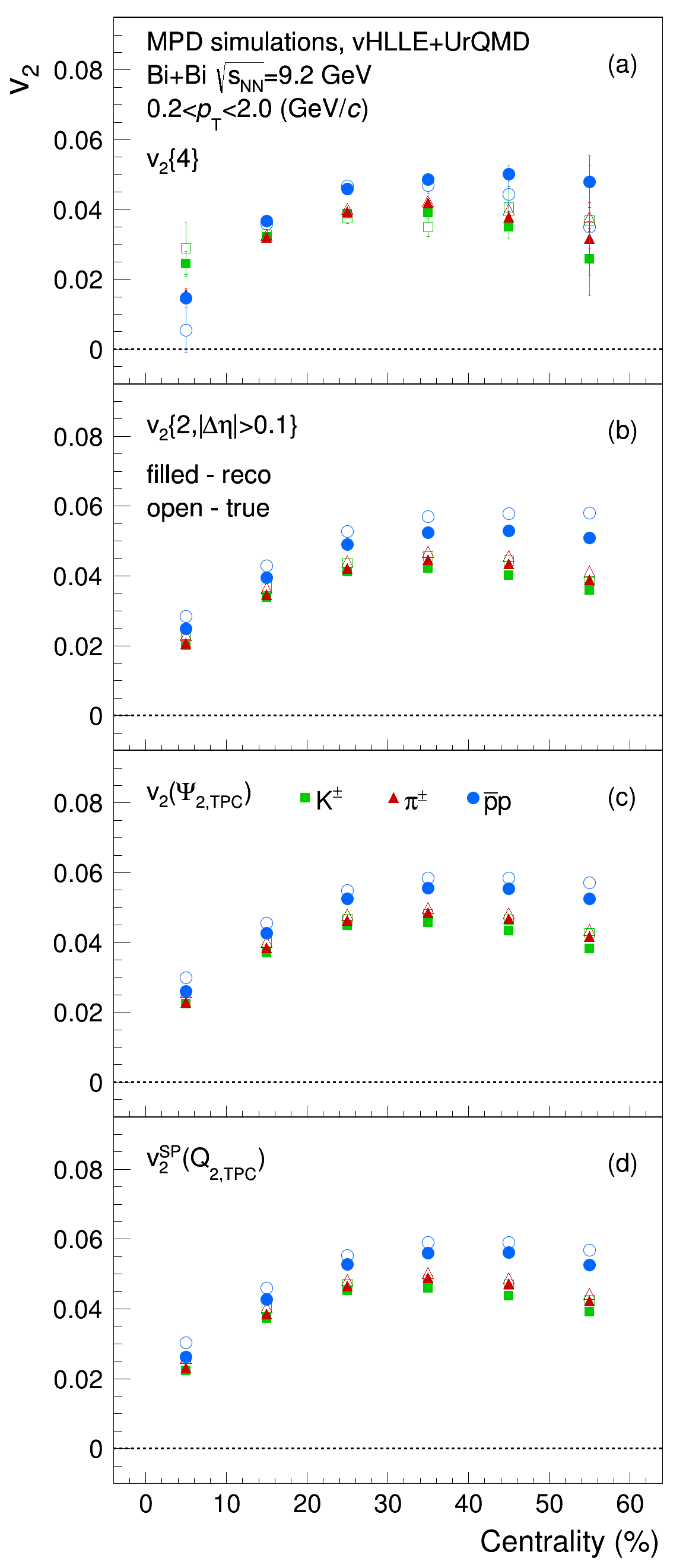}
\caption{Elliptic flow $v_2$  of identified charged hadrons as a function of centrality in \BiBi collisions
at $\snn = 9.2$~\GeV for different methods of flow analysis
of fully reconstructed  events (filled markers) and generated vHLLE+UrQMD model events (open markers).
\label{fig:pid_v2cent_req32}}
\end{figure}

Different methods of flow  measurements have different degrees of sensitivity to the $v_2$ fluctuations $\sigma_{v_2}$: $\sigma_{v_2}^2 = \langle v_2^2 \rangle - \langle v_2 \rangle^2$.
For a Gaussian model of fluctuations,   one can expect ~\cite{flow1}:
$v_2\{2\} = \left\langle v_2 \right\rangle + 0.5\sigma_{v_2}^2/\left\langle v_2 \right\rangle,\ v_2\{4\} = \left\langle v_2 \right\rangle - 0.5\sigma_{v_2}^2/\left\langle v_2 \right\rangle$.
Our previous work demonstrates that the participant eccentricity fluctuations, in the initial geometry of the overlap
region of two colliding nuclei, come mainly from $v_2$ flow 
fluctuations for colliding heavy-ion systems (Au+Au or \BiBi) at $\snn$ > 7~\GeV \cite{flowmpd2,flowmpd4}. 

\begin{figure}[H]
\centering
\includegraphics[width=1\linewidth]{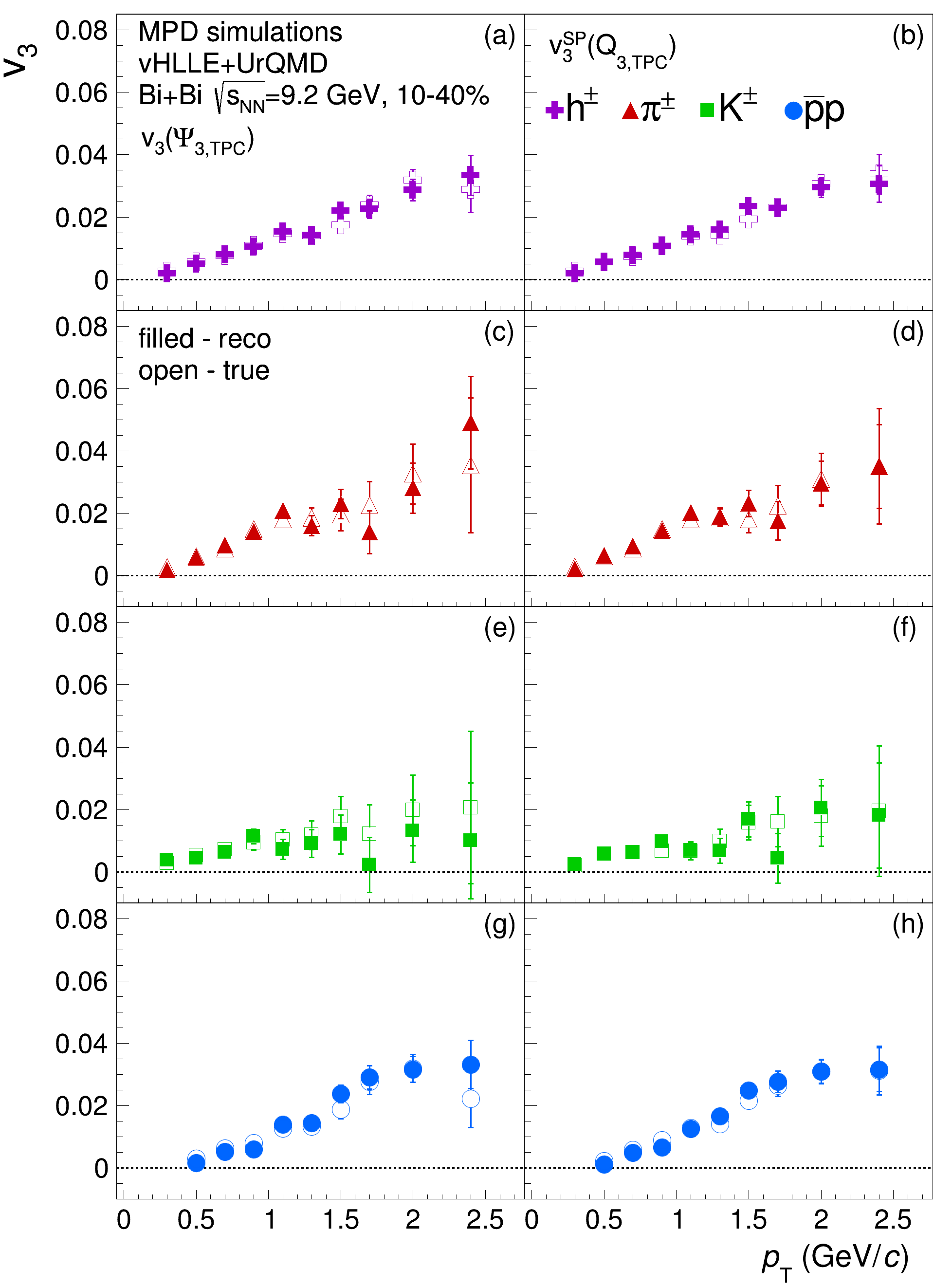}
\caption{Triangular flow $v_3(\pT)$  of identified hadrons as function of transverse momentum in 10-40\% central
  \BiBi collisions at $\snn = 9.2$~\GeV for different methods of flow analysis of fully reconstructed  events (filled markers) and
  generated events with the vHLLE+UrQMD model (open markers). \label{fig:pid_v3_pt_req32}}
\end{figure}

Consequently, the values of $v_2\{\Psi_{1,\FHCAL}\}$ measured with respect  to the first-order 
event plane  $\Psi_{1,\FHCAL}$
will consistently be smaller than the 
values of $v_2\{\Psi_{2,\TPC}\}$  measured in relation to the participant plane  
$\Psi_{2,\TPC}$: $v_2\{\Psi_{1,\FHCAL}\} \simeq \left\langle v_2 \right\rangle, v_2\{\Psi_{2,\TPC}\} \simeq \left\langle v_2 \right\rangle +
0.5\sigma_{v_{2}^2}/\left\langle v_2 \right\rangle$.

Figure ~\ref{fig:pid_v2pt_req32} shows the performance for the measurements of $v_2$ as a function of \pT of identified charged pions (triangles), kaons (boxes) and protons (circles) from 10-40\% central \BiBi collisions for reconstructed and generated vHLLE+UrQMD model events. A good agreement between the $v_2$ results is observed.

\begin{figure*}[t]
\centering
\includegraphics[width=0.48\textwidth]{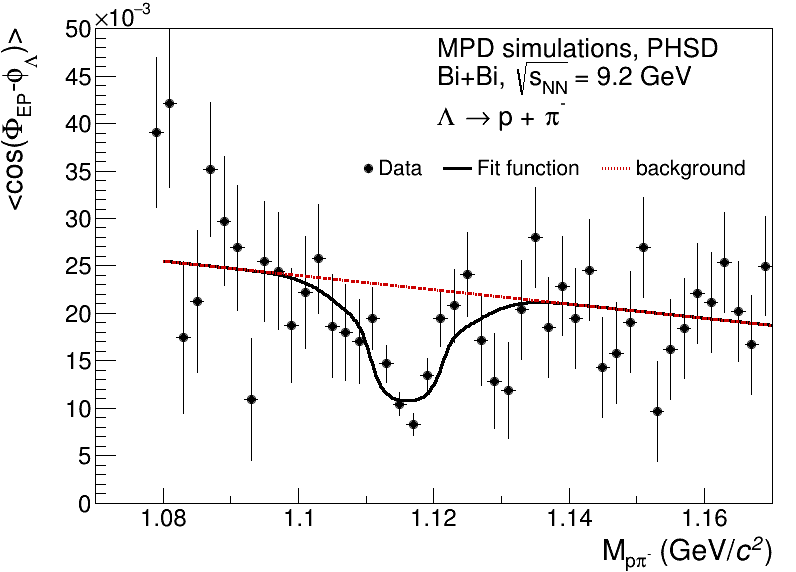}
\hfill
\includegraphics[width=0.48\textwidth]{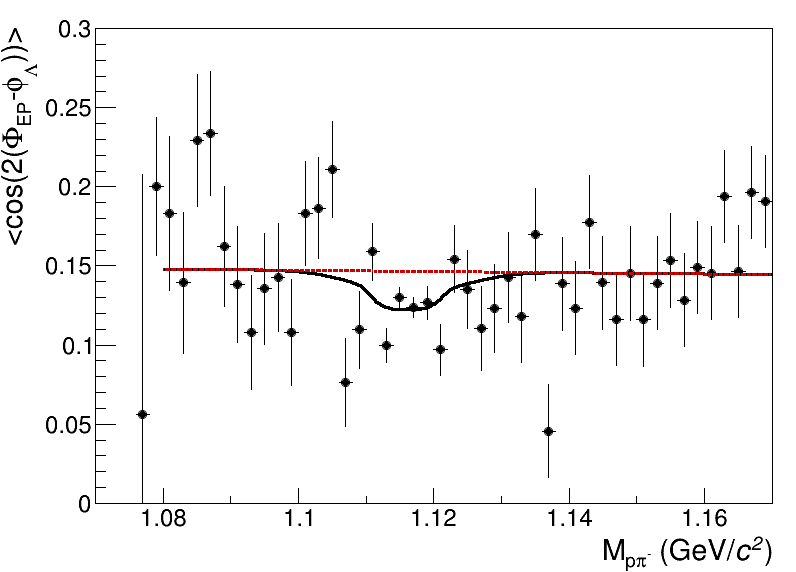}
\caption{The demonstration of the invariant-mass fit method to extract the $v_1$  (left panel) and $v_2$ (right panel) signal for  $\Lambda$  particles produced in 20-50\% central \BiBi collisions at $\snn = 9.2$~\GeV. Reconstructed data are plotted by black symbols, fit results are shown by the colored lines.\label{fig:lambda_invar_v1v2}}
\end{figure*}

\begin{figure*}[b]
\centering
\includegraphics[width=0.48\textwidth]{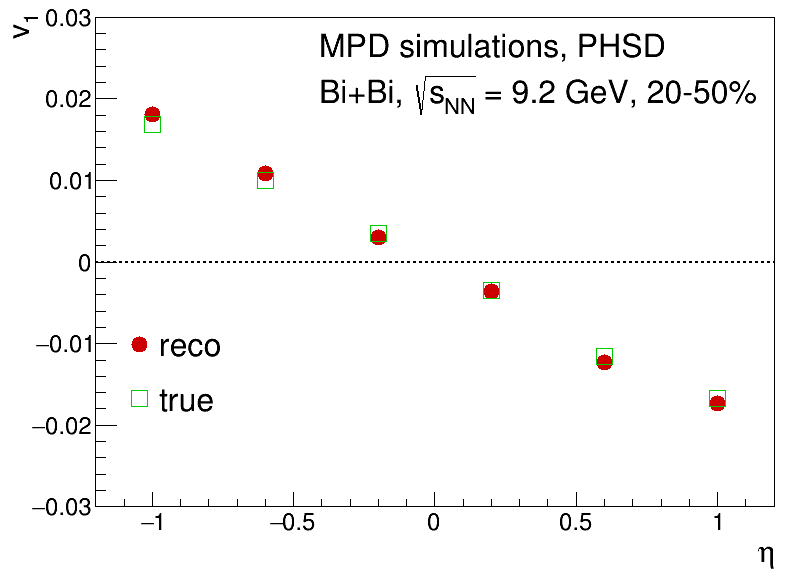}
\hfill
\includegraphics[width=0.48\textwidth]{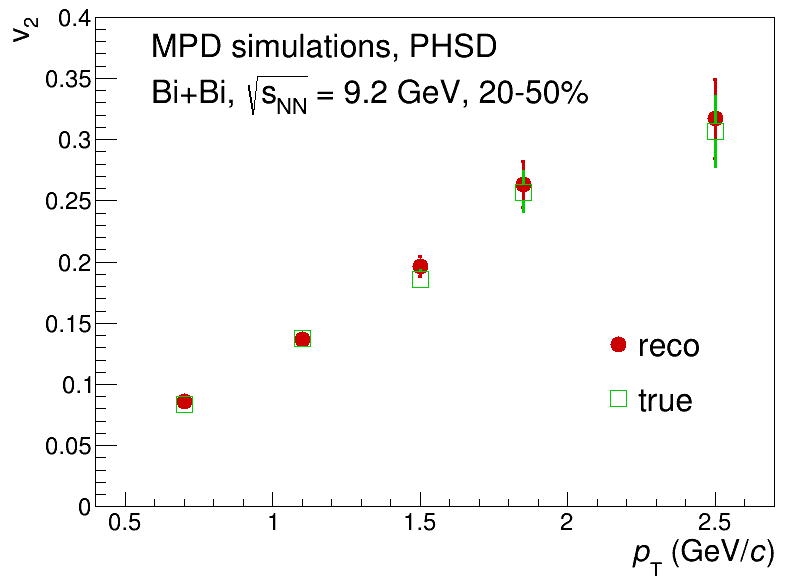}
\caption{ Directed $v_1$ (left) and elliptic $v_2$ (right) flow of \lmb hyperons as a function of pseudorapidity $\eta$ and transverse momentum \pT  in 20-50\% central \BiBi collisions at $\snn = 9.2$~\GeV for the event plane method of the analysis of
fully reconstructed events (filled markers) and generated PHSD  model events (open markers). \label{fig:lambda_v1v2}}
\end{figure*}

Due to the  lack of spectators in the vHLLE+UrQMD model, we can not test the event plane method using the  first-order event plane from spectators $v_{2}^{EP}( \Psi_{1,\FHCAL})$. Figure ~\ref{fig:pid_v2cent_req32} shows the performance for the measurements of the centrality  dependence of the elliptic flow $v_2$ of identified hadrons 
for  different methods of flow analysis. 
The conclusions from the comparison of $v_2$ results are very similar.
The present statistics of 50 M minimum bias events are not sufficient for a statistically
significant four-particle cumulant $v_2\{4\}$ results for 0-10\% central  \BiBi collisions.

The triangular ($v_3$) flow of hadrons is predicted to be more sensitive (than $v_2$) to the viscous damping and may be a good observable to investigate the formation of a QGP and the pressure gradients in the early  phase~\cite{Karpenko:2015xea,v3flow1}. 
The hybrid model calculations show that the hydrodynamically generated $v_3$ signal disappears at low collision energies of
$\snn$ = 5 - 7~\GeV and there is no $v_3$ signal generated  in the hadronic phase~\cite{Karpenko:2015xea,v3flow1}. Figure~\ref{fig:pid_v3_pt_req32} shows
the performance for the measurements of \pT of $v_3$ of identified charged hadrons in 
\BiBi collisions at $\snn = 9.2$~\GeV for different methods of flow analysis of fully reconstructed  events (filled markers) and generated vHLLE+UrQMD events (open markers). The present statistics allows us to check the event plane method using the event plane from the \TPC $v_{3}^\mathrm{EP}( \Psi_{3,\TPC})$.  An overall
good agreement between the $v_3$ results from the analysis of fully reconstructed and generated model data is observed.

For \VZ particles, like $K_{s}^0$ and \lmb, the invariant mass fit method \cite{flowmpd1} can be  applied in order to
separate the $\vn^\mathrm{S}$ value of the signal from the  $\vn^\mathrm{B}$ of combinatorial background.
As an example, Figure~\ref{fig:lambda_invar_v1v2}
demonstrates of the invariant mass fit method to extract the directed $v_1^\mathrm{S}$  (left panel) and elliptic $v_2^\mathrm{S}$ 
(right panel) flow signals for  \lmb  particles produced in 20-50\% central \BiBi collisions at $\snn = 9.2$~\GeV.
The method involves calculating the $v_{n}^{\rm{all}}=\langle \cos n(\Psi_{\rm 1}-\phi_{\rm \Lambda})\rangle(M_\mathrm{p\pi})$ of the same-event distribution as a
function of invariant mass $M_\mathrm{p\pi}$ (denoted by black symbols in Figure~\ref{fig:lambda_invar_v1v2})  and then fitting the resulting
$v_{n}^{\rm{all}}(M_\mathrm{p\pi})$
distribution using
\begin{figure*}[b]
%\begin{center}
\centering
%\begin{multicols}{1}
  \includegraphics[width=\linewidth]{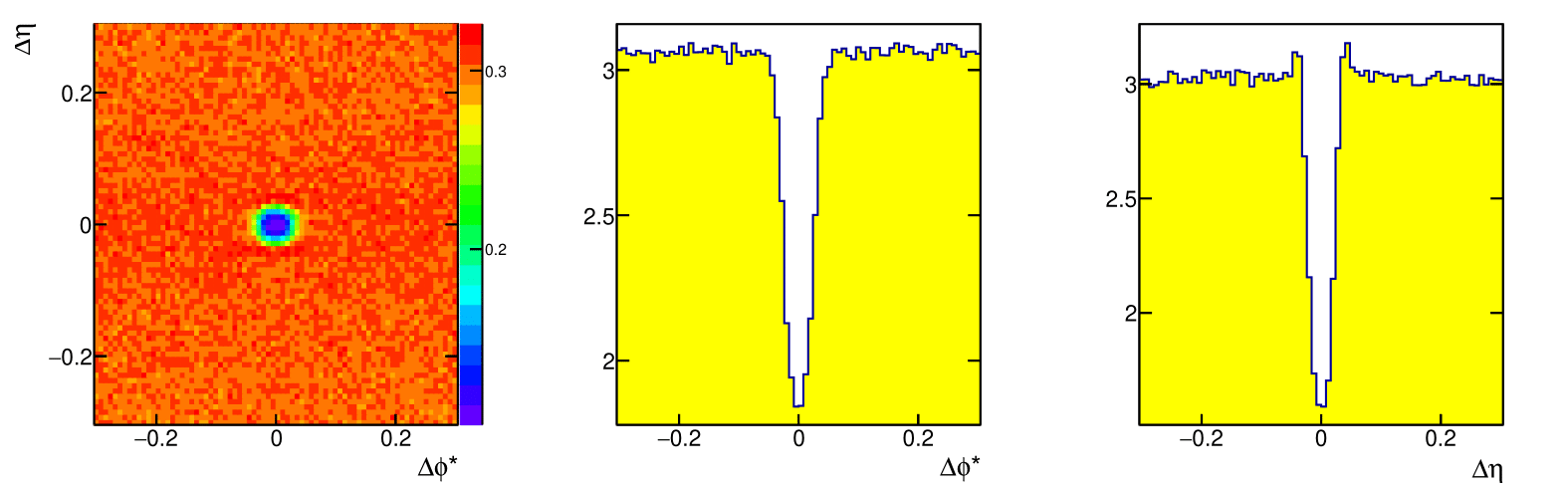}
%\includegraphics[width=127mm]{Figs/EtaPhi.pdf}
%\vspace{-3mm}
%\end{multicols}
\caption{Two-dimensional $\Delta\eta$ $\Delta\phi^*$ distributions for reconstructed tracks (left). 
Projections in $\Delta \phi^*$ (middle) and $\Delta \eta$ (right).}
%\end{center}
\label{femto_fig01}
%\vspace{-5mm}
\end{figure*}

\begin{equation}
\label{eq:InvMassMethod2}
v_{n}^{\rm{all}}(M_\mathrm{p\pi}) =f^\mathrm{B}(M_\mathrm{p\pi})v^\mathrm{B}_{n}(M_\mathrm{p\pi})+f^\mathrm{S}(M_\mathrm{p\pi})v^{S}_{n}~~
\end{equation}
where  $f^\mathrm{B}(M_\mathrm{p\pi})$  and  $f^\mathrm{S}(M_\mathrm{p\pi})$  are the background and the 
signal fractions, respectively.
The background $v^\mathrm{B}_{n}(M_\mathrm{p\pi})$    is parametrized as a linear  function of $M_\mathrm{p\pi}$
 and $v^\mathrm{S}_{n}$ is taken as a fit parameter, see Figure~\ref{fig:lambda_invar_v1v2}.
Figure~\ref{fig:lambda_v1v2} presents the resulting values for directed $v_1$ (left) and elliptic $v_2$ (right) flow of
  \lmb hyperons as a function of pseudorapidity $\eta$ and transverse momentum
  \pT  in 20-50\% central \BiBi collisions at $\snn = 9.2$~\GeV for the event plane method of analysis of
  fully reconstructed  events (filled markers) and generated PHSD  model events (open markers).

The current studies show that
the MPD is able to provide detailed differential
measurements of directed ($v_1$),  elliptic ($v_2$) and triangular ($v_3$) flows of identified hadrons produced in
\BiBi collisions at $\snn$ = 9.2~\GeV with high accuracy.

% \end{multicols}
%============================= Fig. 1 ================================
%============================= Fig. 1 ================================
% \begin{multicols}{2}

\subsection{Femtoscopy and correlations}

Femtoscopy serves as a tool for measuring the spatio-temporal dimensions of the systems created in particle or nuclear collisions.
%, achieving a precision of approximately 1 fm. 
These measurements are made possible by the effects of quantum statistics and final-state interactions, which induce momentum correlations between two or more particles at small relative momenta in their center-of-mass system. By studying the shape of the fireball formed during heavy-ion collisions, valuable insights into the nature of the transition between the hadron phase and the quark-gluon plasma can be gained~\cite{Pratt:1986cc, Bertsch:1988db, Batyuk:2017smw}. Given that pions are among the most copiously produced particles in high-energy reactions, femtoscopic studies concentrate mainly, although not exclusively, on correlation studies of these particles. In this section, we present feasibility studies for two-pion correlation functions performed using UrQMD simulations.

\subsubsection{Femtoscopic correlations of charged pions}

% KM version was:
% One of the main goals of current and future experiments on heavy ion collisions in the energy range from several to a few hundreds of GeV is to study the QCD phase diagram of the  strongly interacting matter~\cite{CBM:2016kpk,Golovatyuk:2019rkb,STAR:2005gfr}. Femtoscopy is a tool for measuring the spatiotemporal characteristics of small and short-lived systems created in particle or nuclear collisions with an accuracy of 1 fm. The possibility of such measurements is due to the effects of quantum statistics and final state interactions which create the momentum correlations of two or more particles at small relative momenta in their center-of-mass system. The possibility of gaining knowledge about the type of phase transition between the hadron phase and the quark-gluon plasma promises to be obtained by studying the shape of the fireball formed during the collision of heavy ions~\cite{Pratt:1986cc, Bertsch:1988db, Cheremnova:2022fsd}. From the theoretical point of view the correlation function (CF) is defined as a ratio two particle production cross-section to the product of the single particle ones. In the experiment the CF defined as $ C(q) = A(q)$/${B(q)}$, where $A(q)$ is the measured distribution of real pairs from the same event.

%%% Grigory's suggestion

From a theoretical perspective, the correlation function (CF) is defined as the ratio of the two-particle production cross-section to the product of the single-particle cross-sections. Experimentally, the CF can be measured as the ratio
$ C(q) = A(q)/B(q)$,
where $A(q)$ is the distribution of pairs from the same event and $B(q)$ represents the reference distribution of pairs from mixed events~\cite{Kopylov:1972uf,Kopylov:1972qw}. The quantity $q_{\mathrm{inv}}$ denotes the Lorentz-invariant momentum difference, defined as
$q_{\mathrm{inv}} = \sqrt{q_0^2-\mathbf{q}^2}$.
%where $\mathbf{q} = \mathbf{p_1} - \mathbf{p_2}$ is the pair relative momentum and $q_0 = E_1 - E_2$ the pair energy difference. 
% Theoretically, the correlation function is given by the Koonin-Pratt equation~\cite{Koonin:1977fh,Pratt:1990zq}:
% \begin{equation}
%   \begin{aligned}
%     C(\mathbf{P},\mathbf{q}) = \int S(\mathbf{r}^*) | \Psi(\mathbf{q},\mathbf{r}^*) |^2 d^3 r^*,
%   \end{aligned}
%   \label{EqFitEXPO}
% \end{equation}
% where $\mathbf{P}$ represents the total momentum, and $\mathbf{q}$ and $\mathbf{r}^*$ denote the relative momentum and particle separation in the pair's reference frame, respectively. Here, $\Psi$ is the two-particle wave function. For simplicity, the final-state interactions (Coulomb and strong) are neglected in $\Psi$, and only quantum statistical effects are considered.

One-dimensional (1D) analyses of pion femtoscopy are challenging because of the non-Gaussian nature of the source, caused by long-lived resonance contributions. Therefore, an exponential Bowler-Sinyukov function (neglecting the Coulomb interaction) is commonly employed to fit the pion CF~\cite{ALICE1dExpoFit}
\begin{equation}
  \begin{aligned}
    C(q) = 1 + \lambda \exp(-Rq),
  \end{aligned}
  \label{EqFitEXPO}
\end{equation}
where $\lambda$ indicates the correlation strength and $R$ the one-dimensional source radius.
More general L\'evy shapes have also been recently explored~\cite{CMS:2023xyd,Schegelsky:2018tit,Csanad:2024hva,Ayala:2023sbb}.

In three-dimensional (3D) analyses performed in the Longitudinally Co-Moving System (LCMS)~\cite{Pratt:1986cc,Bertsch:1988db}, information about the size and shape of the particle-emitting source can be extracted using the 3D Bowler-Sinyukov formula that, for a Gaussian-like source and ignoring the Coulomb correction, takes the form~\cite{Bowler:1991vx,Sinyukov:1998fc}
\begin{equation}
    C(q_{\text{out}}, q_{\text{side}}, q_{\text{long}}) = 1 + \lambda e^{( -q_{\text{out}}^2 R_{\text{out}}^2
    - q_{\text{side}}^2 R_{\text{side}}^2 - q_{\text{long}}^2 R_{\text{long}}^2)}.
  \label{EqFit3d}
\end{equation}
In the LCMS, the vector $q$ is decomposed into three components: $q_{\mathrm{out}}$ (in the direction of the average transverse pair momentum), $q_{\mathrm{long}}$ (in the direction of the beam) and $q_{\mathrm{side}}$ (perpendicular to both directions). This parameterization allows us to measure all three independent combinations of four space-time dimensions of the source.

Here we analyze MC data obtained from the centralized production 1 in   Table~\ref{EvGens}, using the UrQMD model. We discuss the effects that influence femtoscopic correlations from the 
\begin{figure}[H]
\centering
\includegraphics[width=0.99\linewidth]{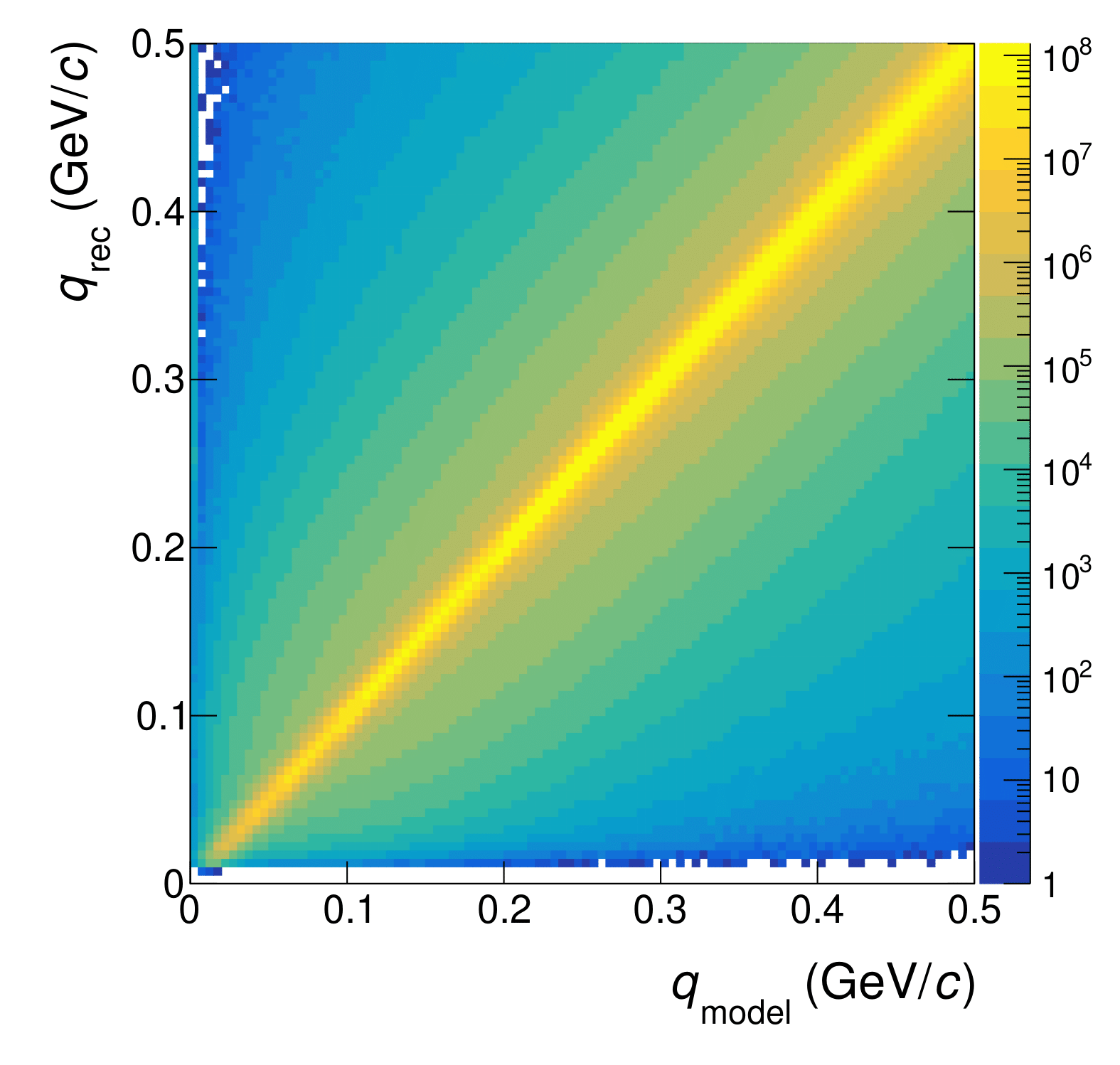}
\includegraphics[width=0.9\linewidth]{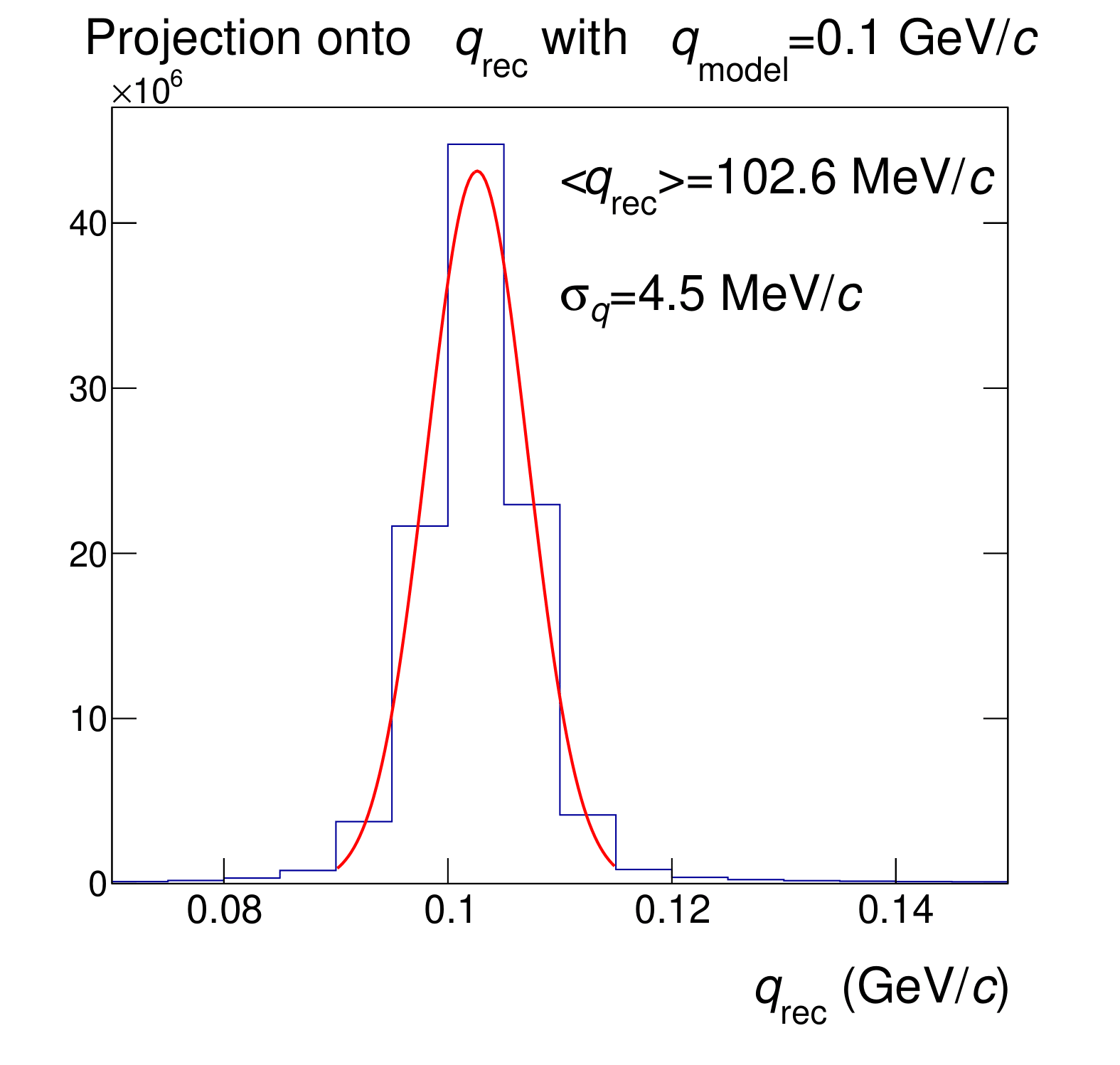}
%\includegraphics[width=65mm]{Figs/MR_MPD.pdf}
%\vspace{-3mm}
\caption{Effect of finite momentum resolution for the two particle relative momentum $q$. The upper panel shows, with different colors, the number of correlated pair relative momenta, quantified in the vertical scale on the right side of the plot. The lower panel shows the projection on the $q_{\mathrm rec}$ axis and corresponds to a distribution with a width of 4.5 \MeVc.}
\label{femto_fig02}
%\vspace{-5mm}
\end{figure}
\noindent experimental point of view. The most significant factors in this context are the two-track effects and the momentum resolution.

In femtoscopic studies of two identical charged particles, track pairs with similar momenta and emission angles from the reaction region are subject to specific reconstruction effects. Track merging occurs when two spatially close tracks are incorrectly reconstructed as one, leading to inefficiency in the reconstruction of close pairs. Conversely, track splitting occurs when a single track is erroneously reconstructed as two tracks, which are very close to each other. This results in a false enhancement of close pairs in the correlation func-
\begin{figure}[H]
\vspace{-0.4cm}
\centering
\includegraphics[width=1.05\linewidth]{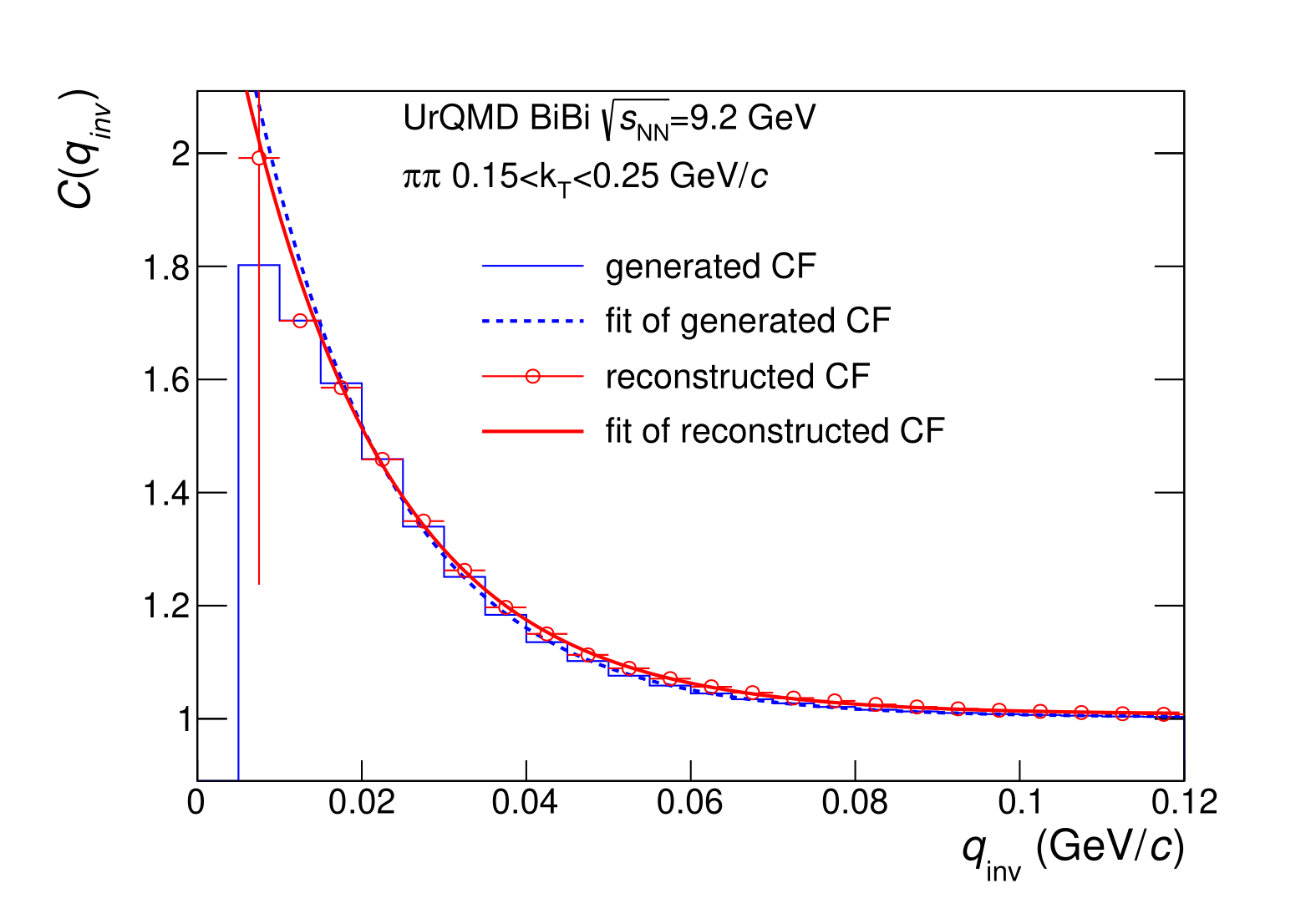}
\caption{Example of simulated pion CFs fitted as function of
the invariant pair
relative momentum $q_{\mathrm{inv}}$.
%The solid blue line is a true UrQMD model correlation fuction.
The CFs were fitted using Eq.~(\ref{EqFitEXPO}).}
\label{femto_fig03}
\end{figure}
\noindent tion, particularly in the region of femtoscopic effects at small momentum differences. Consequently, the extracted radii and $\lambda$ parameters can be affected.

Since two-track effects occur at small angular distances, restrictions on the azimuthal angle $\Delta \phi^*$ and the polar angle 
$\Delta \eta$ between tracks are typically applied~\cite{ALICE:2022boj}. The angle $\phi^*$ is defined as the azimuthal angle $\phi$ of a particle with transverse momentum $\pT$ and charge $ze$ at some radius~$\mathcal{R}$ within the TPC in a magnetic field $\mathcal{B}$, \begin{equation}
  %\phi* = \phi + \arcsin(z e  \mathcal{B}_Z  \mathcal{R} 
  \phi* = \phi + \arcsin(z e  {B}_Z  \mathcal{R} / 2 \pT ).
  \label{EqPhiStar}
\end{equation}

% text to figure 2 (Momentum Resolution)
%\begin{multicols}{2}

% KM version was:
% The   $\Delta \eta$($\Delta \phi^*$) distribution of the pion pairs normalized to an mixed event sample is shown in left panel of Fig.~\ref{femto_fig01}. The $\Delta \phi^*$ projection is shown in the middle panel of Fig.~\ref{femto_fig01}. And the $\Delta \eta$ projection is shown in the right panel of Fig.~\ref{femto_fig01}. The appearance of the inefficiency region due to the two-track effects is clearly seen at low $\Delta \phi^*$ and $\Delta \eta$. The width of inefficiency region depends on the detector geometry and the two-track reconstruction efficiency~\cite{Cheremnova:2022fsd}. 
%
% Finite track momentum resolution is the reason why the reconstructed relative momentum of a pair differs from the true one. This can be taken into account in the theoretical function using the response matrix \cite{KpKmALICE} generated by UrQMD model. The example of such matrix are shown in Fig.~\ref{femto_fig02}. The width of the smearing effect ($\sigma_q$) is estimated to be about 4.5 MeV/c at the region of the femtoscopic effect.

%%% Grigory's suggestion

The $\Delta \eta$ $\Delta \phi^*$ distribution of pion pairs, normalized to a mixed event sample, is shown in the left panel of Fig.~\ref{femto_fig01}. The $\Delta \phi^*$ projection is shown in the middle panel and the $\Delta \eta$ projection is shown in the right panel. A region of inefficiency due to two-track effects is clearly visible at low $\Delta \phi^*$ and $\Delta \eta$. The width of this inefficiency region depends on the detector geometry and the two-track reconstruction efficiency~\cite{Cheremnova:2022fsd}.

Finite track momentum resolution causes the reconstructed relative momentum of a pair to differ from the true value. This can be taken into account in the theoretical function using the response matrix~\cite{KpKmALICE}. An example of such a matrix, 
that correlates the UrQMD generated relative momentum $q_{\rm model}$ with the reconstructed relative momentum $q_{\rm rec}$, is shown in the upper panel of Figure~\ref{femto_fig02}. The width of the smearing effect ($\sigma_q$) is
estimated to be about 4.5~\MeVc in the region of the femtoscopic effect, this is shown in lower panel of Fig.~\ref{femto_fig02}.

The 1D CFs were studied in three intervals of pair transverse momentum \kT ($\kT = |\mathbf{p}_{\mathrm{T,1}} + \mathbf{p}_{\mathrm{T,2}}|/2$): 0.15–0.25, 0.25–0.35, and 0.35–0.45~\GeVc, as 
well as three centrality classes: 0–10\%, 10–30\%, and 30–50\%. The fits were performed using Eq.~(\ref{EqFitEXPO}). Figure~\ref{femto_fig03}
shows the pion CFs as a function of the invariant pair relative momentum $q_{\mathrm{inv}}$. The solid blue line represents the CF with particle momenta from
the UrQMD model. The open circles correspond to the 
\begin{figure}[H]
\centering
\includegraphics[width=1.\linewidth]{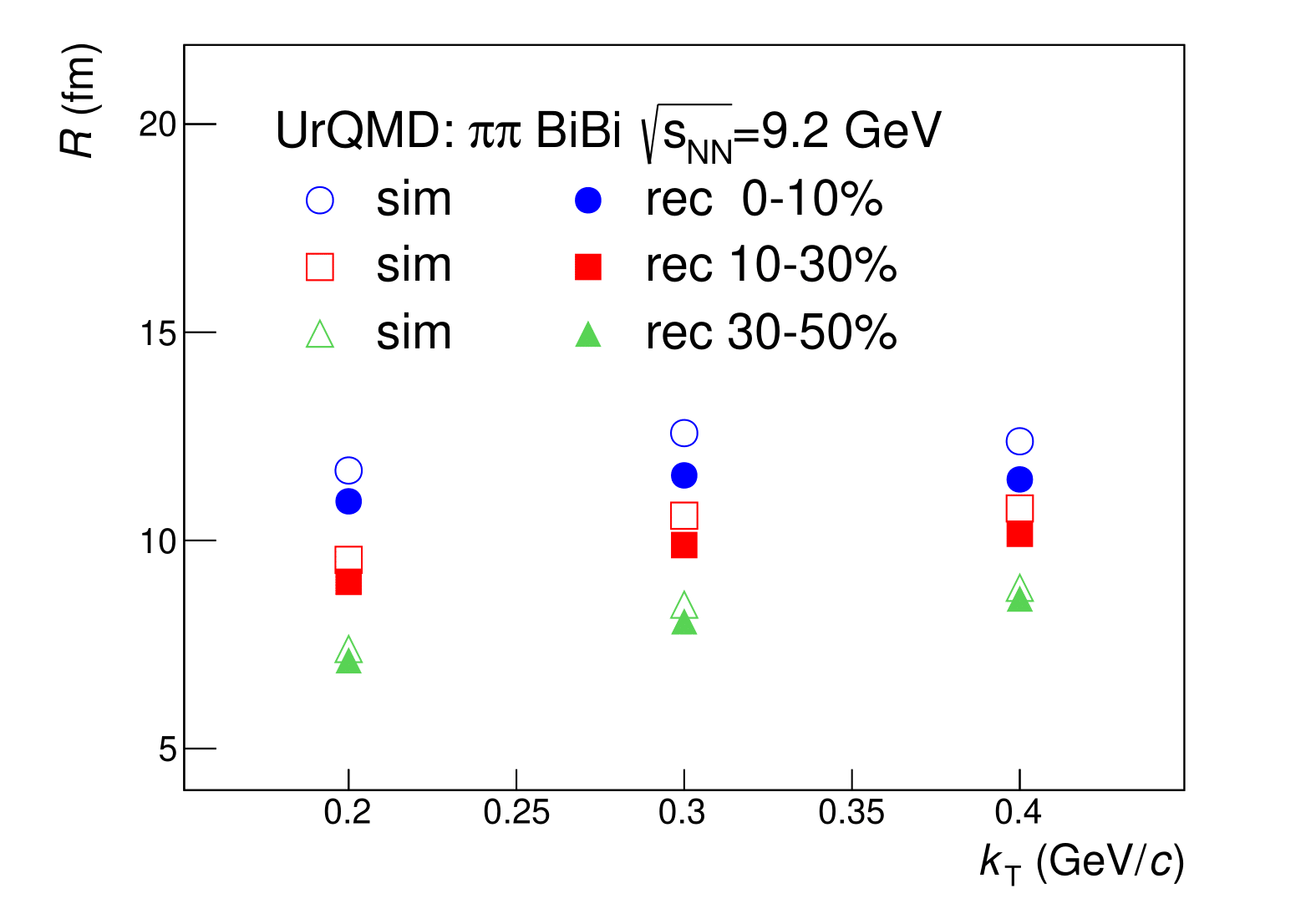}
\caption{The one-dimensional radii extracted from the CFs for charged identical pions versus \kT. Empty and full symbols show results for the simulated and reconstructed CFs.}
\label{femto_fig04}
\end{figure}
\noindent 
CF with reconstructed pion track momenta for tracks with a number of hits greater than or equal to 40. Both CFs were obtained with cuts to exclude two-track inefficiency effects: $|\Delta\eta|<0.07$ and $|\Delta\phi^*|<0.07$, as determined from Fig.~\ref{femto_fig01}. Notice that there is some disagreement between the generated and reconstructed CFs in Fig.~\ref{femto_fig03} in the region $q_{\rm inv} < 0.01$~\GeVc, attributed to the two-track cut effects. The curves in the figure are for fits to the CFs using Eq.~(\ref{EqFitEXPO}). The radius of the reconstructed correlation function is approximately 6\% smaller than that of the 
ideal initial CF due to the distortion caused by momentum resolution.
Figure~\ref{femto_fig04} shows the extracted radii, $R$, as a function of $k_{\mathrm{T}}$ for the 0–10\%, 10–30\%, and 30–50\% centrality intervals. The fit used to obtain the values of $R$ was performed for both the reconstructed correlation function (solid symbols) and the true UrQMD model correlation function (open symbols). The exponential radius is almost flat as a function of $k_{\mathrm{T}}$. The variation of the radius with centrality is consistent with the geometric interpretation of the collisions. The maximum deviation between the reconstructed radii and the model radii  
\begin{figure}[H]
\centering
\includegraphics[width=\linewidth]{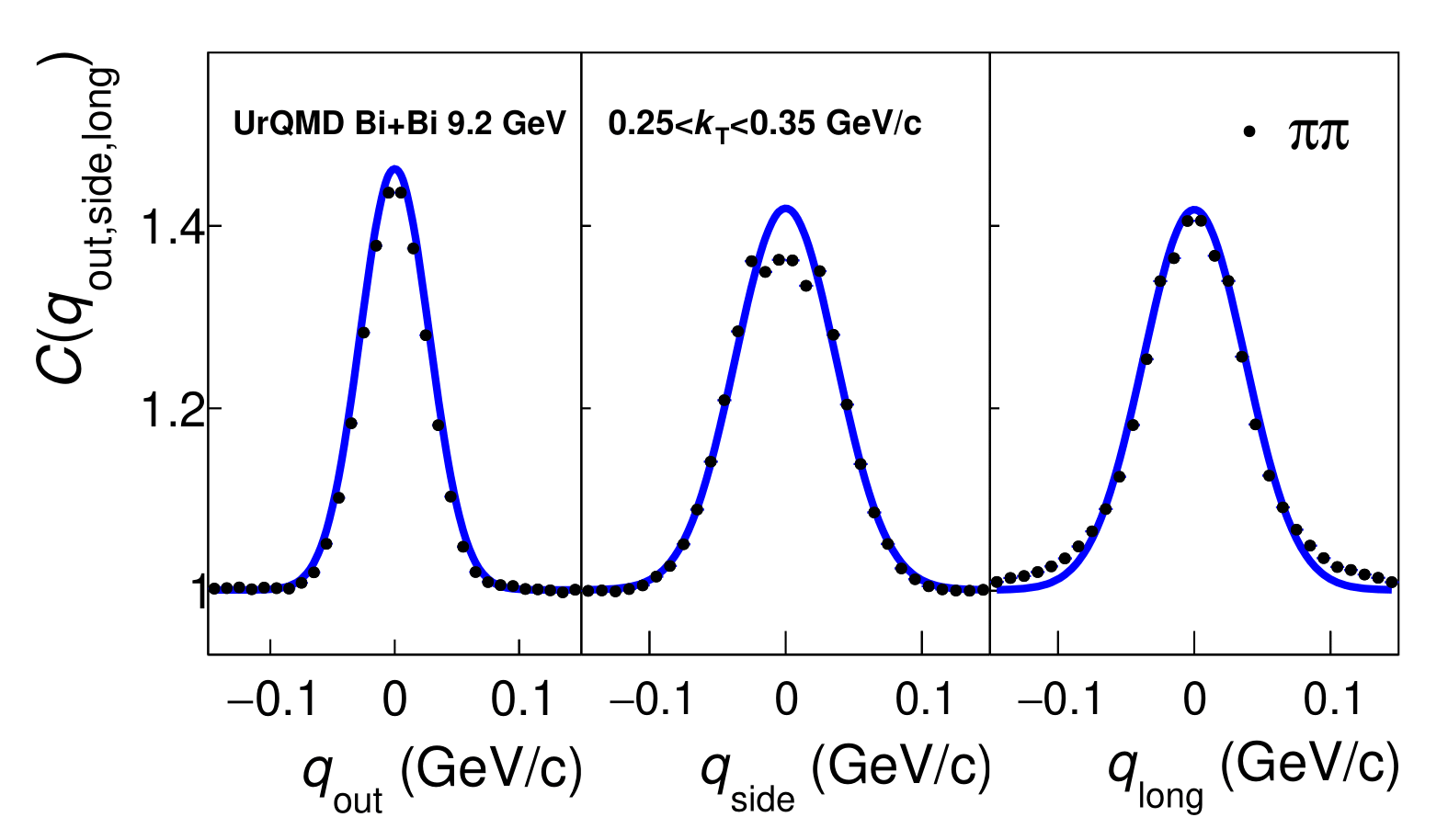}
\caption{Three-dimensional two-pion correlation function projections 
onto the out (left), side (middle), and long (right) directions with $0.25< \kT <0.35$~\GeVc 
for 0--10~\% central \BiBi collisions at \snn=9.2~\GeV. 
Solid lines represent projections of the three-dimensional fit with Eq.(\ref{EqFit3d}) on the corresponding axis.
}
\label{femto_fig06}
\end{figure}
\begin{figure}[H]
\centering
\includegraphics[width=1.02\linewidth]{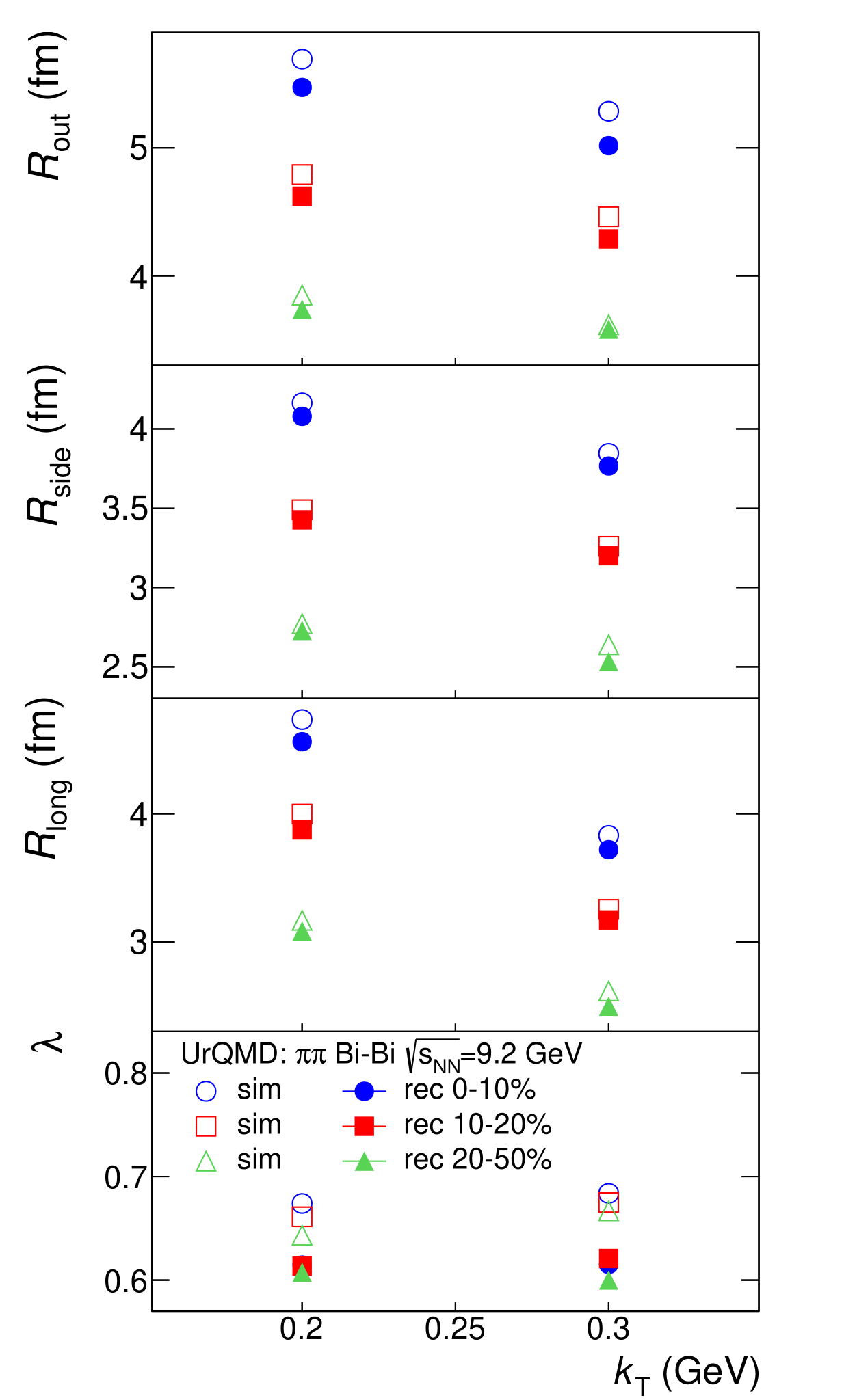}
\caption{The terms $R_\mathrm{out}$, $R_\mathrm{side}$, and $R_\mathrm{long}$ and $\lambda$ versus \kT
for 0--10~\%, 10--30~\%, 30--50~\% central  Bi+Bi collisions at \snn=9.2~\GeV. Empty and full symbols show results for the simulated and reconstructed CFs.}
\label{femto_fig07}
\end{figure}
\noindent
is observed to be approximately 8\%, while the minimum deviation is around 3\%. The reduction in the reconstructed radii, compared to the model ones, is primarily attributed to the effects of momentum resolution.

The 3D $\pi\pi$ correlations were fitted for two \kT intervals: 0.15–0.25 and 0.25–0.35~\GeVc, as well as for three centrality classes: 0–10\%, 10–30\%, and 30–50\%. The fits were performed using Eq.~(\ref{EqFit3d}). Figure~\ref{femto_fig06} shows the 3D CF projections for the first \kT interval onto the out (left), side (middle), and long (right) directions. These correlation functions
were obtained for 0–10\% central Bi-Bi collisions at \snn = 9.2~\GeV, as simulated in the UrQMD model. The projections of the fitted function, according to Eq.~(\ref{EqFit3d}), are also shown in the figure. Deviations of the CF from the fit function at small relative momenta are associated with the application of two-track cuts.

%Very bad idea to move this fig from Femto section, but otherwise latex does not compile....
\begin{figure*}[t]
\centering
% \begin{multicols}{2}
    \includegraphics[width=0.49\textwidth]{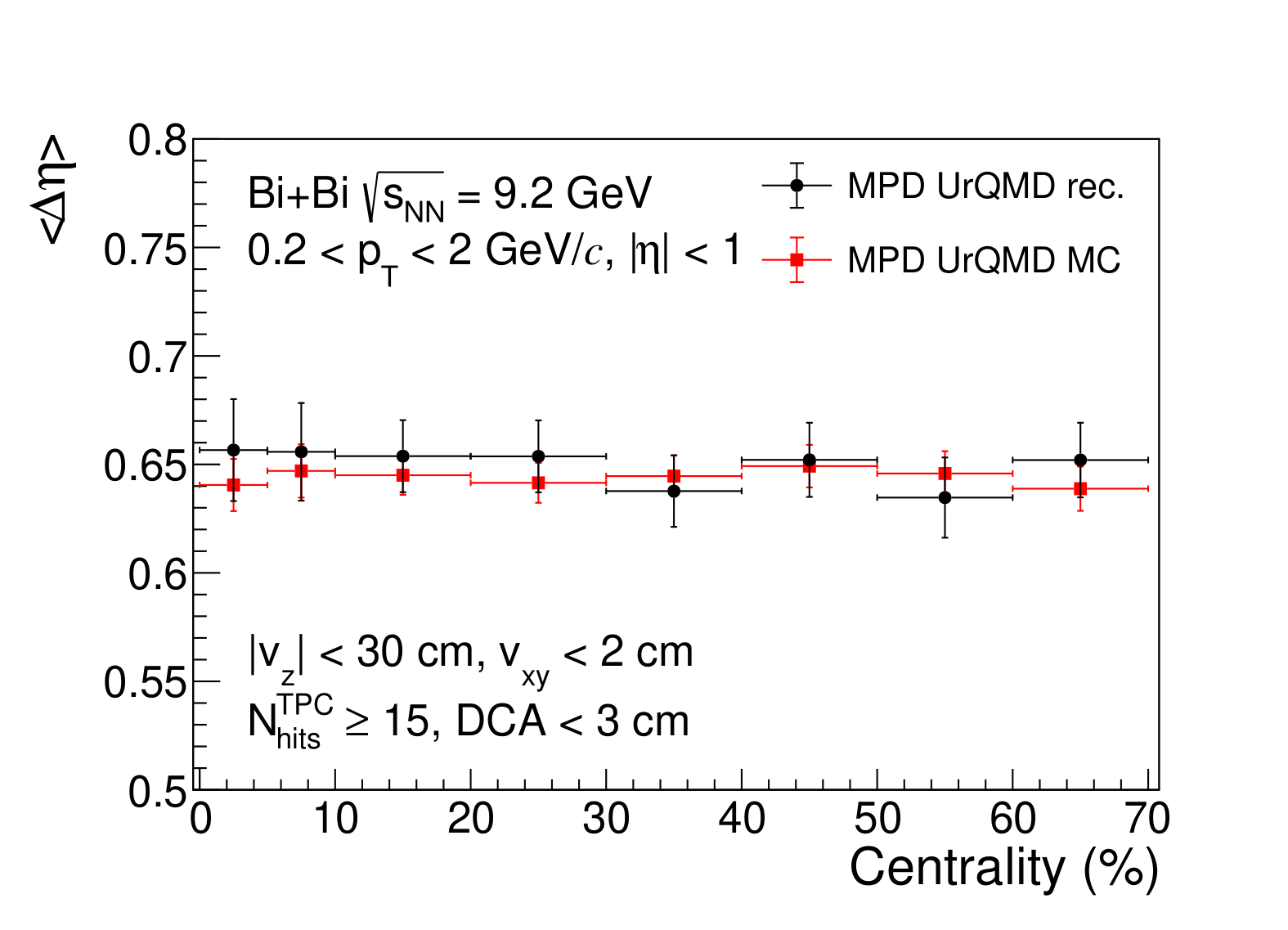}
    \hfill
    \includegraphics[width=0.49\textwidth]{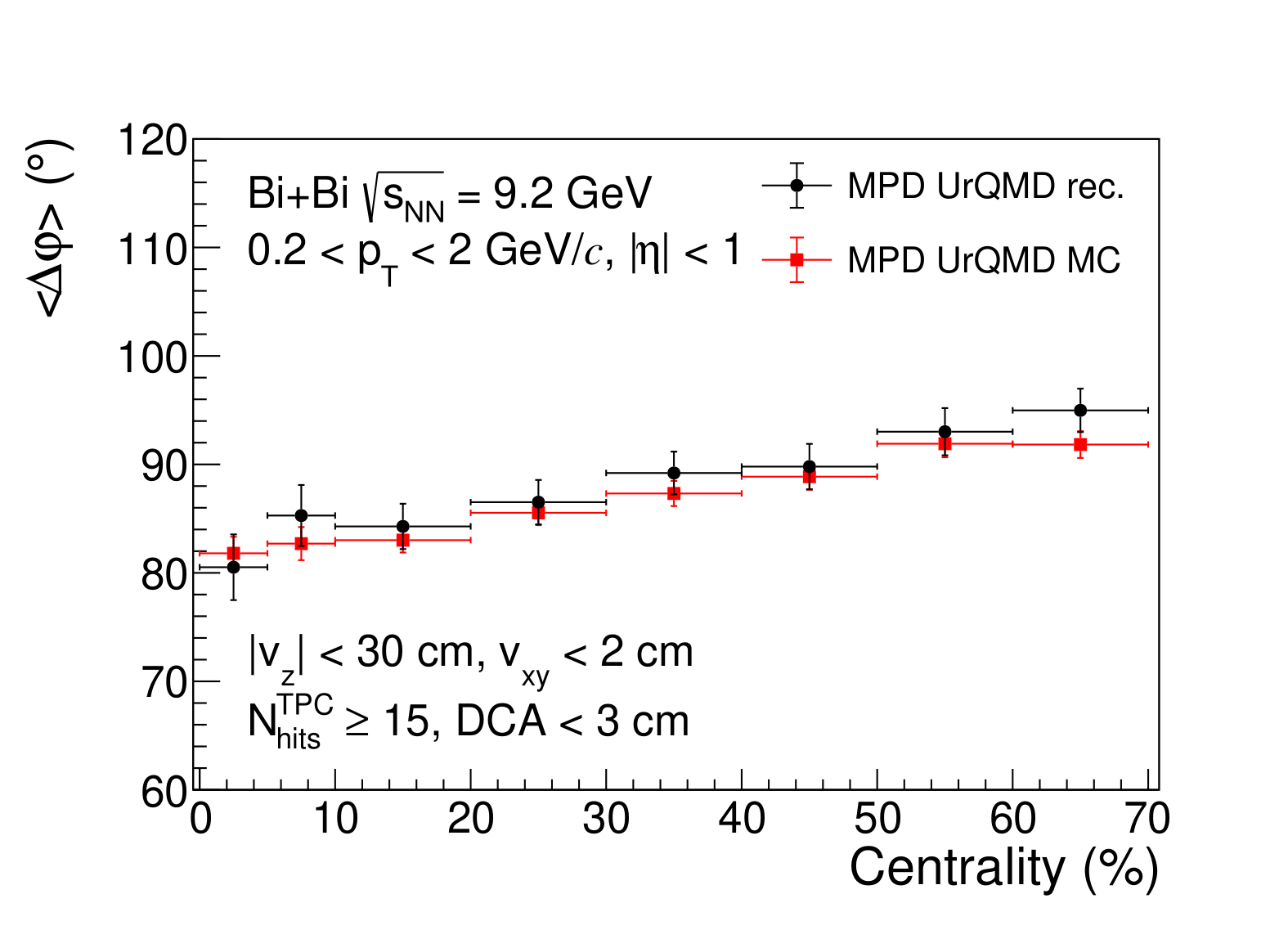}
    % \end{multicols}
\caption{The pseudo-rapidity (left) and azimuthal (right) charge balance function width for inclusive charged hadrons. Black circles represent the widths obtained from reconstructed events whereas red squares represent the widths obtained at generator level (UrQMD data).}
\label{fig:wetawphi}
\end{figure*}

Figure~\ref{femto_fig07} shows the extracted out-side-long radii of pions for two different \kT intervals: (0.15–0.25) and (0.25–0.35)~\GeVc, along with three centrality classes: 0–10\%, 10–30\%, and 30–50\%. The fit was performed for both the reconstructed correlation function (solid symbols) and the true UrQMD model correlation function (empty symbols). The reconstructed radii are smaller than the model ones, primarily due to finite momentum resolution. It is evident from Fig.~\ref{femto_fig07} that the radii in all directions decrease with increasing transverse momentum of the pair. This behavior can be attributed to the presence of radial flow~\cite{Akkelin:1995gh, STAR:2004qya}.

The centrality dependence of the out-side-long radii is related to a simple geometric picture of ion collisions.
The parameter $\lambda$ equals unity in the ideal case of a Gaussian spherical source consisting only of primary particles emitted randomly from the source. The correlation strength $\lambda$ is less than 0.7 for the model, which could be due to the influence of long-lived resonances and a non-ideal Gaussian source distribution. The value of the parameter $\lambda$ for the reconstructed CF is lower than that for the model CF, primarily due to finite momentum resolution and distortion of the CF resulting from two-track cuts.

\subsubsection{Charged balance function}
%\vspace{-0.09cm}

The charge balance function (CBF) has been proposed as a convenient measure of the correlation between oppositely charged particles~\cite{Bass:2000az}. It provides valuable insight into the charged particle production mechanism and can address the fundamental question concerning the hadronization process in nuclear collisions at relativistic energies~\cite{Pratt:2021xvg}. The final degree of correlations is reflected in the balance function and consequently in its width. It is defined as
\begin{equation}
\begin{aligned}
B(\Delta y) = \frac{1}{2}\left\{\frac{\langle N^{+-}(\Delta y) \rangle - \langle N^{++}(\Delta y) \rangle}{\langle N^{+} \rangle}\right. \\
+ \left.\frac{\langle N^{-+}(\Delta y) \rangle - \langle N^{--}(\Delta y) \rangle}{\langle N^{-} \rangle}\right\},
\end{aligned}
\end{equation}
where $\langle N^{+-}(\Delta y) \rangle$ is the average number of opposite-charge pairs with particles separated by a relative rapidity  $\Delta y$, and similarly for $\langle N^{-+}(\Delta y) \rangle$, $\langle N^{++}(\Delta y) \rangle$, and $\langle N^{--}(\Delta y) \rangle$. $\langle N^{+} \rangle$ and $\langle N^{-} \rangle$ are the numbers of positively and negatively charged particles in the rapidity interval, over all events. The charge balance function $B(\Delta\varphi)$, as a function of the relative azimuthal angle $\Delta\varphi$, is defined similarly~\cite{Bass:2000az}. $\langle N^{+-}\rangle$ and $\langle N^{-+}\rangle$ are equal for inclusive CBFs, however, they may differ for partial CBFs. The analysis for partial CBFs is currently outside the scope of the present study. The width of the balance function distribution is defined as
\begin{equation}
\langle \Delta y \rangle = \frac{\sum_\mathrm{i} B_\mathrm{i}\Delta y_\mathrm{i}}{\sum_\mathrm{i} B_\mathrm{i}},
\end{equation}
where $B_\mathrm{i} = B(\Delta y_\mathrm{i})$ is the balance function value for each bin, with the sum running over all bins. The CBF width is sensitive to the duration of electric charge separation, and thus provides information on the hadronization time and may be used to extract information about the space-time characteristics of the particle emitting source. In a hydrodynamic approach, the width is proportional to the inverse strength of the collective radial flow in the system, allowing to estimate collective effects as well.

CBFs for heavy-ion collisions were experimentally studied at SPS~\cite{NA49:2004vzs}, RHIC~\cite{STAR:2010plm,STAR:2015ryu}, and LHC~\cite{ALICE:2013vrb,ALICE:2018jco,Pan:2018dsq}. Two interesting experimental observations were made: the balance function width increases with the increase of the centrality, and the width decreases while the energy of the beam increases.

The CBF modeling for MPD conditions was performed using UrQMD-based production number 1 from Table~\ref{EvGens}. The tracks were selected according to cuts similar to those used in the analysis of the STAR experiment \cite{STAR:2015ryu}: $0.2 < \pT < 2$~\GeVc and $|\eta| < 1$. The tracks were required to have at least 15 hits in the \TPC and to be matched to the primary vertex with DCA < 3 cm. The primary vertex was restricted to be positioned within 30 cm along the beam axis and within 2 cm in the transverse direction. Both, rapidity and azimuthal CBFs for inclusive and identified charged hadrons, were analyzed at \snn = 9.2~\GeV in \BiBi collisions in the 0-80\% centrality class. Figure~\ref{fig:wetawphi} shows the pseudo-rapidity and azimuthal charge balance function width for inclusive charged hadrons, where black circles represent the widths obtained from reconstructed events and red squares the generator level UrQMD data. Notice that the CBFs shown in Fig.~\ref{fig:wetawphi} are not significantly affected neither by the finite momentum resolution nor by particle identification
effects. This observable is considered robustly resistant to common detector inefficiencies due to the fact that only the correct
determination of the electric charge is essential, which is done with
very good accuracy.
\begin{figure*}[b]
\includegraphics[width=0.49\textwidth]{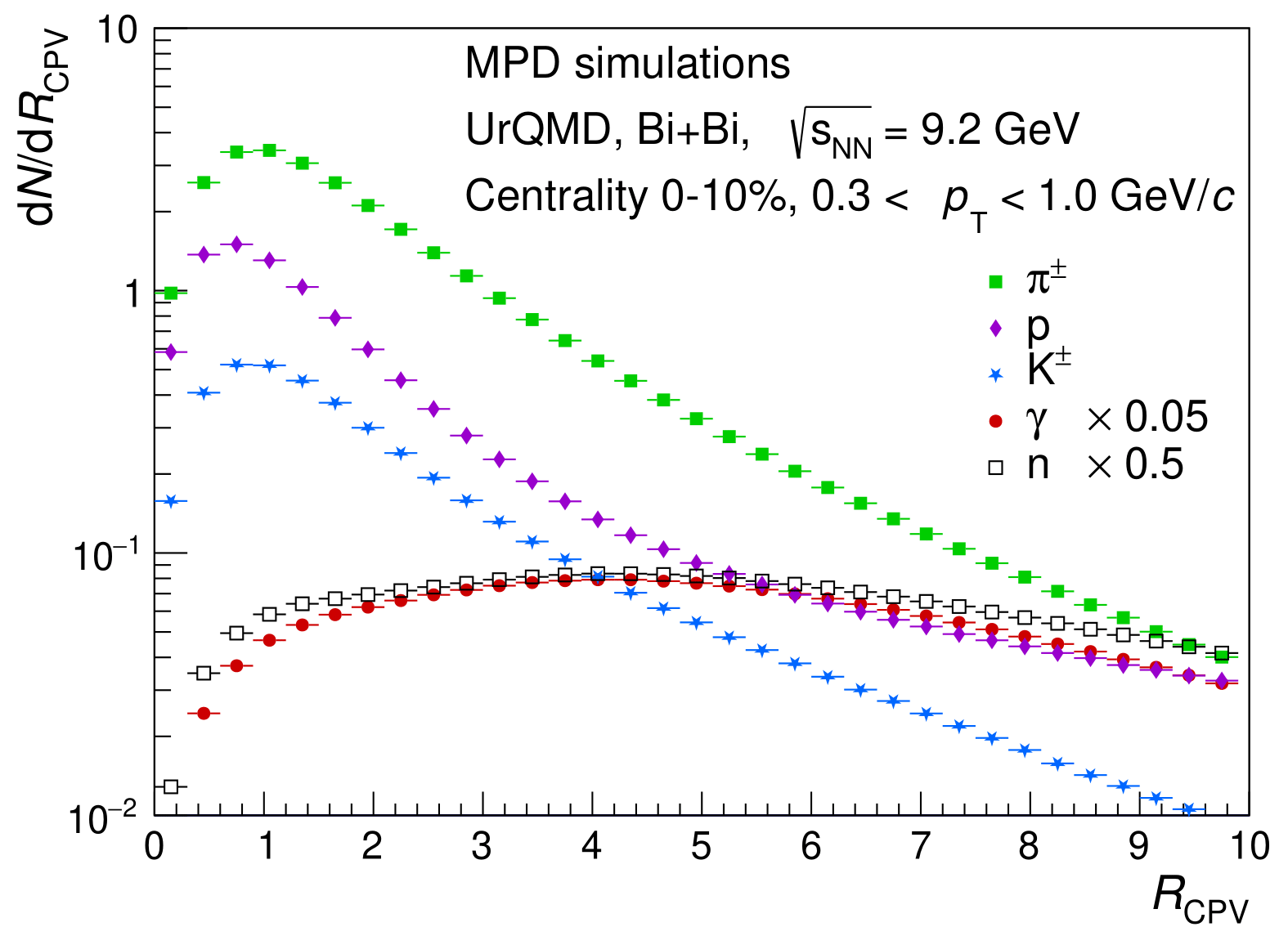}
\hfill
\includegraphics[width=0.49\textwidth]{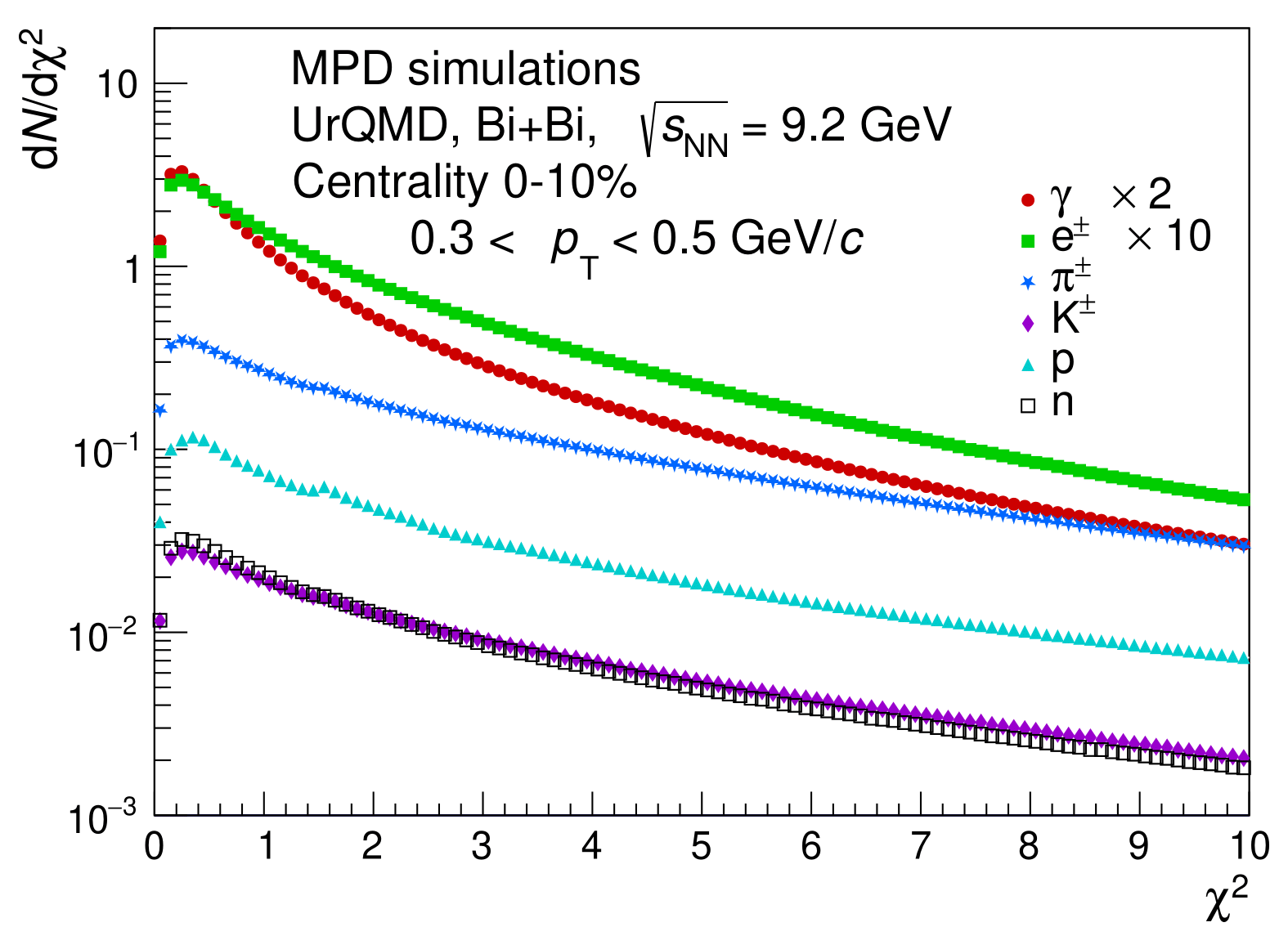}
\caption{Left: distance to closest track in units of standard deviations for clusters produced by different  particles. Right: shower shape fit parameter distribution for different kinds of clusters.
\label{fig:CPV}
}
\end{figure*}

The CBFs were corrected to account for the charge imbalance that is present (due to the finite
values of baryon, strangeness and isospin chemical potentials)  at NICA energies, using the event mixing technique~\cite{STAR:2015ryu}. This technique requires to calculate an additional set of CBFs composed of tracks that are selected from different events. These mixed CBFs can be subtracted from the same-event CBFs, to remove distortions due to charge imbalance. To estimate the reconstruction efficiency, the reconstructed widths were compared to those obtained at the generator level.
%\textbf{Given that rapidity and azimuthal CBFs widths are in acceptable agreement with those obtained on generator level for inclusive as well as foridentified charged hadrons ($\pi^+\pi^-$, $K^+K^-$, $\pi^+K^-$ and $K^-\pi^-$ pairs), the recosntruction efficiency is ... ??\%}
%This shows a good progress towards a better understanding of the hadronization processes dominating within the NICA energy ranges.

In summary, femtoscopic and correlation studies are useful tools to reveal the space-time properties of the particle emitting source in relativistic heavy-ion collisions. We have shown that the MPD momentum resolution allows to carry out this kind of studies, providing an agreement within statistical uncertainties between the reconstructed and model parameters.

%\subsubsection{Factorial moments}
%\input{factorial_moments}

\subsection{ Electromagnetic signals and neutral mesons }
Electromagnetic signals -- photons and electrons -- provide the possibility to measure spectra and correlations of neutral mesons, direct photons and dilepton pairs. Neutral mesons can be reliably identified in a wide momentum range and complement measurements of charged identified hadrons. 
Direct photons are the photons not originated from decays of final state hadrons, but produced in electromagnetic interactions in the course of the collision. Direct photons escape the hot fireball and deliver information about temperature, development of the collective flow and space-time dimensions of 
the system  at all stages of the collision, including the hottest one.
Dileptons similar to (real) direct photons allow us to probe the hot matter, but in addition, reflect in-medium modifications of vector meson properties. This makes them sensitive to both the deconfinement and chiral symmetry restoration phase transitions.
In this section, we review the \MPD capabilities for the measurements of photons, neutral mesons and dielectrons in \BiBi collisions at $\snn = 9.2$~GeV.

\subsubsection{Photons}
\label{sec:thermal}

\begin{figure*}[b]
\includegraphics[width=0.45\textwidth]{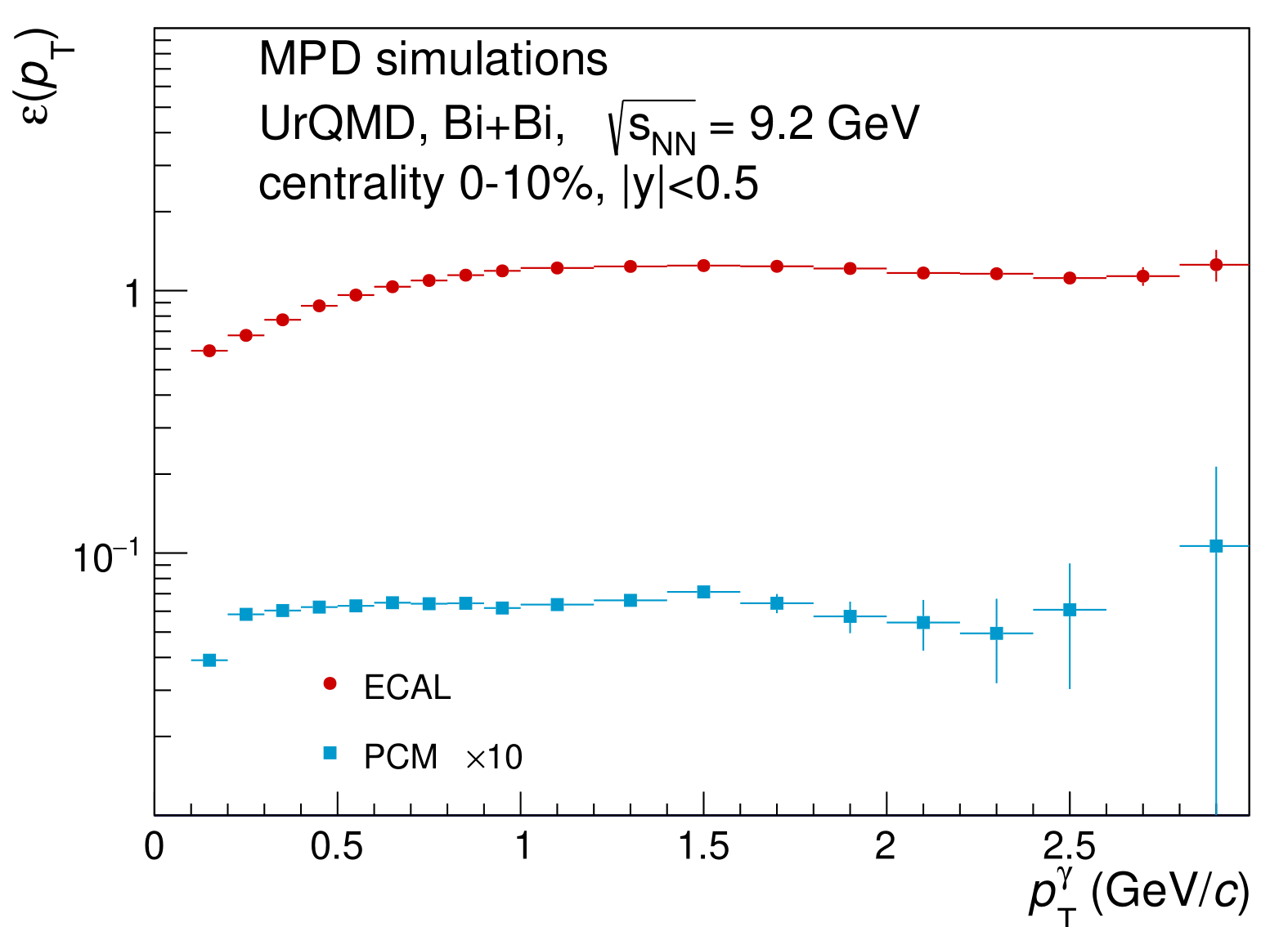}
\hfill
\includegraphics[width=0.45\textwidth]{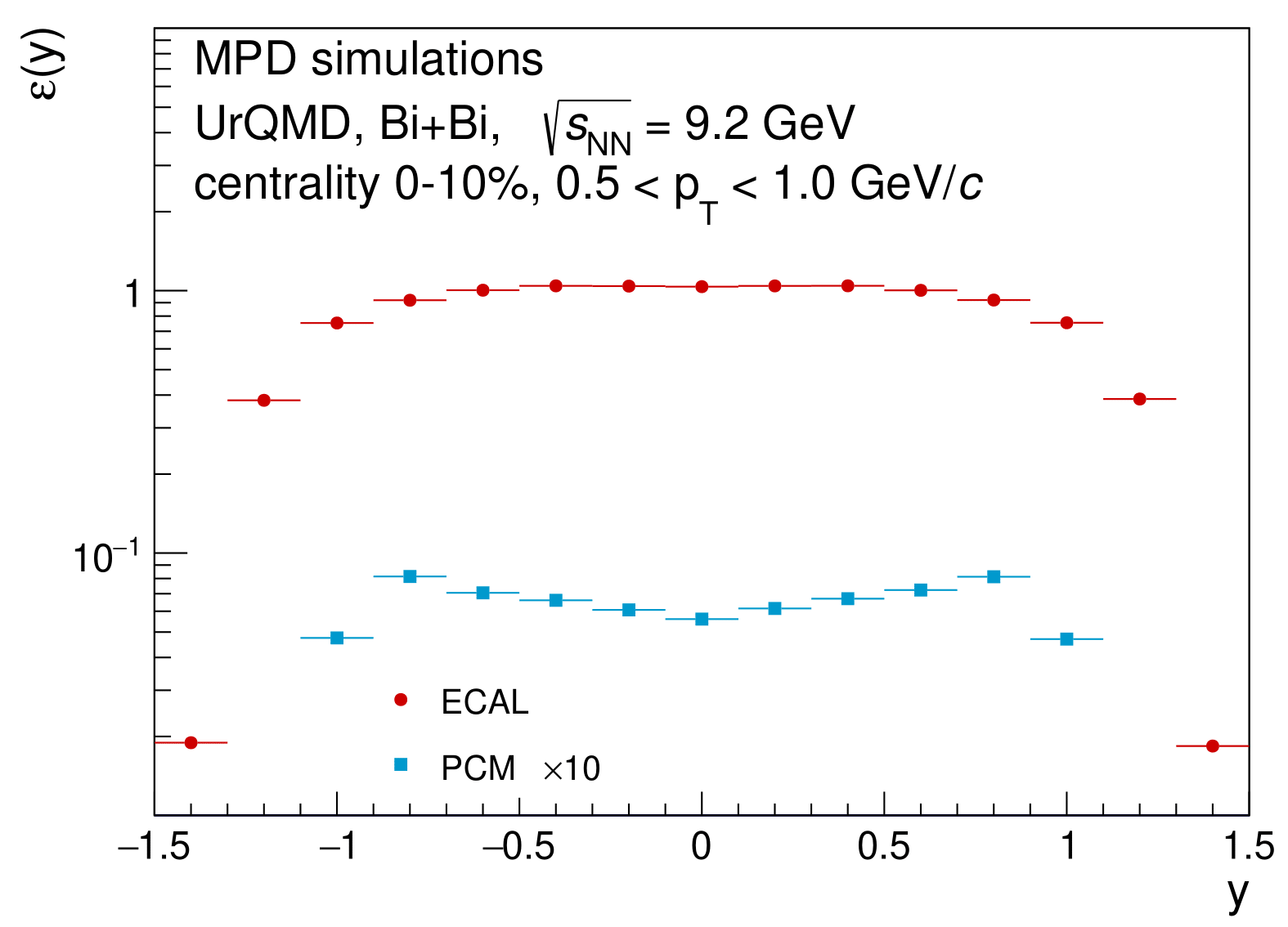}
\caption{Left: inclusive photon reconstruction efficiency in the \ECAL and PCM method as a function of photon $\pT^{\gamma}$. 
Right: inclusive photon reconstruction efficiency in the \ECAL and PCM method as a function of rapidity.
\label{fig:GammaEffPt}
}
\end{figure*}

Direct photons can be emitted either in hard processes involving partons of incoming nucleons ({\it prompt} direct photons) or as the thermal emission of hot quark or hadron matter ({\it thermal} direct photons).
Prompt photon production at NICA energies probes nucleon structure functions in a high $x_\mathrm{Bj}$ region where they are relatively poorly constrained \cite{Blau:2023bvi}. 
%The uncertainties of the pQCD calculations of the prompt photon yield at NICA energies are summarized in \cite{Blau:2023bvi}, where results of Jetphox NLO pQCD calculations  with several recent parameterizations of proton or nuclei structure functions are presented.
%At low $\pT<1$~\GeVc, the difference between calculations with different structure functions is modest and reaches a few tens of percent, but it rapidly increases with $\pT$ and can reach a factor of $\sim 10$ at $\pT\sim 1.5$~\GeVc. Therefore, the measurements of prompt photons at NICA energies will provide strong constraints for structure functions at high $x_\mathrm{Bj}$.
%
Predictions for the thermal direct photon yields in heavy-ion collisions at NICA energies are very scarce. One of them is based on hydrodynamic calculations combined with the UrQMD model~\cite{Blau:2023bvi}. Another one is based on the phenomenological extrapolation of available experimental results \cite{Kryshen:2021yml}. The two approaches provide similar predictions. 
%The comparison of thermal, prompt and decay photon contributions are presented in Figure~\ \ref{fig:gDir-spectra}. 
The expected contribution of direct photons to the inclusive spectrum is on the level of 5-10\% at $\pT\sim 1$~\GeVc which makes their reconstruction challenging, but yet a realistic experimental task.
% As the probability to emit direct photons rapidly increases with temperature, they provide a unique possibility to explore development of collective flow in heavy ion collisions on the early stages of the collision. 
% The dependence of the directed collective flow of direct photons $v_1$ on rapidity in Au-Au collisions at $\snn = 11$ is shown in Figure \ref{fig:photon_v1}. Directed flow $v_1$ is found to have similar dependence on the rapidity to the one predicted for charged hadrons (e.g. protons) at the same center-of-mass energies \cite{Parfenov:2019pxf}. The blue band represents RMS of event-by-event variations of flow due to fluctuations in the initial state. Variations are found to be larger in collisions with smaller energies \cite{Blau:2023bvi}. One of the interesting measurement at NICA would be to explore correlations between direct photon and hadron directed and elliptic flow to trace how fluctuations in the initial state influence hadron collective flow.

Photons in the \MPD can be reconstructed in two ways, either in the electromagnetic calorimeter (\ECAL) or converted in the material of the beam pipe or inner walls of the \TPC and reconstructed as a pair of $e^+e^-$ tracks in the tracking system.

To reconstruct photons in \ECAL, a clusterization procedure is used. It selects a seed cell with the energy above the threshold $E_\mathrm{seed}=30$~MeV and adds all cells with common side and energy exceeding a minimal energy threshold of 5~MeV. If the cluster has more than one local maximum, 
%(cell with energy higher than all surrounding cells at least by $E_\mathrm{locMax}=5$~MeV) 
an unfolding procedure is applied based on the fitting energy depositions in all cells with electromagnetic shower shapes with local positions and energies considered as free parameters. The energy of a cluster is calculated as a sum of the energies of 
the cells. The coordinates of a cluster both in $z$ and $\phi$ directions are assigned to the "centers of gravity" calculated with logarithmic weights, similar to e.g. calorimeters in the ALICE experiment \cite{ALICE:1999ozr}
\begin{equation}
\langle x\rangle = \frac{\sum w_{i} x_i}{\sum w_i}, \quad w_i = \max\left (0,\log\left (\frac{E_i}{E}\right)+5.5\right ),
\label{EMweight}
\end{equation}
where the cutoff parameter 5.5 is chosen as large as possible with expected electronic noise.

Photon identification in the \ECAL is performed based on three independent criteria: time-of-flight, neutrality and 
shower shape. The time-of-flight is based on the good time resolution of the \ECAL which was estimated in beam tests~\cite{Semenov:2020glg} to reach about 250 ps at $E_\mathrm{clu}>500$ \MeV. %This time resolution was parameterized and implemented in MpdRoot package. 
The neutrality of a cluster is estimated by calculating the distance to the closest track reconstructed in the \TPC and extrapolated to the \ECAL surface. The width of this distribution is parametrized and the distance between cluster and extrapolated track \RCPV~is provided in units of $\sigma$, see Figure~\ \ref{fig:CPV}, left. Clusters, associated with charged particles have maxima at $\RCPV\sim 1$, while photon and neutron clusters have wider distributions from random associations between clusters and tracks. 

The third photon identification criterion is based on the shape of the cluster: hadrons produce either cluster with very small dispersion in the case of minimum ionizing particles, or clusters with large dispersion in the case of strong hadronic interaction. Photons and electrons, in contrast, produce compact clusters. Quantitatively, the comparison can be done in two approaches, either by evaluating eigenvalues of the dispersion matrix
\begin{equation}
  M_\mathrm{ij} = \frac{\sum\limits_\mathrm{k} (x_\mathrm{i,k} - \langle x_\mathrm{i}\rangle )( x_\mathrm{j,k} - \langle x_\mathrm{j}\rangle) w_\mathrm{k}}{\sum\limits_\mathrm{k} w_\mathrm{k}}
\end{equation}
where $x_\mathrm{i,k}$ is the i-th coordinate  in the \ECAL surface of the cell with number k and $w_\mathrm{k}$ is the logarithmic weight, the same as in Eq.~(\ref{EMweight}). An alternative approach calculates the result of the fit of the energy distribution within clusters with the expected electromagnetic shower shape. It returns the $\chi^2$ which can also be used to separate photon and hadron showers, as can be seen in Figure~\ \ref{fig:CPV}, right. Non-electromagnetic clusters have wider distributions, a feature that is used for photon or electron selection.

The second method of the photon reconstruction in the MPD is the Photon Conversion Method (PCM). It is based on reconstruction of $e^+e^-$ pairs created in photon conversion in the material of the beam pipe or of the inner vessels of the \TPC. Electron and positron tracks are identified in the \TPC, requiring the measured specific ionization losses $\dEdx$ to be within $3\sigma^{e}_\mathrm{TPC}(\pT)$ from the values expected for electrons. If tracks are matched to the \TOF, their measured velocities are required to be consistent with electron signals within $3\sigma^{e}_\mathrm{TOF}(\pT)$. Two identified tracks are then  combined with a Kalman Filter for a \VZ particle. A set of topological selections are considered and used to select true conversion pairs: $\chi^2<10$, the DCA of two tracks (DCA$\ < 2.8$ cm), the Cosine of Pointing Angle (CPA) between pair momentum and direction from conversion vertex to the primary vertex (CPA$\ >0.98$), the angle between perpendicular to the pair plane and the magnetic field ($|\psi|<0.275$).

\begin{figure*}[b]
\centering
\includegraphics[width=0.49\textwidth]{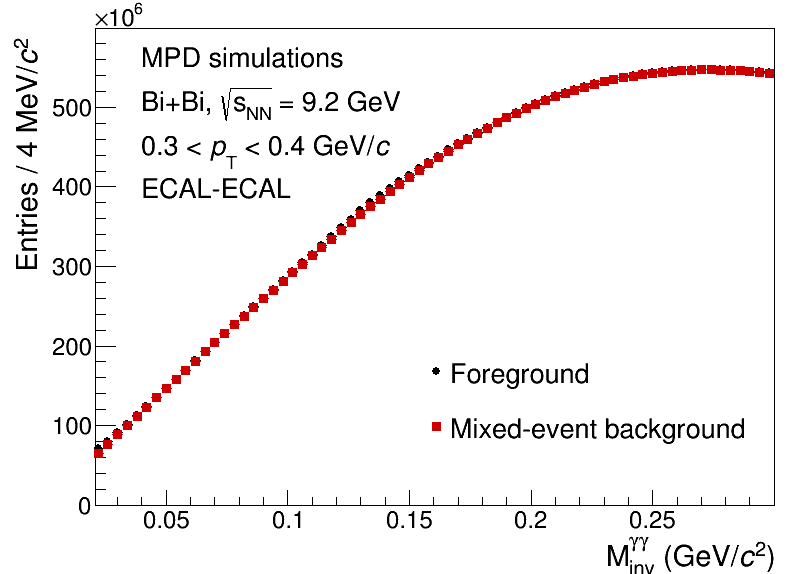}
\hfill
\includegraphics[width=0.49\textwidth]{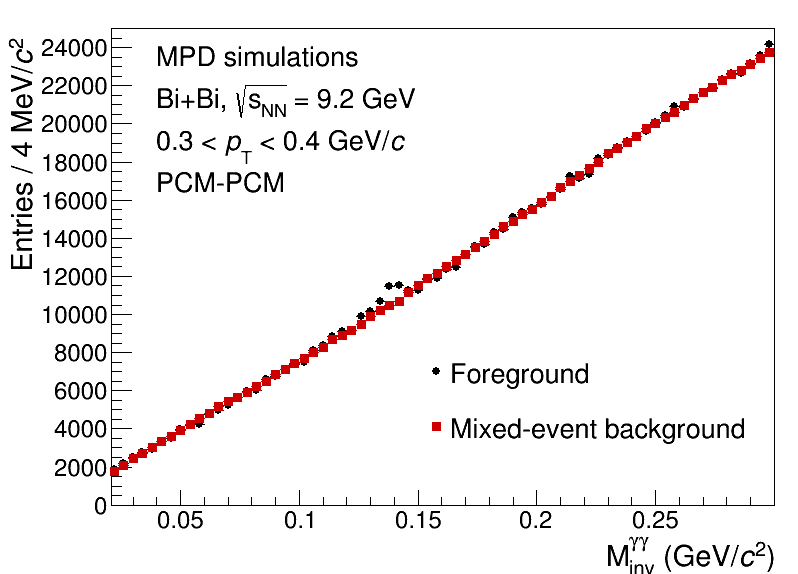}
\includegraphics[width=0.49\textwidth]{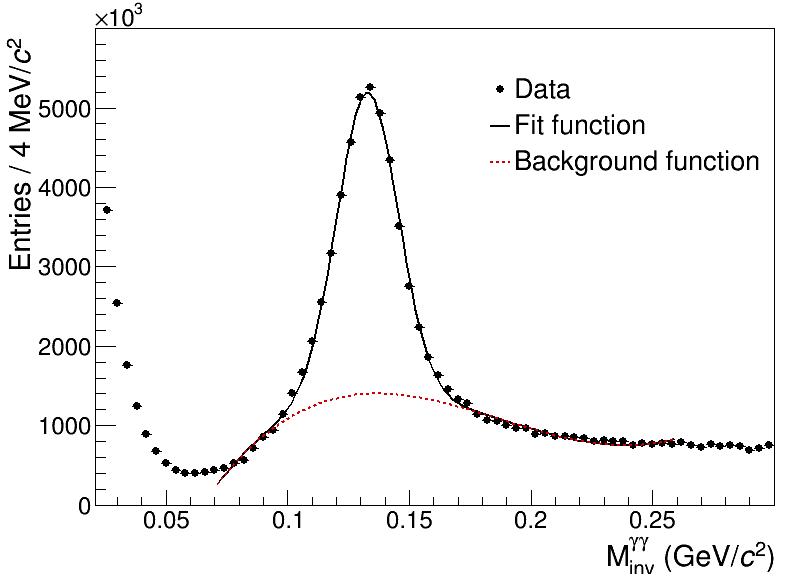}
\hfill
\includegraphics[width=0.49\textwidth]{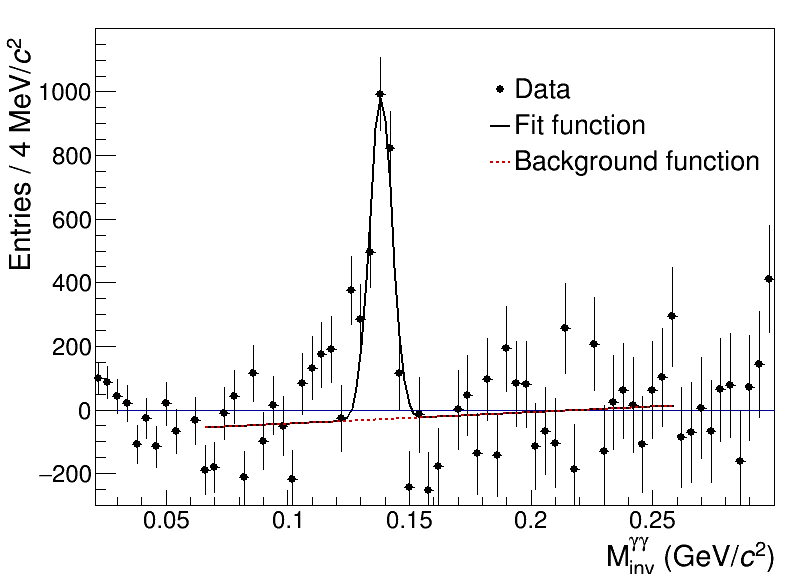}
\caption{Invariant mass distributions for $\gamma\gamma$ pairs before (top) and after (bottom) subtraction of the mixed-event background. The plots on the left and right are for \ECAL-\ECAL and PCM-PCM combinations, respectively. Examples are shown for minimum bias \BiBi collisions at $\sqrt{s_{\rm NN}}=9.2$ GeV. Solid and dashed red curves represent fits to the function described in the text.\label{fig:pi0eta_minv}}
\end{figure*} 

A comparison of the photon reconstruction efficiency for the two methods, as a function of transverse momentum and rapidity is shown in Figure\ \ref{fig:GammaEffPt}. The photon reconstruction efficiency in the \ECAL is close to unity at sufficiently large $\pT$ and decreases to $\sim70$\% at $\pT=0.1$ \GeVc. At $\pT\sim 1$ \GeVc the reconstruction efficiency even exceeds unity due to the finite energy resolution and the shape of the inclusive photon spectrum. The efficiency of PCM method is approximately 100 times smaller (take note of the scale factor for the PCM case) due to the small conversion probability up to the middle of \TPC and relatively strict selection criteria. With a primary vertex selection within $|\zvtx|<50$ cm used in this analysis, the \ECAL allows to reconstruct photons within rapidity $|y|<1$ with almost constant efficiency and up to $|y|<1.3$ with reduced efficiency. The efficiency of the PCM method shows some rapidity dependence due to \TPC acceptance and allows for a photon reconstruction within $|y|<1$.

\subsubsection{Differential \pT spectra for $\pi^{0}$ and $\eta$ mesons}

Spectra of neutral $\pi^0$, $\eta$ and other mesons can be measured with high precision via their two-photon decay channels.

%=======================================================

Neutral meson spectra help to test establishing of the thermal and chemical equilibrium in the hot fireball, its radial collective expansion and other general properties of the system. In addition, combining neutral mesons with charged tracks provides a way to reconstruct short-lived hadronic resonances and to study strangeness production. Furthermore, increased fluctuations of the relative yield of neutral and charged mesons may indicate the presence of a pion Bose-Einstein condensate \cite{Begun:2006gj}, or of the Critical End-Point \cite{Stephanov:2001zj}. 
%The femtoscopy of neutral mesons on one hand opens possibility to measure space-time dimensions of hot matter in complimentary to the charged pion interferometry manner, and on the other hand -- to test another isospin state in the final state hadron interactions \cite{final-state-interactions-HBT}.

Production of $\pi^{0}$ and $\eta$ mesons was measured in the $\pi^{0}(\eta) \rightarrow \gamma + \gamma$ decay channel at mid-rapidity $|y| < 0.5$ in \BiBi collisions at $\sqrt{s_{\rm NN}}=9.2$ GeV using the data of mass production 1 from Table~\ref{EvGens}. The main detector subsystems used in this analysis are the \ECAL, the \TPC and the \TOF detectors. Only events with a reconstructed vertex lying within |$\zvtx$| < 100 cm and centrality in a range 0-90$\%$ were accepted. The number of analyzed minimum bias events is equal to about 4 $\times$ $10^{7}$ collisions. The available statistics is sufficient only to measure the centrality-dependent production of $\pi^{0}$ mesons in fine momentum bins and to estimate the $\eta$ meson production in minimum bias collisions.

\begin{figure*}[t]
\centering
\includegraphics[width=0.49\textwidth]{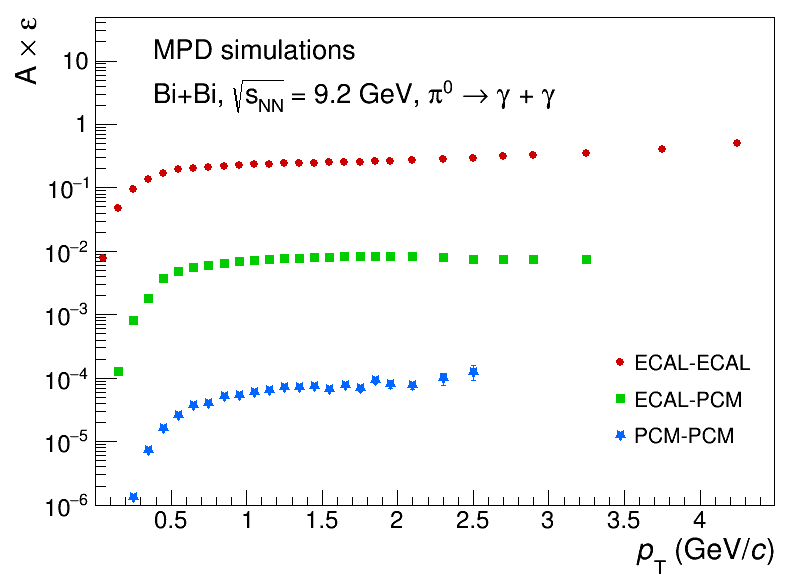}
\hfill
\includegraphics[width=0.49\textwidth]{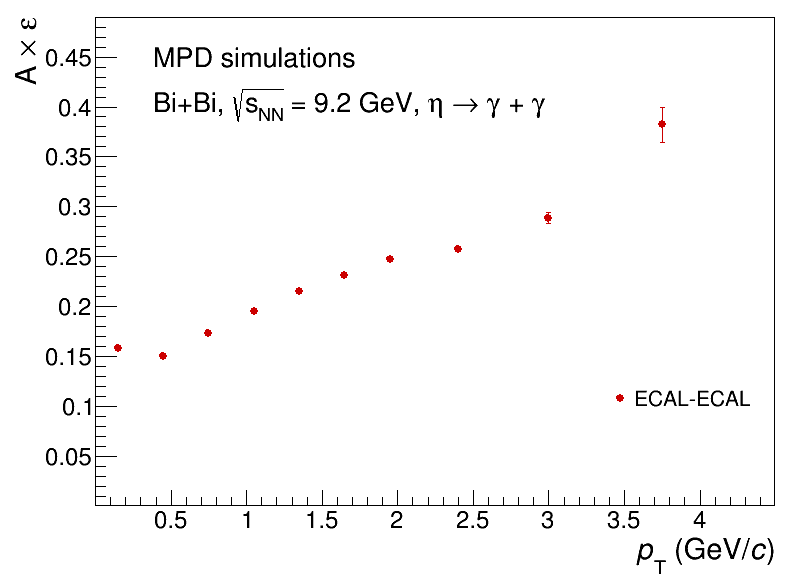}
\caption{Reconstruction efficiency $A\times\varepsilon$ evaluated for $\pi^{0}$ and $\eta$  mesons in the $\pi^{0}(\eta) \rightarrow \gamma + \gamma$ decay channel in \BiBi collisions at $\sqrt{s_{\rm NN}}=9.2$ GeV. \label{fig:pi0eta_receff}}
\end{figure*}

\begin{figure*}[b]
\centering
\includegraphics[width=0.49\textwidth]{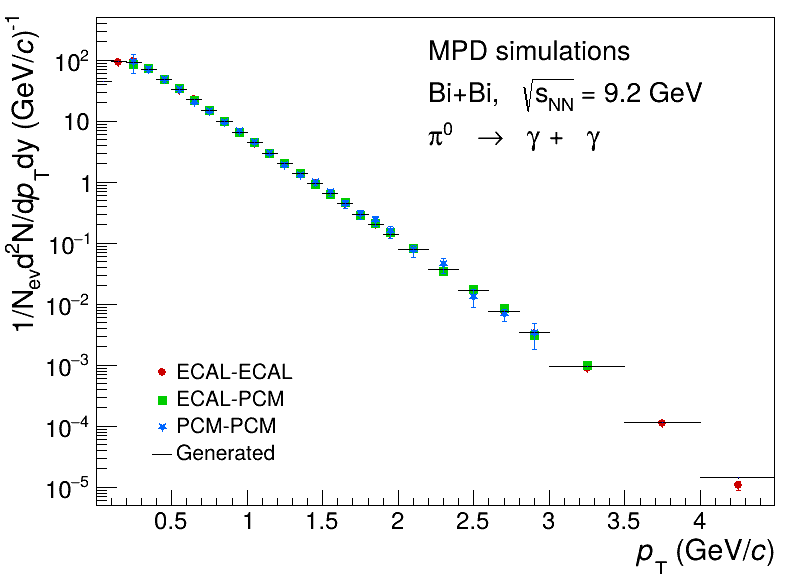}
\hfill
\includegraphics[width=0.49\textwidth]{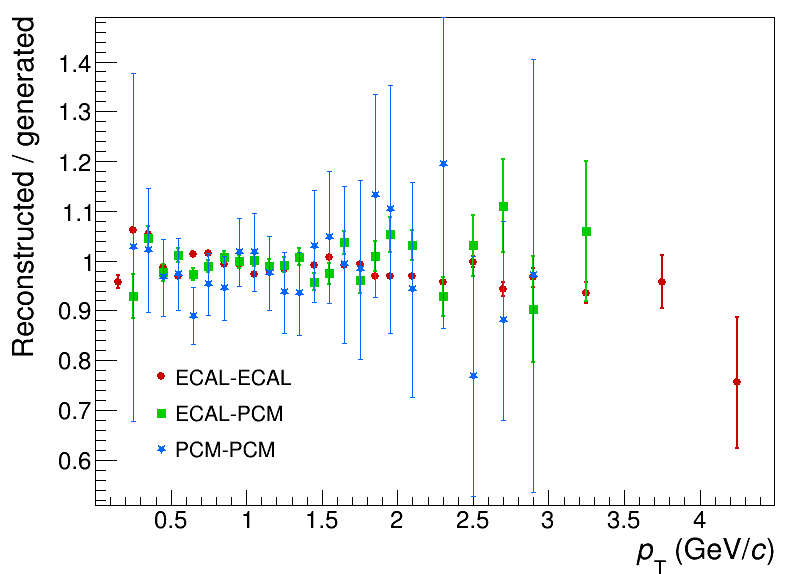}
\caption{Differential production spectra (left) and their ratio to the truly generated one (right) for $\pi^{0}$ mesons in minimum bias \BiBi collisions at $\sqrt{s_{\rm NN}}=9.2$ GeV. Results are shown for different photon selections: \ECAL-\ECAL, \ECAL-PCM and PCM-PCM, see text for details. \label{fig:pi0_diffmeth}}
\end{figure*} 

The two approaches described above were used for the reconstruction of photons:  photon measurements in the \ECAL or photon conversion method. Clusters reconstructed in the calorimeter were selected as photon candidates if they satisfied minimum selections: $E_{\gamma}$ > 0.075 GeV, the number of towers in the cluster is larger than one, the shower shape is consistent with the shape expected for electromagnetic signal, $\chi^{2}$/NDF < 4, the time-of-flight is less than 2 ns. Photon conversion pairs were selected as  described in the previous section.
%Electron tracks were reconstructed in the \TPC with the requirement that the track has more than 10 hits out of maximum 53 and momentum $\pT$ > 0.05 \GeVc. To be identified as electrons or positrons, reconstructed tracks need to have a specific ionization energy loss $\langle \dEdx \rangle$ measured in the \TPC within 2$\sigma_{\TPC}^{e}(\pT)$ of the expected value. Particles with a signal in the TOF subsystem are additionally identified by requiring the time of flight to be within 2$\sigma_{TOF}^{e}(\pT)$ of the expected value. 
%The $\sigma_{\TPC}$ is about 8.5$\%$, while the typical value of $\sigma_{TOF}$ is about 85 ps for central \BiBi collisions. 
%Two oppositely charged tracks are combined in pairs, which should additionally pass selections on the quality of the secondary vertex reconstruction, maximum distance between the $e^{+}$ and $e^{-}$ tracks in the secondary vertex, pointing angle, which is an angle between the line connecting the primary and secondary vertices and the momentum vector of the pair, invariant mass of the pair and the minimum distance to the secondary vertex. Each selection was parameterized as a function of pair momentum to select > 95$\%$ of pairs from true photon conversions. With multiple selections, the purity of the reconstructed photons with PCM method exceeds 95$\%$ in most central collisions.

%
%

\begin{figure*}[b]
\centering
\includegraphics[width=0.465\textwidth]{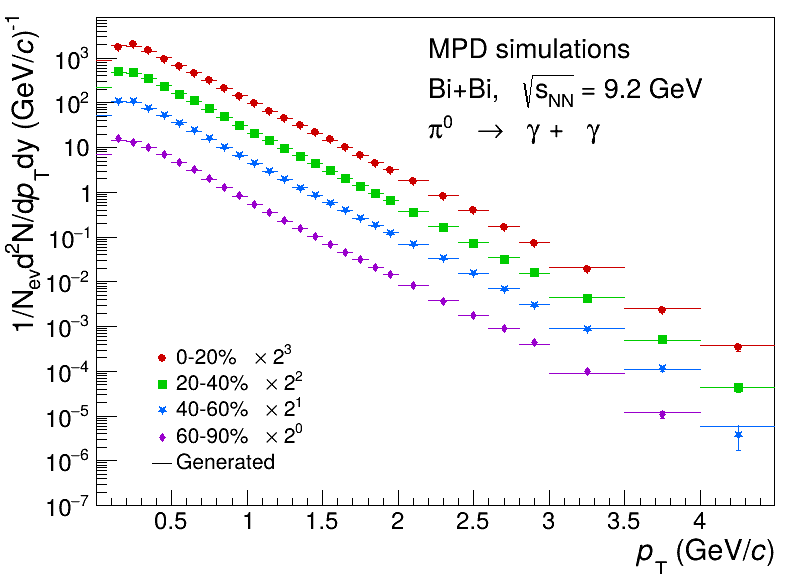}
\hfill
\includegraphics[width=0.465\textwidth]{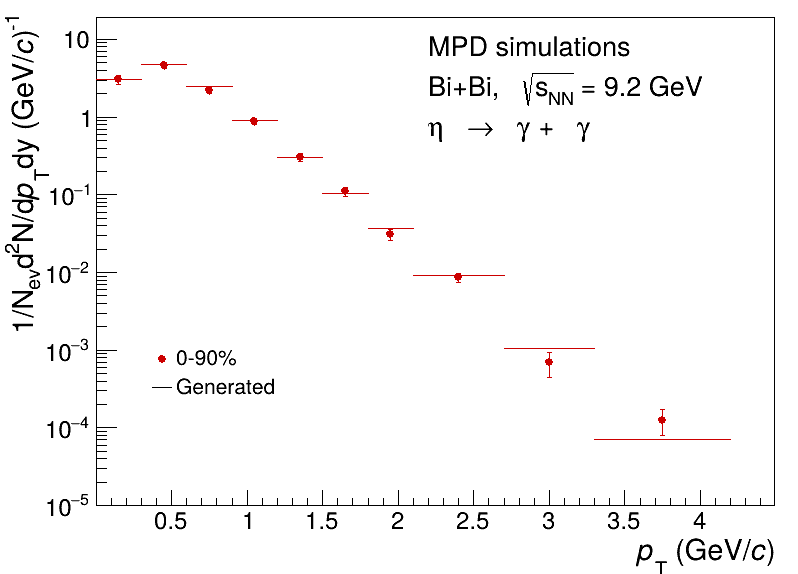}
\caption{Differential production spectra for $\pi^{0}$ and $\eta$  mesons in \BiBi collisions at $\sqrt{s_{\rm NN}}=9.2$ GeV. Results for $\pi^{0}$ meson are shown in different centrality intervals. The measured points are compared to the true ones shown with histograms. \label{fig:pi0eta_spectra}}
\end{figure*}

Yields of $\pi^{0}$ and $\eta$ mesons for each $\pT$ and centrality interval are measured by calculating the invariant mass distributions of photon pairs at midrapidity $|y| < 0.5$ in different combinations: \ECAL-\ECAL, \ECAL-PCM, PCM-PCM. The combinatorial background is estimated using a mixed-event method, when one of the photons is taken from the current event and the second is taken from another event with similar topology (the difference in $\zvtx$ and event centrality does not exceed 20 cm and 10$\%$, respectively). The mixed-event invariant distributions are scaled to the same event distributions at high masses where the contribution of correlated pairs to be minimum. Examples of invariant mass distributions before and after subtraction of the mixed-event background are shown in Figure~\ref{fig:pi0eta_minv}. After subtraction, the resulting distributions contain the remaining correlated background from mini-jets and pairs from misreconstructed hadronic decays, which have a smooth dependence on the mass.  The remaining background is parametrized with a polynomial, while contributions from decays of neutral mesons are described with a Gaussian function. Parameters of the Gaussian and polynomial functions are kept free in fits to the invariant mass distributions. The extracted values of mass and width for $\pi^{0}$ and $\eta$ mesons are found to be consistent with the expected values within uncertainties. Examples of the fits are presented in the same figure. Meson yields are estimated either as integrals of Gaussian functions or by bin counting in the mass range $|m - M_\mathrm{rec}| < 3 \sigma_\mathrm{rec}$ followed by subtraction of the polynomial integral in the same range. The values of $M_\mathrm{rec}$ and $\sigma_\mathrm{rec}$ are the mass and width of the neutral meson extracted from the fit.

The same data sample was used to evaluate the reconstruction efficiencies for $\pi^{0}$ and $\eta$  mesons in the $\pi^{0}$ or $\eta \rightarrow \gamma + \gamma$ decay channel as well as to estimate the expected masses and widths of the reconstructed signals. For each analyzed $\pT$ and centrality interval, the efficiencies  $A\times\varepsilon$ are calculated as the ratio $N_\mathrm{rec}/N_\mathrm{gen}$, where $N_\mathrm{rec}$ is the number of reconstructed particles in the $\gamma + \gamma$ channel after all event and track selection cuts and $N_\mathrm{gen}$ is the number of generated mesons within $|y| < 0.5$ decaying in the $\pi^{0}(\eta) \rightarrow \gamma + \gamma$ channel. Examples of efficiencies evaluated for $\pi^{0}$ and $\eta$ mesons for minimum bias \BiBi collisions as a function of transverse momentum are shown in Figure\ \ref{fig:pi0eta_receff}. The difference at low-$\pT$ between the efficiencies for $\pi^{0}$ and $\eta$ mesons, reconstructed using the same photon selections, is due to the different masses of the particles and hence mean energies of decay photons at the same $\pT$ of parent mesons.  Quite a big difference is observed for $\pi^{0}$ reconstruction efficiencies with different methods explained by the rather small probability of photon conversion in the detector materials with a total radiation length of $X/X_{0} \sim 4.5\%$. The evaluated efficiencies show rather modest dependence on event centrality.

Fully corrected yields evaluated according to Eq.~(\ref{eq:CorrYield}) for $\pi^{0}$ meson in minimum bias \BiBi collisions at $\snn=9.2$ GeV with three different reconstruction methods are shown in Figure~\ref{fig:pi0_diffmeth}. The spectra agree with each other and with the truly generated one within uncertainties. The momentum coverage for the measured spectra is comparable. Figures~\ref{fig:pi0eta_minv} and \ref{fig:pi0eta_receff} clearly demonstrate the difference between the methods. The \ECAL-\ECAL method has the highest efficiency, but measurements at low momenta are characterized by a rather poor energy resolution and a significant hadronic and combinatorial background. In contrast, the PCM-PCM approach takes advantage of the much better energy resolution of the tracking system and the superior purity of photon reconstruction at low momenta, resulting in much narrower reconstructed peaks and lower background. However, the method suffers from low efficiency due to small photon conversion probability. The hybrid \ECAL-PCM method occupies an intermediate position, sharing the advantages and disadvantages of the above two methods. Measurements with the \ECAL-\ECAL and \ECAL-PCM methods allow us to study the dependence of the $\pi^{0}$ production on centrality. The statistics of the PCM-PCM method does not allow such a detailed study with the available dataset. Measurements with the \ECAL-\ECAL have smaller statistical uncertainty and are used hereafter by default. Nevertheless, measurements with \ECAL-PCM and PCM-PCM are important, especially at low momentum, to study the performance and systematic effects in the calorimeter. The available statistics is sufficient to measure only the centrality-integrated $\eta$ meson production using the \ECAL-\ECAL method.

The differential yields measured for $\pi^{0}$ and $\eta$ mesons as a function of transverse momentum in centrality differential \BiBi collisions at $\sqrt{s_{\rm NN}}=9.2$ GeV are shown in Figure~\ref{fig:pi0eta_spectra}. The measurements span a wide $\pT$ range from 0.1 to 4.5 \GeVc with the accumulated statistics. The reconstructed spectra are compared to truly generated ones shown with histograms. Reconstructed spectra match the generated ones within statistical uncertainties. Additional photon selections, such as a cluster neutrality and/or a higher minimum energy of clusters with $E_{\gamma}$ > 0.2 GeV were optionally used  to further suppress the hadronic background and optimize the reconstructed peak shapes.

The fully corrected spectra obtained using different selections were compared and found to agree within 5-10$\%$, with a tendency for a larger discrepancy at lower momenta. Since statistical uncertainties in such comparisons are highly correlated, the observed discrepancies serve as a rough estimate for the signal extraction systematic uncertainty.

\subsubsection{Collective flow of inclusive photons and neutral mesons}

\begin{figure*}[t]
\includegraphics[width=0.49\textwidth]{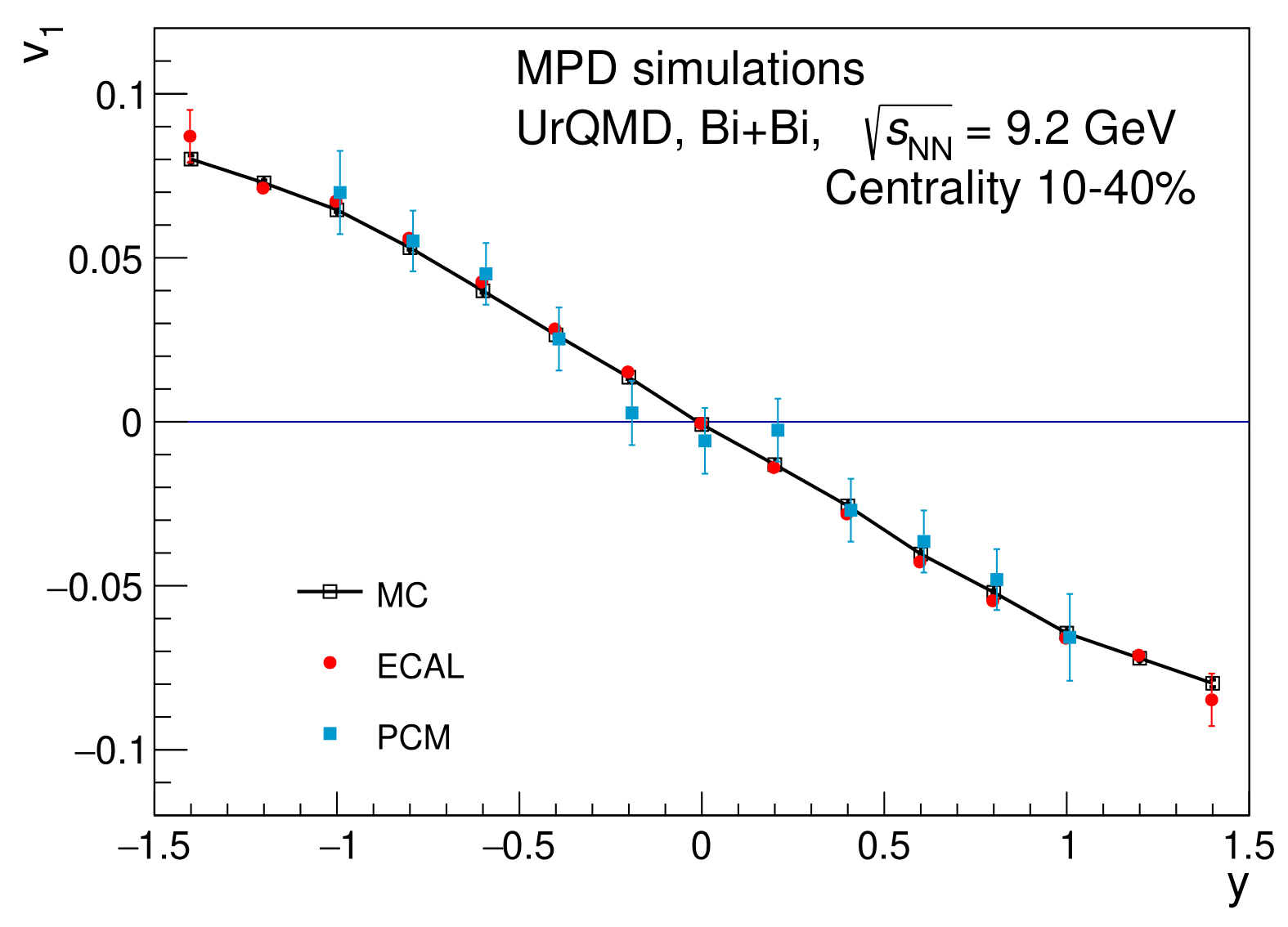}
\hfill
\includegraphics[width=0.49\textwidth]{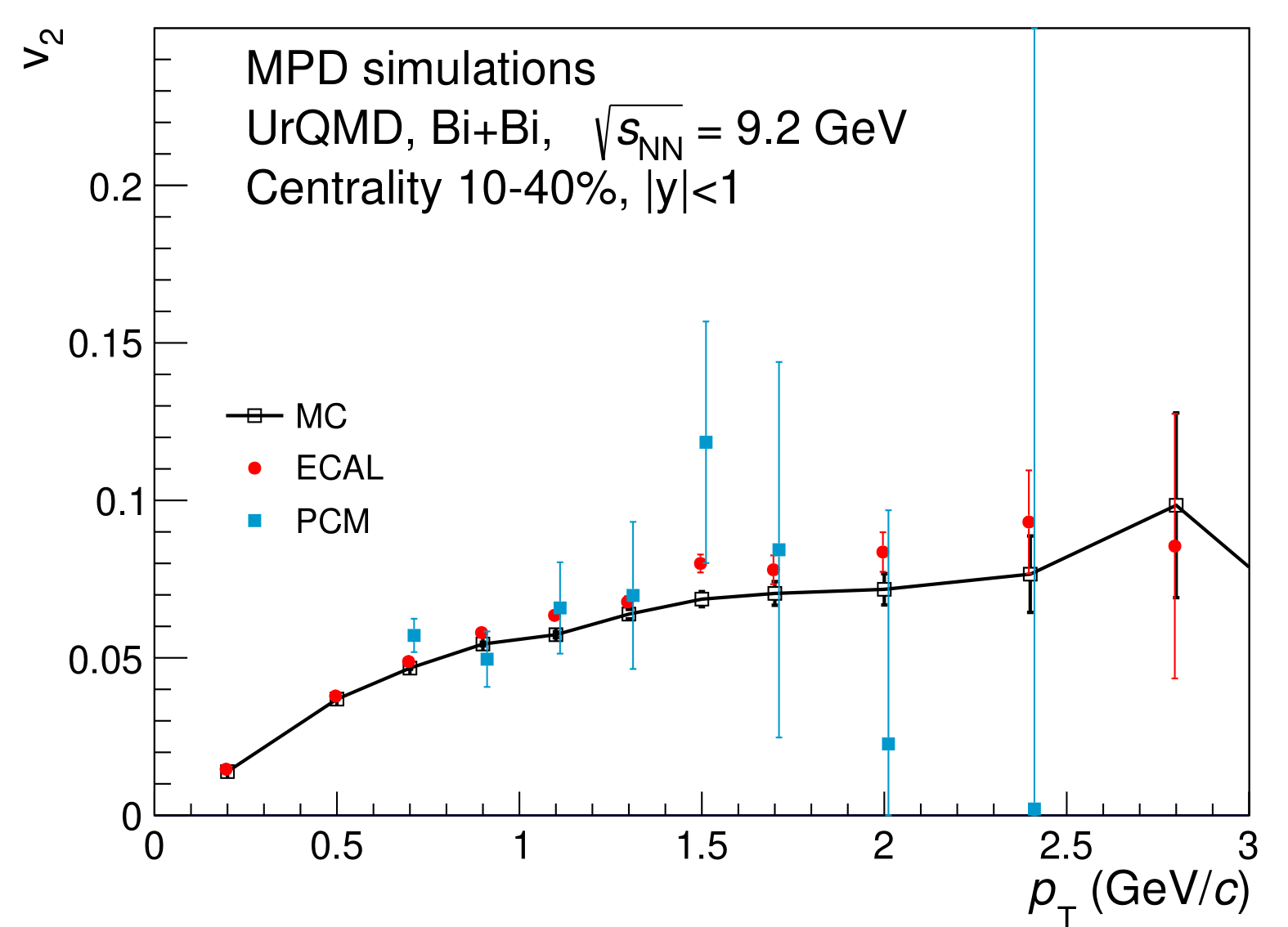}
\caption{Left: inclusive photon directed collective flow vs. rapidity. Right: Inclusive photon elliptic collective flow vs. \pT.
\label{fig:gammaV1Y}
}
\end{figure*}

\begin{figure*}[b!]
\includegraphics[width=0.49\textwidth]{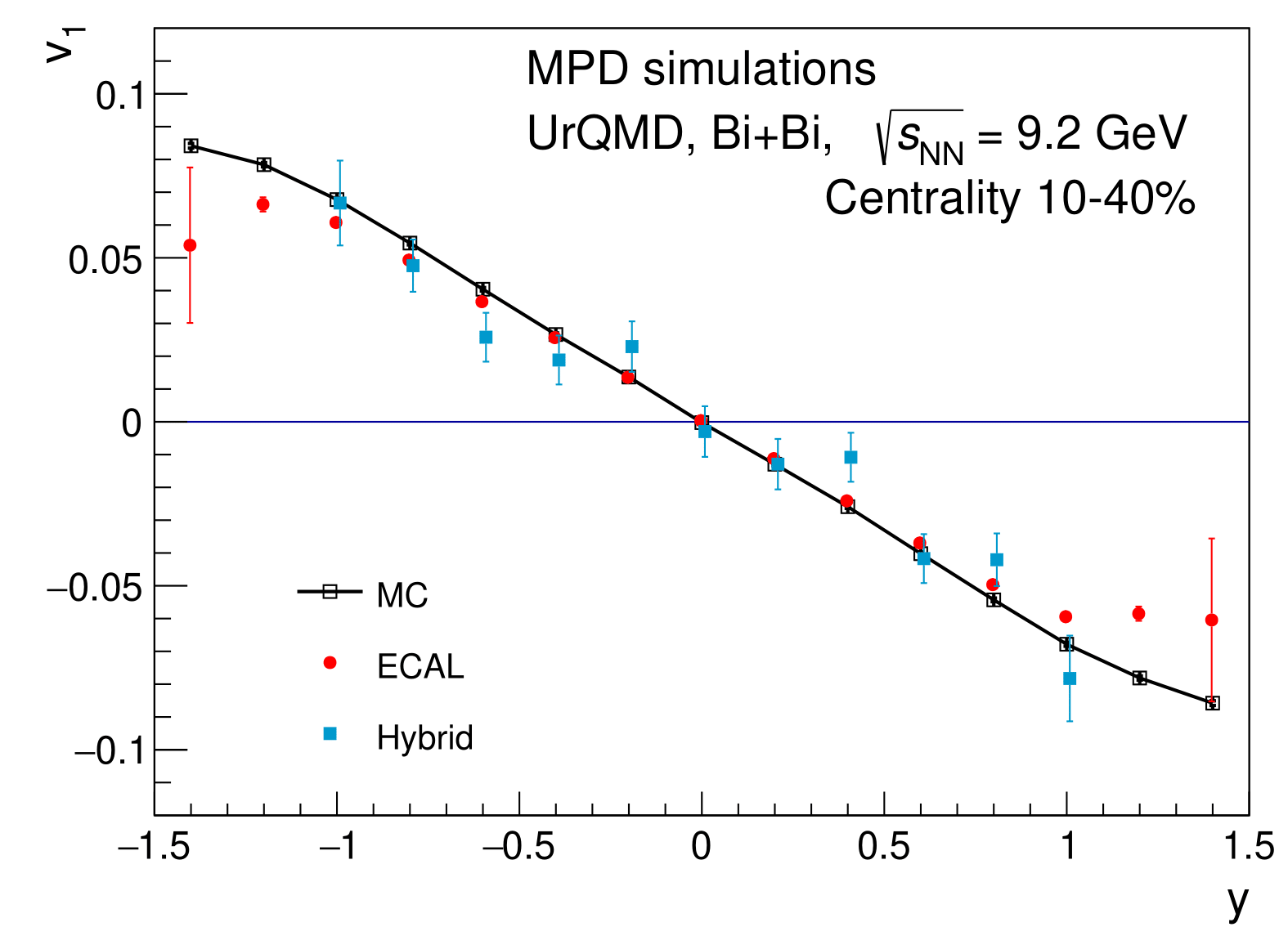}
\hfill
\includegraphics[width=0.49\textwidth]{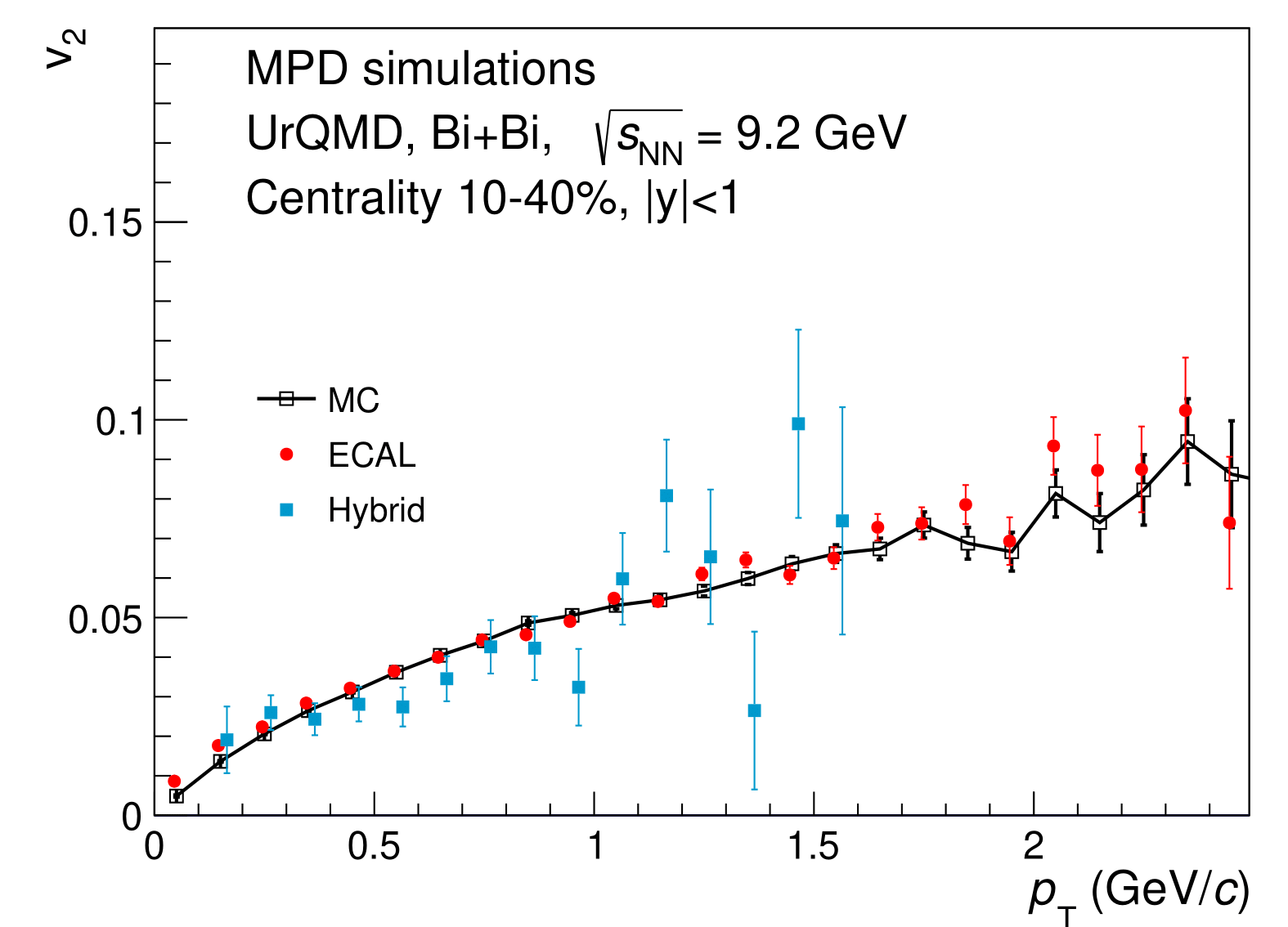}
\caption{Left: Neutral pion directed collective flow vs. rapidity. Right: Neutral pion elliptic collective flow vs. $\pT$.
\label{fig:pi0V1Y}}
\end{figure*}

The measurement of the collective flow of inclusive photons is a necessary ingredient for the extraction of the direct photon flow. The latter is measured as a difference of the inclusive photon flow $v_{2}^{\gamma. \mathrm{incl}}$ and flow of decay photons $v_{2}^{\gamma. \mathrm{dec}}$, estimated from the neutral meson flow
\begin{equation}
  v_\mathrm{n}^{\gamma. \mathrm{dir}} =  \frac{ v_\mathrm{n}^{\gamma. \mathrm{incl}}R_\gamma -  v_\mathrm{n}^{\gamma. \mathrm{dec}}}{R_\gamma-1},\quad R_\gamma = \frac{N^{\gamma, \mathrm{incl}}}{N^{\gamma, \mathrm{dec}}}.
\end{equation}

%The spectra in the measurement ranges sample $\sim 90\%$ of the total particle yields with the remaining $\sim 10\%$ lying outside of the measurement range at lower and higher momenta. The unmeasured yields can be estimated by extrapolating the measured spectra with different functions such as Levy, Boltzmann Gibbs blast-wave, $m_{T}$-exponential and power-law functions. Variation of the results will provide an estimate of the corresponding systematic uncertainty for integrated yield and mean transverse momentum from the extrapolation. 

We compare the reconstructed  directed and elliptic flow of inclusive photons with the truly generated signals in Figure~\ref{fig:gammaV1Y}, left plot. The inclusive photon directed flow $v_1$, integrated over $\pT$, measured with the \ECAL, reproduces the inclusive photon flow calculated at the generator level in the range $|y|<1.5$. The PCM method also reproduces the generated flow, though with larger uncertainties within $|y|<1$.

The dependence of the elliptic collective flow $v_2$ of inclusive photons on the transverse momentum is presented in Figure~\ref{fig:gammaV1Y}, right plot. The simulation was performed using approximately 1 million Minimum Bias events after event selection. With available statistics, one can measure the elliptic flow of inclusive photons with reasonable accuracy up to $\pT\sim2.5$ \GeVc with the \ECAL and up to $\pT\sim 1$ \GeVc with the PCM method.

In Figure~\ref{fig:pi0V1Y}, we present a comparison of the neutral pion directed flow as a function of rapidity and the  elliptic flow $v_2$ as a function of $\pT$.
All three methods can potentially be used to extract the neutral pion flow. However, the PCM method lacks statistics and  does not produce any reasonable result  at this point. We  found that both the \ECAL and the hybrid methods produce consistent results and reproduce the flow of primary generated neutral pions shown with the MC curve. Similar to inclusive photons, the collective flow can be measured up to $|y|<1.5$ in rapidity and analyzed statistics of 1 million Minimum Bias events after event selection allows the reconstruction of $v_2$ up to $\pT\sim2.5$ \GeVc.

\subsubsection{ Dielectrons }
%Electron pairs, $e^{+}e^{-}$, and dileptons in general  are a powerful tool for characterizing the strongly interacting matter produced in relativistic heavy-ion collisions. They are sensitive signals to search for the conjectured critical point and first order phase transitiion advocated by some models and expected to occur within the energy range of $\snn=$ 2.4-11 GeV covered by the NICA facility, (4-11 GeV in the collider mode and 2.4-3.5 in the fixed-target mode of operation). Dileptons are also well suited to search for the onset of the deconfinement and chiral symmetry phase transitions expected to occur within the same energy range. However, the measurement of electron pairs remains challenging due to the large combinatorial background (B) primarily arising from $\pi^{0}$-Dalitz decays and photon conversions in the detector material and the large physics background arising from the known hadronic source decays, the so-called cocktail. Minimizing the B together with a precise knowledge of the cocktail are necessary to identify the thermal source of interest.

Dielectrons ($e^+e^-$ pairs) open another set of possibilities in exploring the properties of hot matter.
As they add another variable -- mass of the virtual photon -- they provide the possibility to measure the temperature of hot matter without blue
\begin{table}[H]
\centering
\caption{Selection cuts for electron track reconstruction and eID. The signals from \TPC, TOF and \ECAL, i.e. $\langle \dEdx \rangle$ in \TPC, time-of-flight in TOF, and $E/p$ and time-of-flight in \ECAL, respectively, are expressed in units of standard deviations from the signals expected for true electrons. Similar expressions are used for \TOF and \ECAL matching variables, d$\phi$ and d$z$.}
\label{tabI}
%\label{tabII}
\vglue4mm
%\begin{tabular}{lclc} \hline
\begin{tabular}{cccccccc} \hline
Variable & Cut \\ \hline \hline
%$|\eta|$ & $|\eta|$ $<$ 0.7 $\&$ 0.7 $<$ $|\eta|$ $<$ 1.0 \\
${n_{\rm \TPC}^{\rm hits}}$ & 39 \\
DCA & $<$ 3$\sigma$ \\
\TPC \dEdx &      $n_{\sigma, e} < 2 \sigma$,\quad\quad $\pT <0.8$ \GeVc\\
  & $-1 < n_{\sigma, e} < 2 \sigma$, $\pT > 0.8$ \GeVc\\
%TPC \dEdX PID & TPC n$\sigma$ $< $ 2$\sigma$ at p = 0\\
% & and -1 to 2$\sigma$ for p $>$ 800 \MeVc   \\
\TPC-\TOF match. & ${n_{\sigma, e}^{\rm d\phi}}$ and ${n_{\sigma, e}^{\rm dz}}$ $<$ 2$\sigma$  \\
TOF eID & $|{n_{\sigma, e}^{\rm ToF}}|$ $<$ 2$\sigma$  \\
\TPC-\ECAL match. &  ${n_{\sigma, e}^{\rm d\phi}}$ and ${n_{\sigma, e}^{\rm dz}}$ $<$ 3$\sigma$ \\
\ECAL $E/p$ eID &      ${n_{\sigma, e}^{\rm E/p}}$ $<$ 2$\sigma$  \\
\ECAL $m^{2}$ eID &  $|{n_{\sigma, e}^{\rm ToF}}|$ $<$ 1.5$\sigma$ \\
\hline
\end{tabular}
\end{table}
\noindent shift which appears due to the radial expansion of the fireball in case of real photons. One can expect that, at NICA energies, the heavy flavor decay contribution will be negligible and thermal virtual photon emission will be the dominant source in the intermediate mass region $1<M_\mathrm{ee}<3$~\GeVcsq. 
This will provide access to the temperature of the hot source. Thermal photon emission will also appear in the low mass region, $M_\mathrm{ee}<0.5$~\GeVcsq, where one can relate virtual and real photon yields with the Kroll-Wada formula~\cite{Kroll:1955zu} and calculate the real direct photon yield. Thermal dilepton emission in the low-mass region, $M_\mathrm{ee}<0.7$~\GeVcsq reflects the temperature of the hadron gas formed in the late stages of the collision and conveys information about the in-medium modification of the $\rho$-meson spectral function.

%===================================================================
%Neutral mesons can be reconstructed via their two-photon decay. There are three options: two photons in \ECAL, two photons in conversion and hybrid method. Example of the $\pi^0$ peak after background subtraction obtained with 3 methods is presented in Figure~\ \ref{fig:pi0Peak}. Reconstruction of two photons in \ECAL provides the largest efficiency but also the worst resolution. Note that width of the peak is defined not only by energy resolution, but also by cluster overlap in central collisions.
%as one can conclude from Figure~\ \ref{fig:pi0Mass}. 

%The PCM method provides the best resolution but low efficiency. There is characteristic left wing of the peak due to electron energy loss. Hybrid method provides intermediate results.

%\begin{figure}[H]
%includegraphics[width=\linewidth]{figures/demo_pi0_cen.pdf}
%\caption{Comparison of peak with and peak position.\label{fig:pi0Peak}}
%\end{figure}

% Comparison of width and position of different methods

% Centrality dependence of peak position and width

%\begin{figure}[H]
%\includegraphics[width=\linewidth]{figures/massPi0.pdf}
%\caption{Comparison of peak with and peak position.}
%\end{figure}

%\begin{figure}[H]
%\includegraphics[width=\linewidth]{figures/effPi0.pdf}
%\caption{Comparison of efficiency vs pT for several rapidity bins.}
%\end{figure}

% \begin{figure}[H]
% \includegraphics[width=\linewidth]{figures/EffPi0Y.pdf}
% \caption{Dependence on rapidity for two pT bins for 3 methods.}
% \end{figure}

% Dynamic range for a given statistics

The MPD performance for the measurement of electrons was studied and optimized using a sample of 15 million minimum bias \BiBi collisions at $\snn=9.2$ \GeV generated in mass production 1 from Table~\ref{EvGens}.  
To improve the statistical significance of the dielectron yield in this relatively
small sample of events, the branching ratios of dielectron sources, namely, $\omega\rightarrow e^{+}e^{-}$, $\omega\rightarrow \pi^{0 }e^{+}e^{-}$, $\rho\rightarrow e^{+}e^{-}$, $\phi\rightarrow \eta e^{+}e^{-}$ and $\phi\rightarrow e^{+}e^{-}$, were enhanced by a factor of 20 in the decay table of the generator. The dilepton mass spectra is later scaled down to retrieve the realistic  dielectron yield from these sources. 
Furthermore, the yields and spectral shapes of the vector mesons $\rho^{0}(770)$, $\omega(782)$ and $\phi(1020)$ generated with 
UrQMD were rescaled to match more realistic predictions of the PHSD event generator. 

The MPD is well suited for such measurements. Accurate tracking is provided by the \TPC and electron identification, together with hadron rejection, are achieved by the combined effect of the measurements of the average specific energy loss \dEdx of the track while traversing the \TPC gas, the particle time-of-flight in the \TOF and \ECAL detectors, and the particle energy in the \ECAL. The latter contributes to the electron identification and hadron rejection by requiring the particle $E/p$ ratio to be unity. 

\begin{figure}[H]
\includegraphics[width=\linewidth]{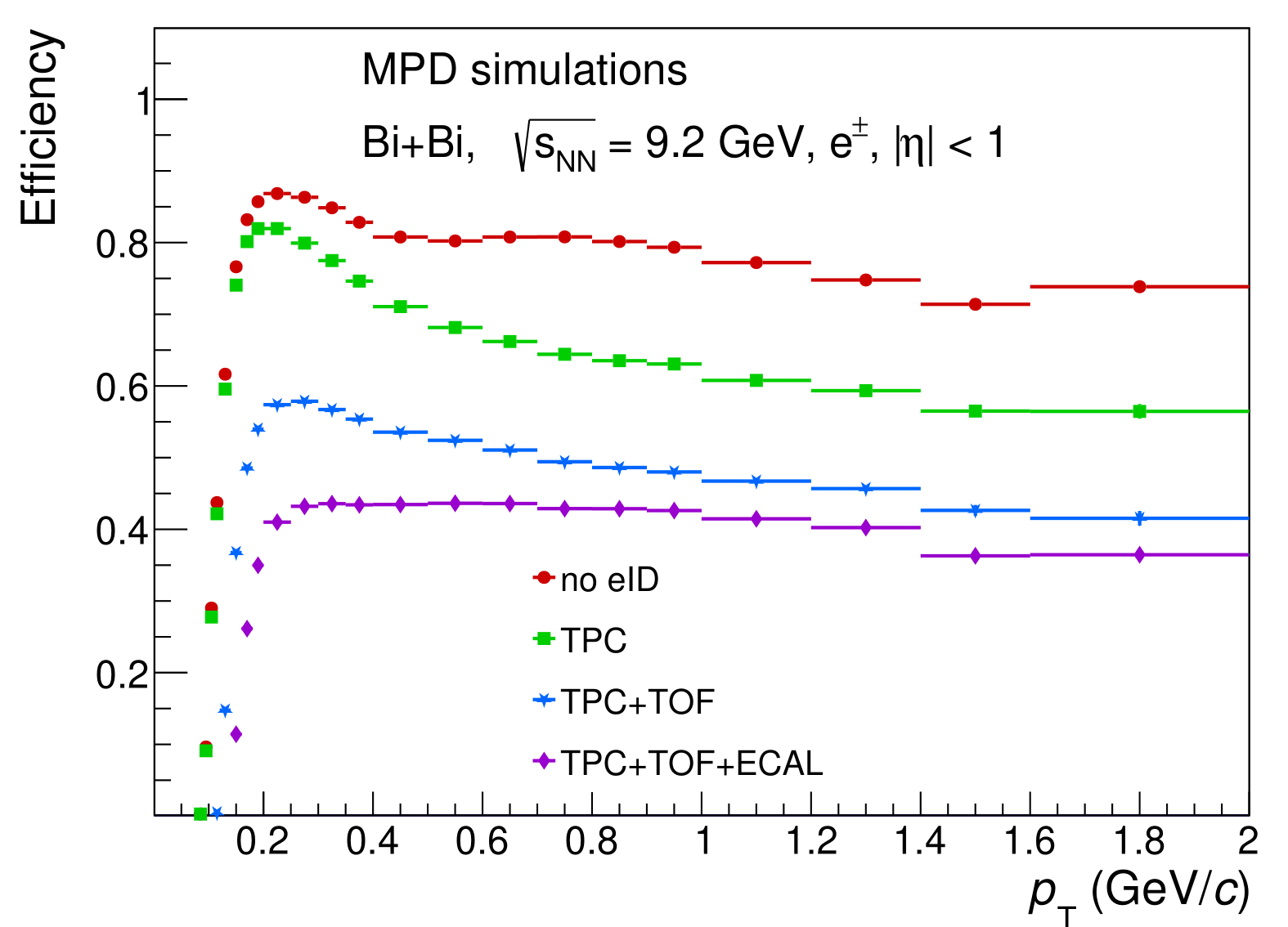}
\includegraphics[width=\linewidth]{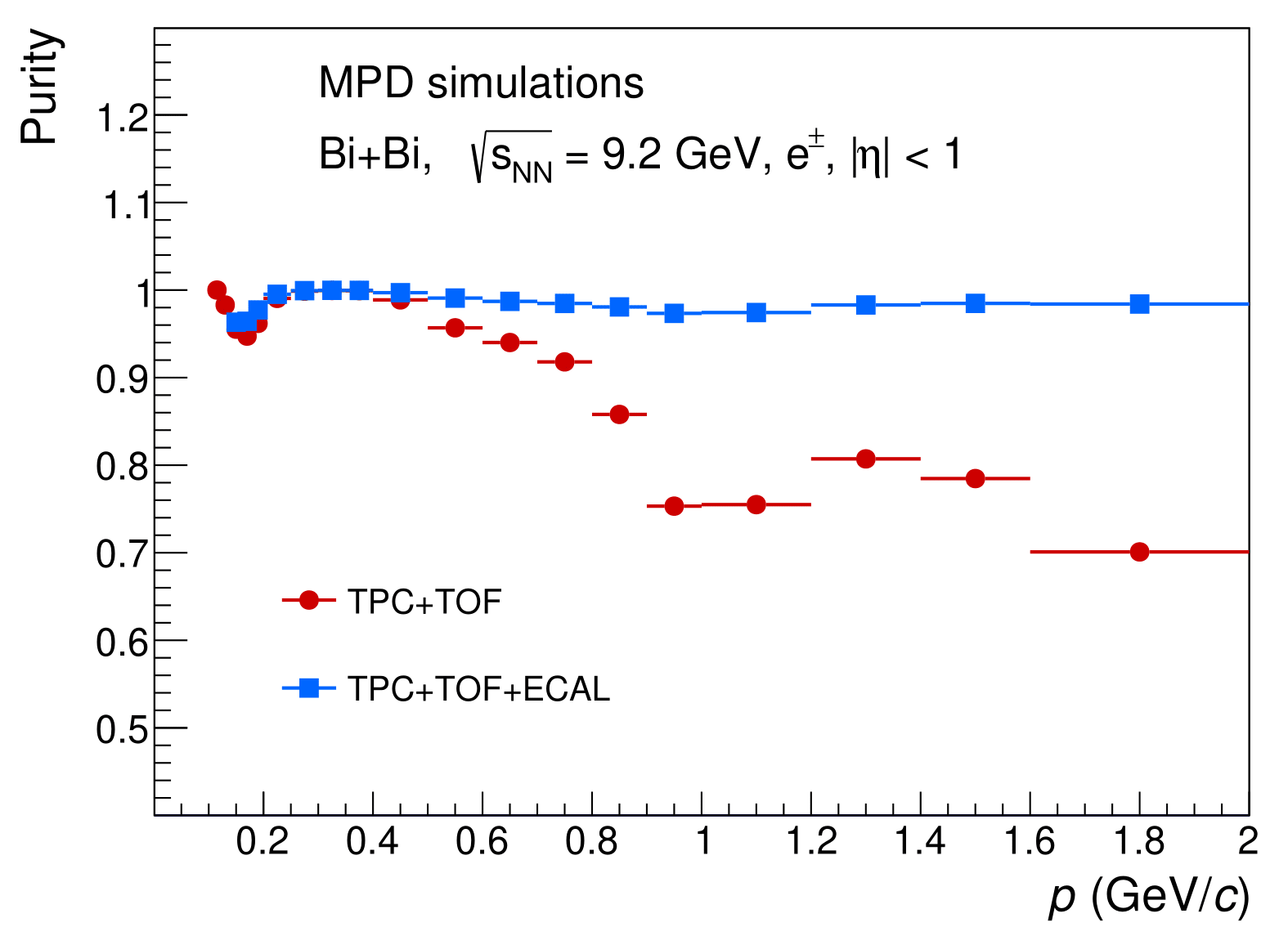}
\caption{Electron track reconstruction and eID efficiency using different detector subsystems as a function of transverse momentum (upper panel) and electron purity (lower panel) achieved with and without \ECAL for eID as a function of total momentum in \BiBi collisions at $\snn=$ 9.2 GeV.\label{fig_eff_purity}}
\end{figure}

Tracks from events having a primary vertex reconstructed within |$\zvtx| <$ 130 cm are reconstructed in the \TPC within the pseudorapidity interval $|\eta| < 1.0$, requiring at least 39 hits out of a maximum of 53 hits, and identified using a momentum-dependent cut on the truncated specific energy loss $\langle \dEdx \rangle $ signal. The tracks are then extrapolated to the
vertex region and a 3$\sigma$ cut is applied on the distance-of-closest-approach (DCA) to the primary vertex. This cut removes nearly 98 $\%$ of the contributions from conversions occurring in the detector material behind the beam pipe. Finally the tracks are extrapolated to the \TOF and \ECAL detectors and matched to hits in these detectors within 2 or 3 $\sigma$ of the extrapolation point both in $z$ and $\phi$ directions.  The time-of-flight measurement of the track is primarily provided by the \TOF detector. The \ECAL also provides a measurement of the track time-of-flight. It has a worse time resolution of 250 ps at high energy, but the measurement is nonetheless useful as it provides electron identification (eID) for those tracks that fall within the inactive area between the modules of the TOF detector. The \ECAL's main benefit is the measurement
\begin{figure}[H]
\begin{center}
\includegraphics[width=\linewidth]{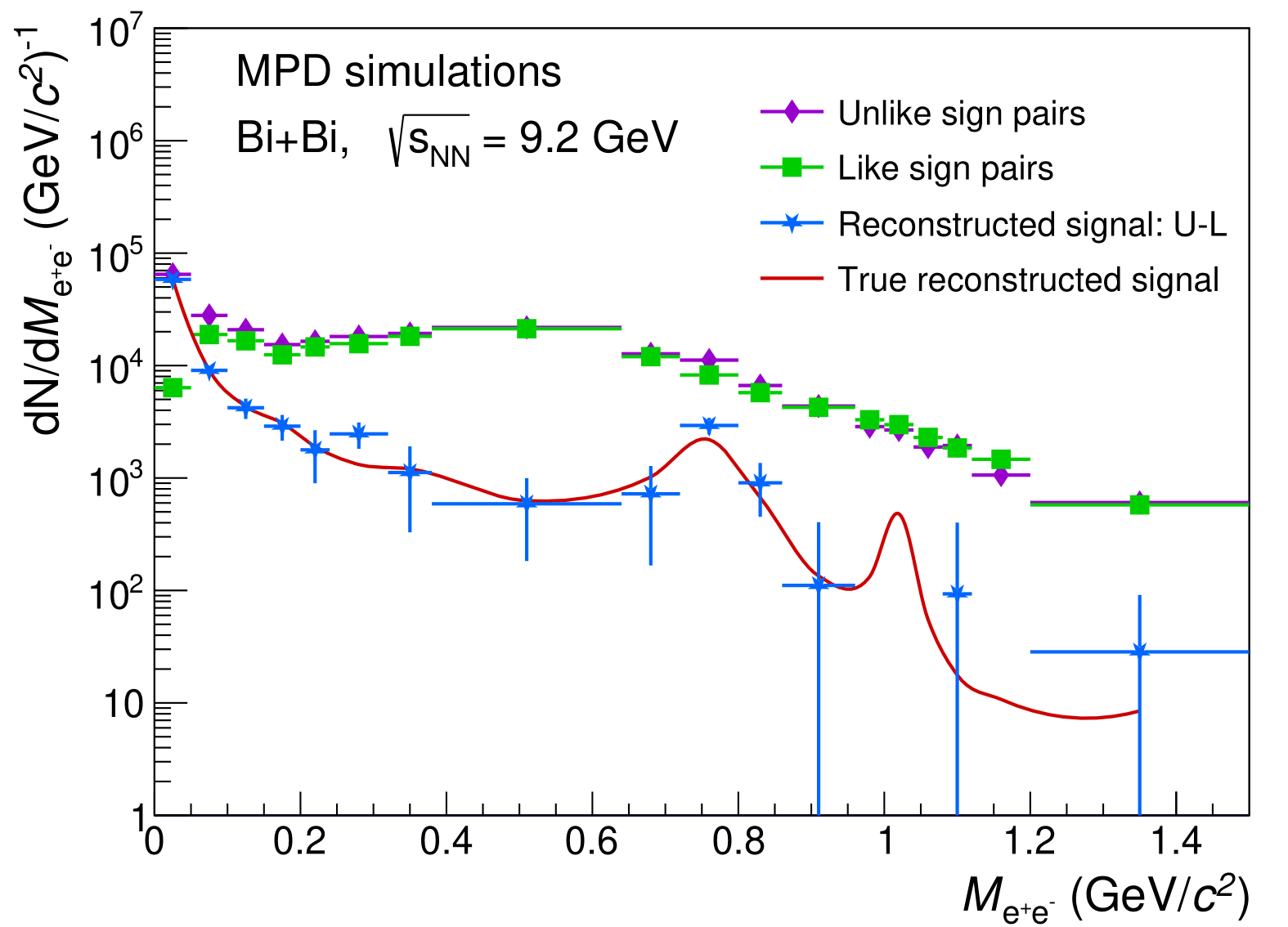}
\includegraphics[width=\linewidth]{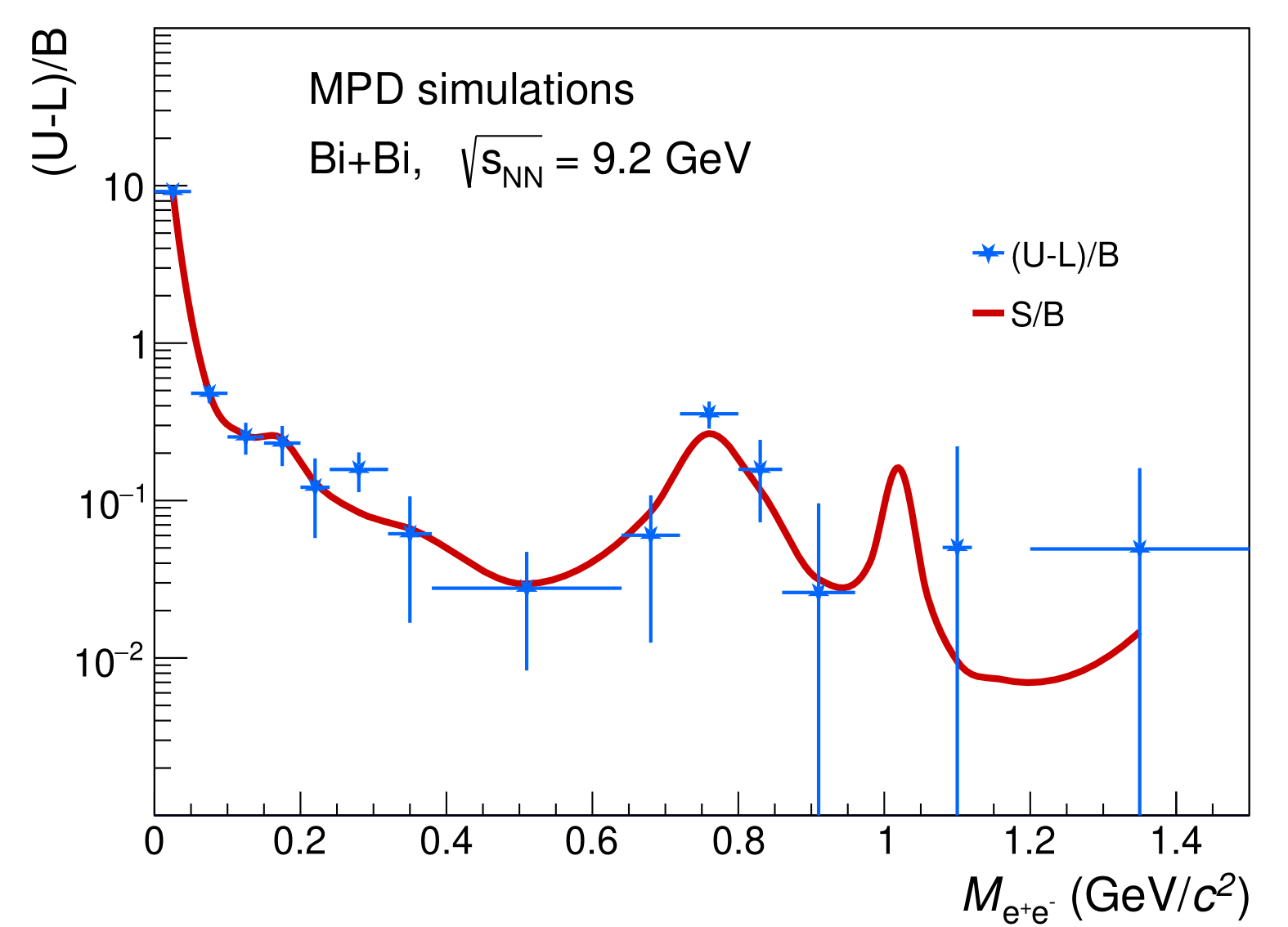}
\caption{Distributions of Unlike sign (U), Like sign (L), measured signal (U-L) and True signal (S) pairs  (upper panel) and measured ((U-L)/B) and true (S/B) signal-to-background ratios (lower panel) in \BiBi collisions at $\snn=$ 9.2 GeV.\label{fig_SB_CTC}}
\end{center}
\end{figure}
\noindent of the particle energy which, coupled with its momentum reconstructed in the \TPC, gives the $E/p$ ratio - a critical discriminant variable for electron-hadron separation.  All the selection cuts applied along the track reconstruction and identification chain are listed in Table~\ref{tabI}.

The selection cuts result in a very good single electron reconstruction efficiency and electron purity, as depicted in Figure~\ref{fig_eff_purity}. The upper panel of the figure shows the gradual decrease of the single electron reconstruction efficiency as the various matching and electron identification cuts are applied. The final single electron reconstruction efficiency of fully reconstructed tracks (identified in \TPC, TOF and \ECAL) with $\pT> 200$ \MeVc amounts to approximately $45\%$. Requirement of \ECAL signal reduces efficiency to zero for $\pT<150$ \MeVc as such tracks do not reach \ECAL. The bottom panel shows an almost 100\% purity of the final electron sample over the entire momentum range. The figure also shows the purity of the electron sample, without the $E/p$ cut enabled by the \ECAL, to be around $80\%$ for $\pT> 1$ \GeVc, highlighting the important role of the \ECAL in reducing the hadronic contamination at high momenta. The reconstruction efficiency drops rapidly for electrons with $\pT <$ 200 \MeVc, reaching 0 at about $\pT =$ 100 \MeVc (for an electron emitted at $y=0$, the minimum momentum to reach the \TOF detector is 110 \MeVc).

A novel pair analysis strategy for the measurement of dileptons at MPD  is being developed aiming at reducing the combinatorial background while keeping a high reconstruction efficiency. To enhance the chances of recognizing electrons originating from $\pi^{0}$ Dalitz decays and gamma conversions, the rapidity phase space of fully reconstructed electrons is divided into a fiducial ($|\eta|< 0.7$) and a veto ($0.7<|\eta|< 1.0$) region. 
% 
% In addition, the total electron sample is categorized in three different sub-samples or pools: Pool-1 includes the fully reconstructed electron tracks within the fiducial area, 
% Pool-2 contains the fully reconstructed electron tracks within the veto area, Pool-3 contains partially reconstructed electron tracks, i.e. electrons reconstructed in the \TPC, and not identified at least in one the outer detectors, the \TOF or \ECAL.
% % %
% The analysis procedure consists of three steps. In the first step, tracks belonging to Pool-1 in a given event are paired with oppositely charged tracks in the same event from Pool-1 and Pool-2. Pairs with invariant mass (\Minv) smaller than 120 \MeVmass are tagged as pairs from $\pi^{0}$ Dalitz or conversions and not used for further pairing. In the second step, the remaining tracks from Pool-1 are paired with oppositely charged tracks  from Pool-3 in the same event and both tracks are removed as a potential Dalitz pair if they have \Minv less than 80 \MeVmass and opening angle, $\theta$, smaller than 5 or 10 degrees. In the third and last step, the remaining tracks from Pool-1 with $\pT>200$ \MeVc are paired among themselves to build the unlike sign (U) and like sign (L) invariant mass spectra. 
%
Fully reconstructed electron tracks in the fiducial area are paired among themselves or with tracks in the veto area. Unlike-sign pairs with $M_\mathrm{ee}< 120$ MeV/$c^2$ are tagged as pairs from $\pi^0$ Dalitz decais or conversions and are not used for further pairing. Furthermore, a proximity cut is applied in the \TPC: fully reconstructed electron tracks in the fiducial area are paired with partially reconstructed electron tracks, i.e. electrons reconstructed in the \TPC, and not identified at least in one of the outer detectors, the \TOF or \ECAL, and both tracks are removed as a potential Dalitz pair if they have $M_\mathrm{ee}<80$ MeV/$c^2$ and opening angle, $\theta< 5^\circ$ or 10$^\circ$. The remaining fully reconstructed electron tracks in the fiducial area, with $\pT > 200$ \MeVc, are paired among themselves to build the unlike sign (U) and like sign (L) invariant mass spectra. 

The combinatorial background B  is approximated by the L sign spectrum and thus the reconstructed signal is obtained as ${\rm S} = {\rm U}-{\rm B} \approx {\rm U}-{\rm L}$, as shown in the upper panel of Figure~\ref{fig_SB_CTC}.  The lower panel shows the differential 
S/B ratio. Currently, a S/B ratio of about $6\%$ is observed over the integrated mass range of $0.2< M_\mathrm{ee}< 1.5$ \GeVcsq. The S/B ratio that is obtained in the same mass range following a standard analysis based on mixing of all tracks from the fiducial region, is about 2.6\%. This demonstrates the advantage that is provided by the adopted analysis strategy.

In summary, the MPD experiment demonstrates a strong capability for comprehensive dielectron measurements, benefiting from excellent electron identification and high electron purity, particularly due to the critical role of the \ECAL in reducing the hadronic contamination. Tools such as machine learning, to further improve the S/B ratio and the signal significance, are currently under development.

\section{Conclusions}
\label{sec:conclusions}
In this work, the physics performance of the MPD experiment was studied in \BiBi collisions at $\snn=9.2$ GeV using large samples of events simulated using UrQMD~\cite{Bleicher:1999xi,Bass:1998ca}, DCM-QGSM-SMM~\cite{Baznat:2019iom}, PHQMD~\cite{Aichelin:2019tnk}, PHSD~\cite{Cassing:2008sv,Cassing:2009vt} and vHLLE+UrQMD~\cite{Karpenko:2013wva,Karpenko:2015xea} event generators.
A wide variety of observables was analyzed, focusing on those expected to be available for an experimental study with the first collected data sets of 50--100 M events. Good MPD performance for the measurement of light flavor hadrons and (hyper)nuclei, photons and (di)electrons is demonstrated. 

The measured differential particle  yields span the phase space in transverse momentum and rapidity, corresponding to $\sim 70\%$ of the total light flavor hadron production cross section. This provides a reduction of systematic uncertainties in the estimation of integrated particle  yields, important for mapping the QCD phase diagram in terms of baryon chemical potential and temperature and for studying particle ratios in the strange sector. Differential \pT measurements cover a wide range from \pT $\sim$ 100 \MeVc to a few \GeVc for most
light hadrons, providing an opportunity to study the dynamics of heavy-ion collisions and to better understand the kinetic freezeout conditions. The ability of the MPD to measure the production of various hadronic resonances over a wide range of lifetimes $\tau \sim 1-45$~fm/$c$ helps to investigate the properties of the late hadronic phase, which may significantly affect the transition and CEP signatures.

%Bla-bla-bla for light nuclei and hypernuclei
The measurements for light nuclei ($d,t$) cover the midrapidity region ($|y|<1$)
and are more restricted in the low $\pT$ range due to losses in the detector material. Nevertheless, accurate reconstruction of the shapes of transverse spectra  and rapidity distributions of nuclei is possible, allowing us to study the freezeout process
and the role of momentum-space correlations in the production of nuclear clusters.

The feasibility studies showed that the measurement of hypertritons is possible with the MPD. The selection criteria for 
$_{\lmb}^3$H reconstruction are optimized for best significance,  the detector efficiency for $_{\lmb}^3$H as a function of $\pT$ is found to vary from 1\% to 7\%  near mid-rapidity. It is shown that the data set volume that could be collected during the first period of data taking is sufficient to obtain enough statistics and to get the yields of hypertritons in several proper time intervals for the measurement of the $_{\lmb}^3$H lifetime.

The  performance of
the MPD has been verified for anisotropic flow
measurements of identified charged pions, kaons,
protons and $\Lambda$  particles as a function of rapidity ($y$) and transverse momentum ($\pT$)
in different centrality classes. A detailed comparison of the results obtained from the analysis of the fully reconstructed data and generator-level data has allowed us 
to conclude that the MPD system will provide detailed differential measurements of 
directed $v_1$, elliptic ($v_2$) and triangular ($v_3$) flows with high efficiency.

Femtoscopic and correlation measurements are important tools to determine the space-time sizes and the hadronization properties of the particle emitting source. The main limitation for an accurate determination of the parameters describing the space-time source of particles is the finite track resolution, which causes a smearing to distinguish single particle tracks. The smearing effect is estimated to be about 4.5 \MeVc and this affects the determination of the femtoscopic parameters within less than 10\% of the generated values. CBFs studies, describing the correlations of oppositely charged particles, were also performed. The rapidity and azimuthal widths of the reconstructed balance functions are shown to coincide within the sample statistics with the corresponding generated functions. 

Photons in the MPD can be reconstructed and identified either in the \ECAL or via the photon conversion method. The first approach provides a reconstruction efficiency close to unity, while the second one ensures purity close to unity. Photons can be used to reconstruct neutral meson yields and correlations. A statistics of 50 M events is sufficient to extract the centrality-dependent neutral pion spectrum in the range $0.1<\pT<4$~\GeVc and an $\eta$-meson yield in minimum bias collisions. The estimated uncertainties of these spectra on the level of a few percent are sufficient to extract the direct photon spectrum. Collective flows of inclusive photons and neutral pions are also extracted and agree with those at the generator level within statistical uncertainties, which at mid-\pT are at percent level for the 50 million events.  

The ability to extract the dilepton spectrum was tested on the example of the UrQMD event generator. Although a sample of 50 million events is not sufficient to extract a high-statistics dilepton spectrum in \BiBi collisions, it provides a realistic estimate of the background levels and the required statistics.

%Although statistics of 50 million events is not sufficient for extraction of dilepton signals in \BiBi collisions, a better understanding of background levels and estimates of the required statistics will be gained.

\section*{Acknowledgments}
\label{sec:ackn}
The authors from the Institute of Physics and Technology, Satbayev University are funded by the Science Committee of the Ministry of Science and Higher Education of the Republic of Kazakhstan, Grant number AP23487706. The authors of Fudan University are supported by
the China Ministry of Science and Technology (MOST) National Key R\&D Program of China Grant number 2024YFA1611002 and the National Natural Science Foundation number 11925502. The authors of the MexNICA Collaboration acknowledge support from UNAM-PAPIIT grant number IG100322, from Consejo Nacional de Humanidades, Ciencia y Tecnolog\'ia (CONAHCyT) grant numbers A1 S 7655 and CF-2023-G-433, from the DAI UAM PIPAIR 2024 projects under Grant number TR2024-800-00744 and the Consejo de la Investigaci\'on Cient\'\i fica de la Universidad Michoacana de San Nicol\'as de Hidalgo (CIC-UMSNH) grant number 18371. The authors of the High School of Economics University acknowledge support from the Basic Research Program at HSE University. The authors of the Kurchatov Institute acknowledge support from a state assignment of the National Research Centre. The work of the NRNU MEPhI group was funded by the Russian Federation Ministry of Science and Higher Education,
project \lq\lq Fundamental and applied research at the NICA (JINR) megascience experimental complex” FSWU-2025-0014. The authors of the North Ossetian State University acknowledge support from grant number FEFN-2024-006. 
The authors of Peter the Great St. Petersburg Polytechnic University (SPbPU) acknowledge support from the Ministry of Science and Higher Education of the Russian Federation (FSEG-2025-0009).
The authors of Saint-Petersburg State University acknowledge support from the University's research project number 103821868. The authors of the VIN\v{C}A Institute of Nuclear Science acknowledge support from the Ministry of Science, Technological Development and Innovation of the Republic of Serbia through the theme 010220.

\section*{References}

%Using BibTeX
\nocite{*}
\bibliographystyle{rmf-style}
\bibliography{main}
%
%%%%%%%%
%
%Introducing references manually
%

%\begin{thebibliography}{99}
%\bibitem{ELK} E. Ley-Koo, Recent progress in confined atoms and molecules: Superintegrability and symmetry breakings, Rev. Mex. Fis. 64 (2018) 326, \url{https://doi.org/10.31349/RevMexFis.64.326}
%
%\bibitem{Griffiths} D.J. Griffiths, Introduction to Electrodynamics, 2nd ed. (Prentice Hall, Englewood Cliffs, NJ, 1989), pp. 331–334.
%
%\end{thebibliography}
\end{multicols}
\end{document}